\documentclass[11pt]{article}
\usepackage{fullpage,amsmath,amssymb,graphicx,natbib,xcolor,todonotes,mathtools,enumitem,graphicx,url}
\usepackage[linesnumbered,ruled,vlined]{algorithm2e}
\usepackage{todonotes}
\SetKwInput{KwInput}{Input}                
\SetKwInput{KwOutput}{Output}              

\begin{document}
\title{\textsc{\normalsize{{\bf AN ECONOMETRIC PERSPECTIVE ON ALGORITHMIC SUBSAMPLING} \bigskip}}}

\author{\textsc{Sokbae Lee}\thanks{Department of Economics, Columbia University and Institute for Fiscal Studies} \and \textsc{Serena Ng}\thanks{Department of Economics, Columbia University and NBER\newline
The authors would like to thank David Woodruff and Shusen Wang for helpful discussions.
The first author would like to thank the European Research Council for financial support (ERC-2014-CoG- 646917-ROMIA) and the UK Economic and Social Research Council for research grant (ES/P008909/1) to the CeMMAP.
The second author would like to thank the National Science Foundation for financial support  (SES: 1558623).}
}
\date{\bigskip\today}
\def\pconv{\smash{\mathop{\longrightarrow}\limits^p}}     
\def\dconv{\smash{\mathop{\longrightarrow}\limits^d}}
\def\iid{\smash{\mathop{\sim}\limits^{iid}}}
\def\argmax{\mbox{argmax}}
\def\argmin{\mbox{argmin}}
\def\max{\mbox{max}}
\def\min{\mbox{min}}
\def\tilde{\widetilde}
\newcommand{\norm}[1]{\left\lVert#1\right\rVert}
\newtheorem{definition}{Definition}
\newtheorem{lemma}{Lemma}
\newtheorem{proposition}{Proposition}
\newtheorem{theorem}{Theorem}
\newtheorem{corollary}{Corollary}

\maketitle
\let\Large=\normalsize

\begin{abstract}

Datasets that are terabytes in size are increasingly common, but computer bottlenecks often frustrate a complete analysis of the data.  While more data are better than less, diminishing returns suggest that we may not need terabytes of  data to estimate a parameter or test a hypothesis. But which rows of data should we analyze, and might an arbitrary subset of rows preserve the features of the original data?  This paper reviews a line of work that is grounded in theoretical computer science and numerical linear algebra, and which finds that an algorithmically desirable {\em sketch}, which is a randomly chosen subset of the data, must preserve the eigenstructure of the data, a property known as a {\em subspace embedding}. Building on this work, we study how  prediction and inference can be affected by data sketching within a  linear regression setup. We show that  the  sketching error is small compared to the sample size effect which a researcher can control. As a sketch size that is algorithmically optimal may not be suitable for prediction and inference, we use statistical arguments to provide  `inference conscious' guides to  the sketch size. When appropriately implemented, an estimator that pools over different sketches can be nearly as efficient as the infeasible one using the full sample.

\end{abstract}



\thispagestyle{empty}
\setcounter{page}{0}
\bibliographystyle{harvard}

\newpage
\baselineskip=18.0pt
\section{Introduction}
The availability of terabytes of data for economic analysis is increasingly common. But analyzing large datasets is time consuming and sometimes beyond the limits of our computers. The need to work around the data bottlenecks was no smaller decades ago when the data were in megabytes than it is today when data are in terabytes and petabytes.  One way to alleviate the bottleneck is to work with a {\em sketch} of the data.\footnote{The term `synopsis' and `coresets' have also been used. See \citet{comrode-etal:NOW}, and \citet{ahv}. We generically refer to these as sketches.} These are data sets of smaller dimensions and yet representative of the original data.  We study how the design of linear sketches affects estimation and inference in the context of the linear regression model. Our formal statistical analysis complements those in the theoretical computer science and numerical linear algebra   derived using different notions of accuracy and whose focus is computation efficiency.

 There are several motivations for forming sketches of the data from the full sample. If the  data are too expensive to store and/or too large to fit into computer memory, the data would be of limited practical use. It might be cost effective in some cases to get a sense from a smaller sample whether an expensive  test based on the full sample is worth proceeding. Debugging is certainly faster with fewer observations. A smaller dataset can be adequate  while a researcher is learning how to specify the regression model, as loading a gigabyte of data is much faster than a terabyte even we have enough computer memory to do so.  With  confidentiality reasons, one might only want to circulate  a subset rather than the full set of data. 

 For  a sketch of the data to  be useful, the sketch must preserve the characteristics of the original data. Early work in the statistics literature used a sketching method known as  `data squashing'. The idea is to approximate the likelihood function by merging  data points with similar likelihood profiles. There are two different ways to  squash the data. One approach is to construct subsamples randomly.  While these methods work well for the application under investigation, its general properties are not well understood. An alternative is to take the data structure into account. \citet{datasquashing}  also forms multivariate bins of the data, but they match low order moments within the bin by non-linear optimization. \citet{owen-squashing}   reweighs a random sample of $X$  to fit the moments using empirical likelihood estimation.  \citet{madigan-squashing} uses likelihood-based clustering to  select data points that match the target distribution. While theoretically appealing,  modeling the likelihood profiles can itself be time consuming and not easily scalable.

Data sketching is of also interest to computer scientists because  they are frequently required to provide  summaries  (such as frequency,  mean, and maximum) of data as they  stream by continuously.\footnote{The seminal paper on frequency moments is \citet{ams:99}. For a review of the literature, he reader is referred to \citet{comrode-etal:NOW}.}   Instead of an exact answer which would be costly to compute,  {\em pass-efficient} randomized algorithms are designed to run fast, requires little storage, and guarantee the correct answer with a certain probability. But this is precisely the underlying premise of data sketching in statistical analysis.\footnote{Pass-efficient algorithms read in data at most a constant number of times. A computational method is referred to as a streaming model if only one pass is needed.} 

Though randomized algorithms are increasingly used for sketching  in a wide range of applications,  the concept remains largely unknown to economists except for a brief exposition  in \citet{ng-worldcongress}.  This paper provides a gentle introduction to these algorithms in Sections 2 to 4.  To our knowledge,  this is the first review on sketching  in the econometrics literature. We  will use the term  {\em algorithmic subsampling} to refer to randomized algorithms designed for the purpose of sketching, to  distinguish them from   bootstrap and subsampling methods developed  for frequentist inference.  In repeated sampling, we only observe one sample   drawn from the population. Here,  the  complete data can be thought of as the population which we observe, but we can only use a subsample.  Algorithmic subsampling does not make distributional assumptions, and  balancing between fast computation and favorable  worst case approximation error often leads to algorithms that are  oblivious to the properties the data. In contrast,   exploiting the probabilistic structure is often an important aspect of econometric modeling. 

Perhaps the most important tension between the algorithmic and the statistical view is that while fast and efficient computation tend to favor sketches with few rows, efficient estimation and inference inevitably  favor using as many rows as possible.  Sampling schemes that are optimal from an algorithmic perspective may not be desirable from an econometric perspective, but it is entirely possible for schemes that are not  algorithmically optimal to be statistically desirable. As there are many  open questions about the usefulness of these  sampling  schemes for  statistical modeling,  there is  an increased interest in these methods within the statistics community. Recent surveys on sketching with a regression focus include \citet{ahfock-astle-richardson} and \citet{geppert-etal:17}, among others.   Each paper offers distinct insights, and the present paper is no exception. 

Our focus is on efficiency of the estimates for prediction and inference  within the context of the  linear regression model.  Analytical and practical considerations confine our focus eventually back to  uniform sampling, and to a smaller extent, an algorithm known as  the countsketch.  The results in Sections 5 and 6  are new. It will be shown  that  data sketching has two effects on estimation, one due to sample size, and one due to approximation error, with the former dominating the later in all quantities of empirical interest. Though   the sample size effect has direct implications for the power of any statistical test, it is at the discretion of a researcher. We show that moment restrictions can be used to guide the sketch size, with fewer rows being needed when  more moments exist. By targeting the power of a test at a prespecified alternative,  the size of the sketch can also be tuned so as not to incur excessive  power loss in hypothesis testing.  We refer to this as the `inference conscious' sketch size.

There is an inevitable trade-off between computation cost and statistical efficiency, but the statistical loss from using fewer rows of data can be alleviated by combining estimates from different sketches. By the principle of `divide and conquer', running several estimators in parallel can be statistically efficient and still computationally  inexpensive. Both uniform sampling and the countsketch are amenable to parallel processing  which facilitates averaging of quantities  computed from different sketches.   We assess two ways of combining estimators: one that averages the  parameter estimates, and one that averages  test statistics.   Regardless of how information from the different sketches are combined,  pooling over subsamples always  provides more efficient estimates and more powerful tests. It is in fact possible to bring the power of a test arbitrarily close to the one using the full sample, as will be illustrated in Section 6.

\subsection{Motivating Examples}

The sketching problem can be summarized as follows. Given an original matrix $A\in \mathbb R^{n\times d}$, we are  interested in  $\tilde A\in \mathbb R^{m\times d}$ constructed  as
\[ \tilde A=\Pi A\]
where  $\Pi\in \mathbb R^{m\times n}$, $m<n$. In a linear regression setting, $A=[y\; X]$ where $y$ is the dependent variable, and $X$ indicates the regressors. Computation of the least squares estimator takes $O(nd^2)$ time which becomes costly when  the number of rows, $n$  is large. Non-parametric regressions fit into this setup if $X$ is a matrix of sieve basis.  Interest therefore arises to use fewer rows of $A$ without sacrificing too much  information. 

To motivate why  the choice of the sampling scheme (ie. $\Pi$) matters,
 consider as an example a $5\times 2$ matrix
\[ A=\begin{pmatrix} 1 & 0 & -.25 & .25 & 0 \\ 0  & 1 & .5 & -.5 & 0 \end{pmatrix}^T.\]
The rows have different information content as the row norm is $(1, 1, 0.559, 0.559,0)^T$.
Consider now three $2\times 2$ $\tilde A$ matrices constructed as follows:
\begin{align*}
 \Pi_1&=\begin{pmatrix} 1 & 0 & 0 & 0 &0\\ 0 & 1 & 0 & 0 &0 \end{pmatrix},
 \quad &\tilde A_1=\Pi_1 A=\begin{pmatrix} 1 & 0 \\ 0 & 1\end{pmatrix}\\
\Pi_2&=\begin{pmatrix} 1 & 0 & 0 & 0 & 0\\ 0 & 0 & 0 & 0 &1\end{pmatrix}, \quad &\tilde A_2=\Pi_2 A=\begin{pmatrix} 1 & 0 \\ 0 & 0\end{pmatrix}\\
\Pi_3&=\begin{pmatrix} 0 & 0 & 1 & 1 & 0\\ 0 & 1 & -1 & 1 &1\end{pmatrix}, \quad &\tilde A_3=\Pi_3 A=\begin{pmatrix} 0 & 0 \\ .5 & 0\end{pmatrix}.
\end{align*}
Of the three sketches, only $\Pi_1$ preserves the rank of $A$. The sketch defined by  $\Pi_2$ fails because it chooses row 5 which has no information. The third sketch is obtained by taking a linear combination of  rows that do not have independent information.   The point is that unless $\Pi$ is chosen appropriately, $\tilde A$ may not have the same rank as $A$. 

Of course, when  $m$ is large, changing rank is much less likely and one may also wonder if this pen and pencil problem can ever arise in practice. Consider now estimation of a Mincer equation which has the logarithm of \textsc{wage} as the dependent variable, estimated using the \citet{ipums} dataset which provides  a preliminary but complete count data for the 1940 U.S. Census. This data is of interest because it was the first census with information on wages and salary income.  For illustration, we use a sample of $n=24$ million white men between the age of 16 and 64 as the `full sample'.  The predictors that can be considered are  years of  education, denoted (\textsc{edu}), and potential experience, denoted (\textsc{exp}). 

\begin{figure}[htbp]
    \caption{Distribution of Potential Experience}
    \label{fig-example}
    \begin{center}
        \includegraphics[width=6in,height=1.750in]{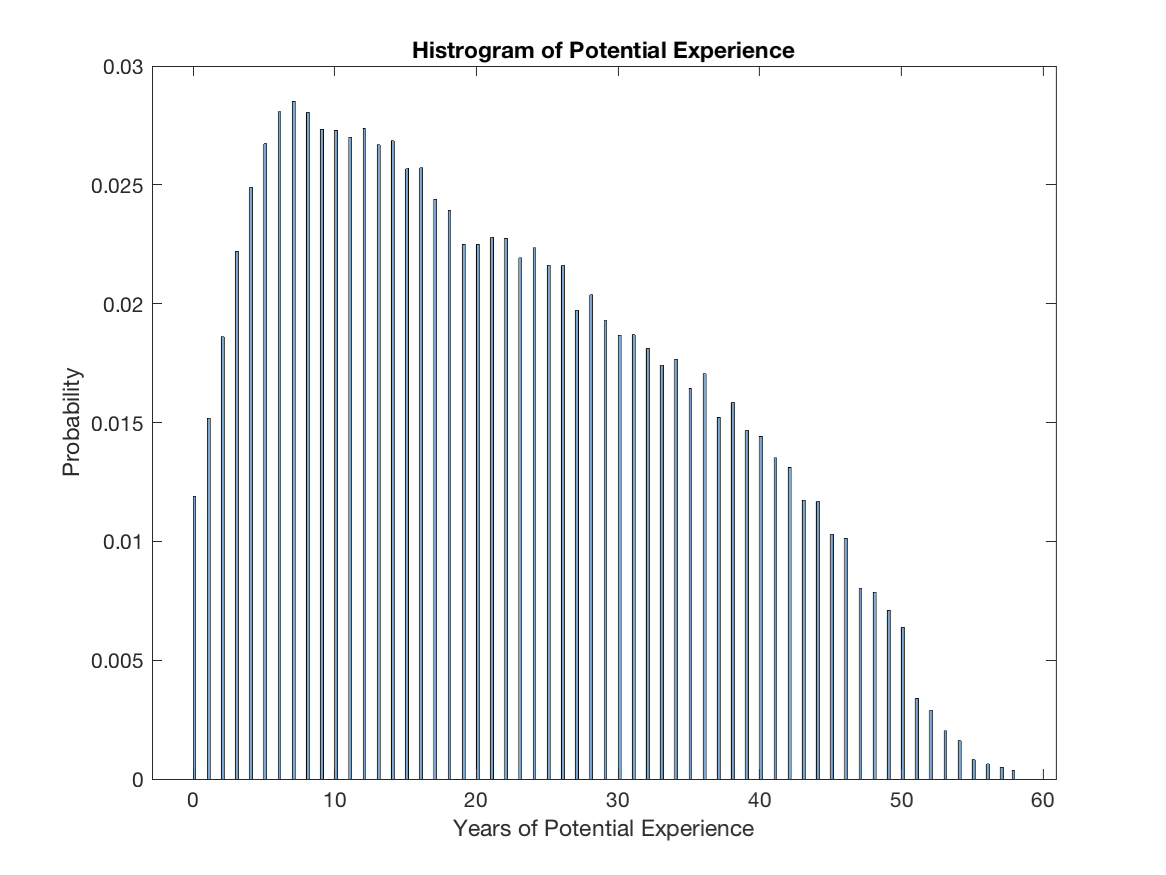}
    \end{center}
\end{figure}

Two Mincer equations with different covariates are considered:
\begin{subequations}
\begin{eqnarray}
\label{eq:mincer1}
\text{log wage} &=&\beta_0 + \beta_1 \text{edu} + \beta_2 \text{exp} + \beta_3 \text{exp}^2 + \text{error}\\
\text{log wage} &=&\beta_0 + \beta_1 \text{edu} + \sum_{j=0}^{11} \beta_{2+j} 1 \{ 5j \leq \text{exp} < 5(j+1) \} + \text{error}.
\label{eq:mincer2}
\end{eqnarray}
\end{subequations}
 Model (\ref{eq:mincer1})  uses \textsc{exp} and $\textsc{exp}^2$ as control variables. Model (\ref{eq:mincer2})  replaces potential experience with indicators of experience in five year intervals. Even though there are three predictors including the intercept, the number of covariates $K$ is 4 in the first model and 14 in the second. In both cases, the parameter of interest is the coefficient  for years of education ($\beta_1$).  The full sample estimate of $\beta_1$ is 0.12145 in specification (\ref{eq:mincer1}) and 0.12401 in specification (\ref{eq:mincer2}).

Figure \ref{fig-example} shows the histogram of $\textsc{exp}$.   The  values of \textsc{exp} range from 0 to 58. The problem in this example arises because    there are few observations with over 50 years of experience.   Hence there is no guarantee that an arbitrary subsample  will include observations with $\textsc{exp}>50$. Without such observations,  the subsampled covariate matrix may not have full rank.  Specification (\ref{eq:mincer2}) is more vulnerable to this problem  especially when $m$ is small.

\begin{center}
\begin{figure}[htbp]
    \caption{Distributions of Estimates from Uniform Sampling without Replacement}
    \label{fig-subsampleU}
        \includegraphics[width=3.0in,height=1.7in]{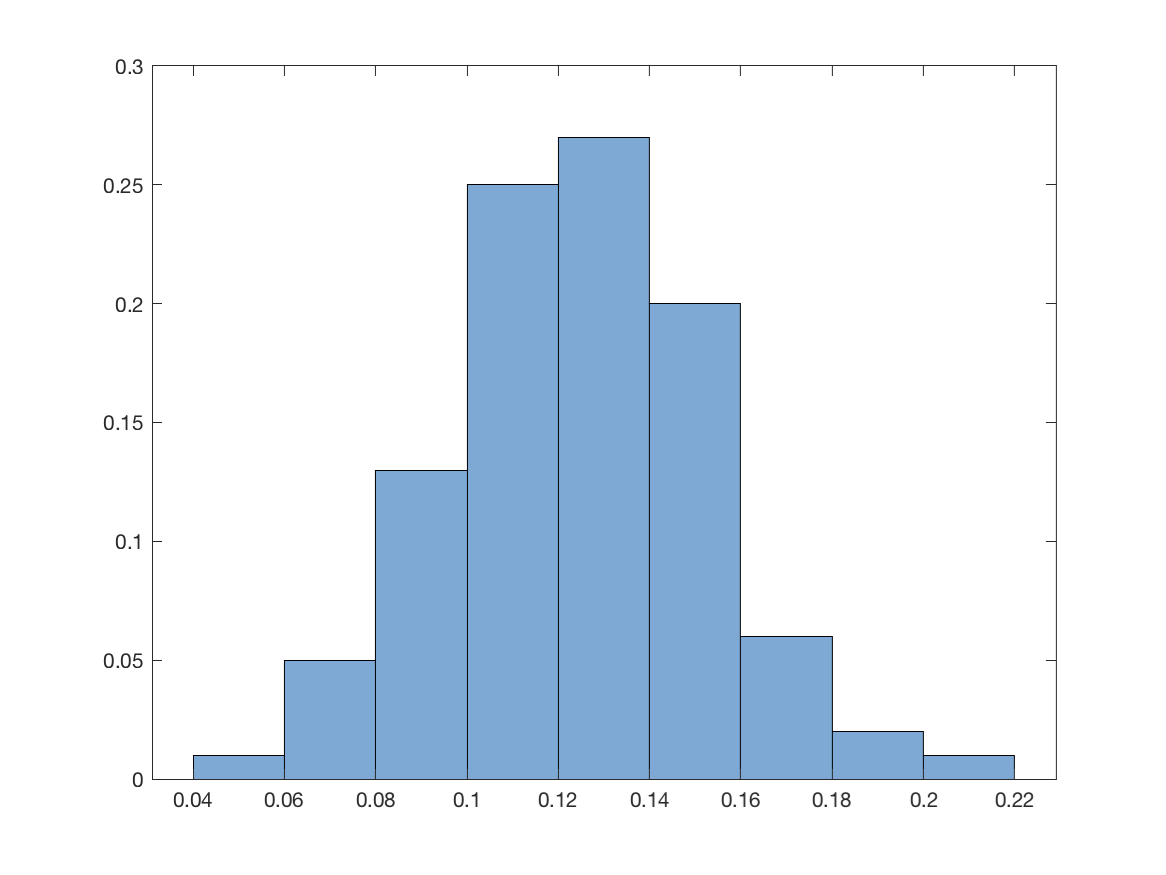}
        \includegraphics[width=3.0in,height=1.7in]{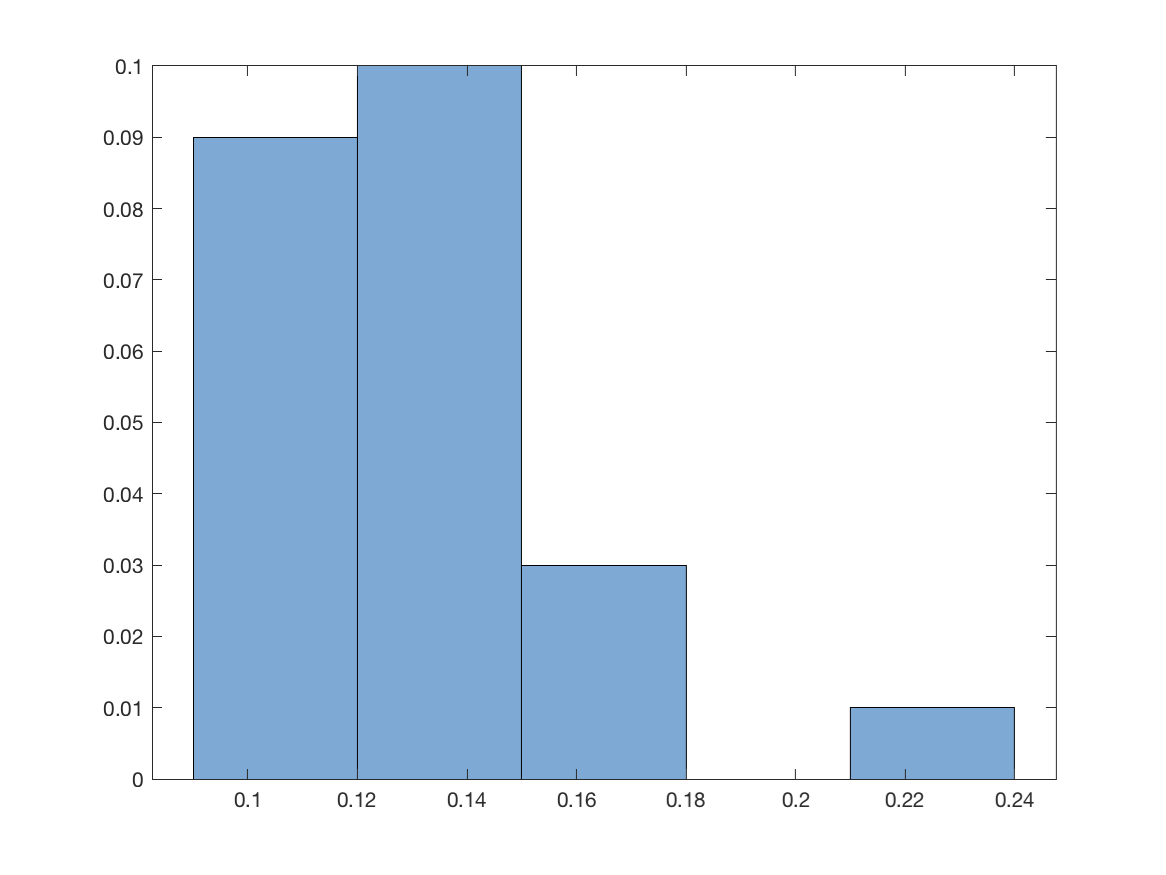}
        \end{figure}
    \end{center}

\begin{center}
\begin{figure}[htbp]
    \caption{Distributions of Estimates from CountSketch Sampling}
    \label{fig-subsampleCS}
        \includegraphics[width=3.0in,height=1.7in]{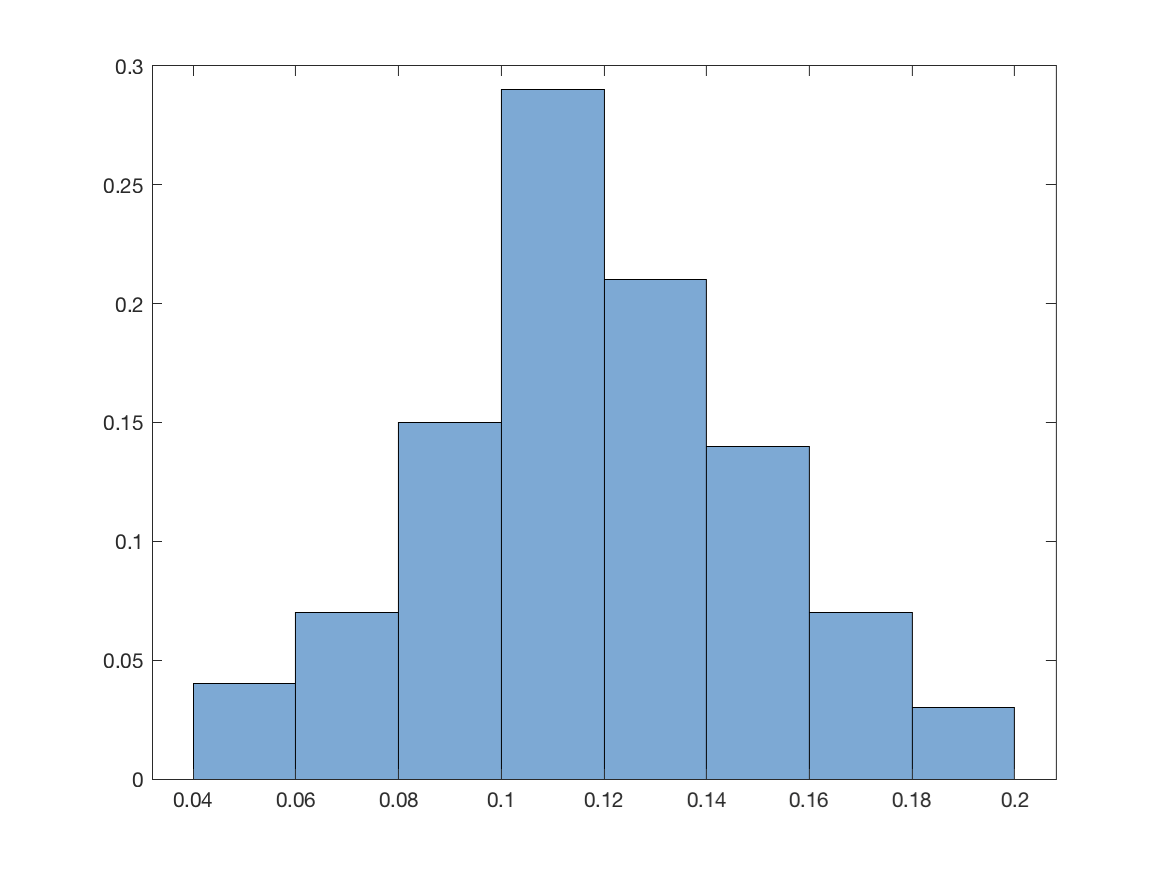}
        \includegraphics[width=3.0in,height=1.7in]{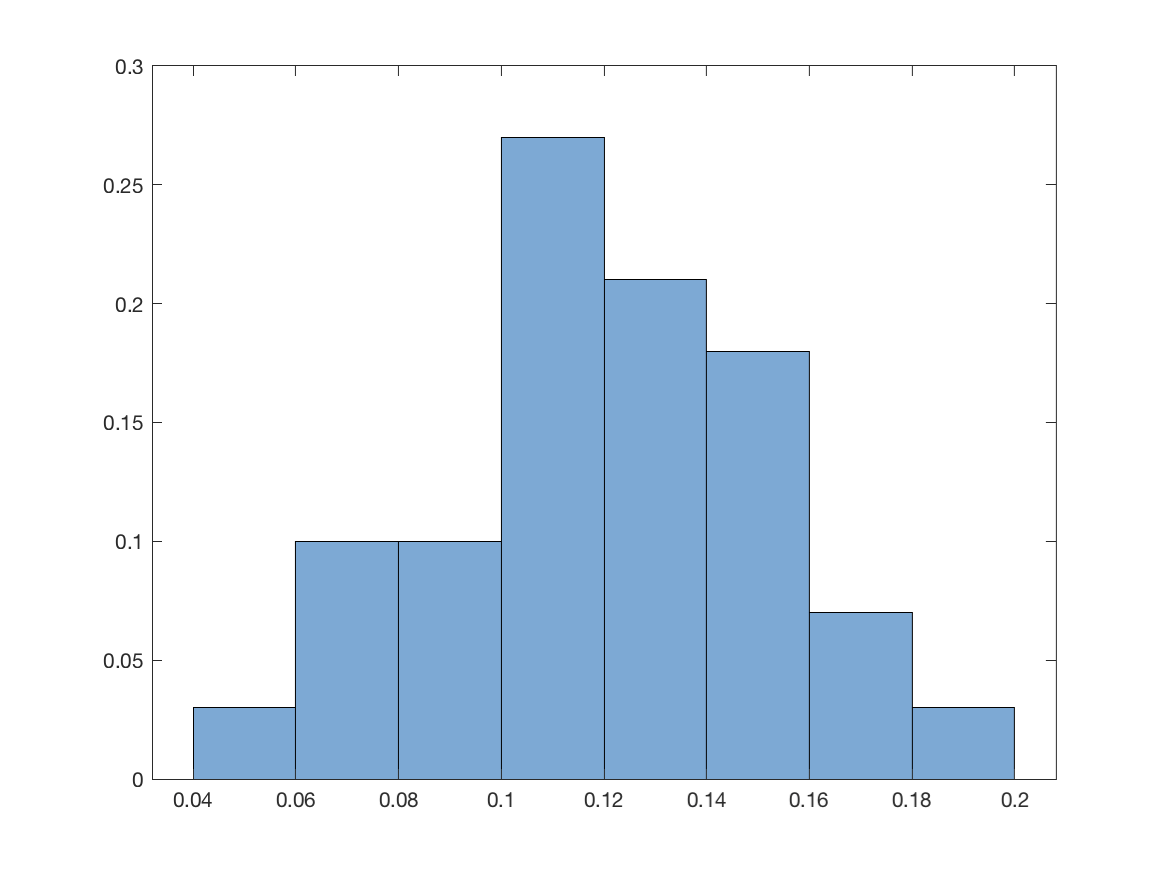}        
\end{figure}
\end{center}

We verify that  rank failure is empirically plausible in a small experiment with sketches of size $m=100$ extracted using two sampling schemes. The first method is uniform sampling without replacement which is commonly used  in economic applications. The second is the countsketch which  will be further explained below.   Figure \ref{fig-subsampleU} and Figure \ref{fig-subsampleCS} show the histograms of subsample estimates for uniform sampling and  the countsketch, respectively. The left panel is for specification (\ref{eq:mincer1}) and the right panel is for specification (\ref{eq:mincer2}). In our experiments,   singular matrices never occurred with specification (\ref{eq:mincer1});  the OLS estimates can be computed using both sampling algorithms and  both performed pretty well. However, uniform sampling without replacement produced  singular matrices for specification (\ref{eq:mincer2}) 77\% of the time. The estimates seem quite different from the full sample estimates, suggesting not  only bias in the estimates, but also that the bias might not be random. In contrast, the countsketch failed only once out of 100 replications. The estimates are shown in the right panel of Figure \ref{fig-subsampleCS}  excluding the singular case.

This phenomenon can be replicated in a Monte Carlo experiment with $K=3$ normally distributed predictors. Instead of $X_3$, it is assumed that we only observe a value of one if $X_3$ is three standard deviation from the mean. Together with an intercept, there are four regressors. As in the Mincer equation, the regressor matrix has a reduced rank of 3 with probability of 0.58, 0.25, 0.076  when $m=200, 500, 1000$ rows are sampled uniformly; it  is always full rank only when $m=2000$.  In contrast, the countsketch never encounters this problem even with $m=100$. The  simple example underscores the point  that the choice of sampling scheme matters. As will be seen below, the issue remains in  a more elaborate regression  with  several hundred covariates.  This motivates the  need to better understand how to form sketches  for estimation and inference.

    \vspace*{-.750in}

\section{Matrix Sketching}

This section presents the key concepts in algorithmic sampling. The material is based heavily on  the  monographs  by \citet{mahoney-book} and \citet{woodruff-book}, as well as the seminal work of \citet{dmm:06}, and subsequent refinements developed in \citet{dmms},  \citet{nelson-nguyen:13}, \citet{nelson-nguyen:14}, \citet{cohen-nelson-woodruff:15}, \cite{wang-gittens-mahoney:18}, among many others.  

We begin by setting up the notation. Consider an $n\times d$ matrix positive-definite $A$. Let  $A^{(j)}$ denote its $j$-th column of $A$ and $A_{(i)}$ be  its $i$-th row. Then
\[ A=\begin{pmatrix}
A_{(1)} \\
\vdots\\
A_{(n)} \end{pmatrix} =\begin{pmatrix} A^{(1)} &  \ldots & A^{(d)}
\end{pmatrix}
\]
and $A^TA=\sum_{i=1}^n A^T_{(i)} A_{(i)}$. The  singular value decomposition of $A$ is $A=U\Sigma V^T$ where
$U$ and $V$ are the left and right eigenvectors of dimensions ($n\times d$) and $(d\times d)$ respectively. The matrix $\Sigma$ is $d\times d$ diagonal with  entries containing the  singular values of $A$ denoted  $\sigma_{1},\ldots,\sigma_{d}$, which are ordered such that $\sigma_{1}$ is the largest.   Since $A^T A$ is positive definite, its $k$-th eigenvalue 
$\omega_k(A^T A)$ equals $\sigma_k(A^TA ) =\sigma_k^2(A)$, for $k=1,\ldots d$.  The best rank $k$ approximation of $A$ is given by
\[A_k=U_{k}U_{k}^T A\equiv P_{U_k}A\]
where $U_{k}$ is an $n\times  k$ orthonormal matrix of left singular vectors corresponding to the $k$ largest singular values of $A$, and $P_{U_k}=U_{k} U_{k}^T$ is the projection matrix.

The   Frobenius norm (an average type criterion) is
$
\| A \|_F=\sqrt{\sum_{i=1}^n\sum_{j=1}^d |A_{ij}|^2}=\sqrt{\sum_{i=1}^k\sigma_{i}^2}.$
 The spectral norm (a worse-case  type criterion) is
 $
\| A \|_2 =\sup_{\|x\|_2=1}   \|Ax\|_2=\sqrt{\sigma^2_{1}},
 $
where $\|x\|_2$ is the Euclidean norm of a vector $x$. The spectral norm is bounded above by the Frobenius norm since
$
\| A \|^2_2=|\sigma_{1 }|^2\le \sum_{i=1}^n\sum_{j=1}^d |A_{ij}|^2=\| A\| ^2_F=\sum_{i=1}^d \sigma_{i}^2.
$

Let $f$ and $g$ be real valued functions defined on some unbounded subset of real positive numbers $n$. We say that
$g(n)=O(f(n))$ if $|g(n)|\le k |f(n)|$ for some constant $k$  for all $n \ge n_0$.
This means that $g(n)$ is at most  a constant multiple of $ f(n)$ for sufficiently large values of $n$.  We say that  $g(n)=\Omega(f(n))$ if $g(n)\ge kf(n)$ for all $n \ge n_0$.
This means that  $g(n)$ is at least $kf(n)$ for some constant $k$.
We say that $g(n)=\Theta(f(n))$ if $k_1 f(n) \le g(n) \le k_2 f(n)$ for  all $n\ge n_0$.
This means that $g(n)$ is at least $k_1f(n)$ and at most $k_2f(n)$.

\subsection{Approximate Matrix Multiplication}
 Suppose we are given two matrices, $A\in \mathbb R^{n\times d}$ and $B\in \mathbb R^{n\times p}$ and are interested in the $d\times p$ matrix $C=A^TB$. The textbook approach  is to compute 
each element of $C$ by summing over dot products:
\[C_{ij}=[A^TB]_{ij}=\sum_{k=1}^n A^T_{ik} B_{kj}.\]
Equivalently, each element is  the inner product of two vectors $A_{(i)}$ and $B^{(j)}$. Computing the entire $C$ entails three loops through   $i\in[1, d]$, $j\in[1, p]$, and $k\in[1, n]$.   An algorithmically more efficient approach is to form $C$ from outer products:
\[ C=\underbrace{A^TB}_{d\times p}  =\sum_{i=1}^n \underbrace{A_{(i)}^T B_{(i)}}_{(d\times 1)\times (1\times p)},\] 
 making $C$  a  sum of
 $n$  matrices each of rank-1. 
 Viewing  $C$ as a sum of $n$ terms  suggests to approximate it by summing $m< n$ terms only. But which $m$   amongst the  $\frac{n!}{m!(n-m)!} $ possible terms to sum?  Consider the following Approximate Matrix Multiplication algorithm (AMM).  Let $p_{j}$ be the probability that row $j$ will be sampled.
\bigskip

{\bf Algorithm AMM:} 
\\
 \begin{algorithm}[H]
\DontPrintSemicolon

\KwInput{ $A\in \mathbb R^{n\times d}$, $B\in\mathbb  R^{n\times p}$,  $m>0$, $p=(p_1,\ldots,p_n)$.}
\For{$s=1:m$}
{sample $k_s\in[1,\;\ldots n]$ with probability $p_{k_s}$ independently with replacement;\\

set $\tilde A_{(s)}=\frac{1}{\sqrt{m p_{k_s}}} A_{(k_s)}$ and
 $\tilde B_{(s)}=\frac{1}{\sqrt{m p_{k_s}}} B_{(k_s)} $}

 \KwOutput{$\tilde C=\tilde A^T \tilde B$.}

\end{algorithm}

\bigskip 
 
 The algorithm essentially produces
 \begin{eqnarray}
 \tilde C
 &=&(\Pi A)^T \Pi B=\frac{1}{m}\sum_{s=1}^{m} \frac{1}{p_{k_s}} A^T_{(k_s)} B_{(k_s)}
 \label{eq:samplingC}
 \end{eqnarray}
 where  $k_s$ denotes the index for the non-zero entry in the $s$ row of the matrix
 \[ \Pi=\frac{1}{\sqrt{m}}
 \begin{pmatrix} 0  & \frac{1}{\sqrt{p_{k_1}}} & 0 & \ldots & 0\\
 \ldots & \ldots & \ldots & \ldots & \ldots & \\
 0 & 0 &\ldots  & \frac{1}{\sqrt{p_{k_m}}} & 0 \end{pmatrix}. \]  The $\Pi$ matrix  only has  only one non-zero element per row, and the $(i,j)$-th entry $\Pi_{ij}=\frac{1}{\sqrt{mp_j}}$ with probability $p_j$.  In the case of uniform sampling with  $p_k=\frac{1}{n}$ for all $i$, $\Pi$ reduces to  a sampling matrix scaled by $\frac{\sqrt{n}}{\sqrt{m}}$.
 
 While $\tilde C$ defined by (\ref{eq:samplingC}) is recognized in econometrics as  the estimator of \citet{horvitz-thompson}  which uses inverse probability weighting to account for different proportions of observations in stratified sampling, $\tilde C$   is a sketch of $C$ produced by the  {\em Monte Carlo} algorithm AMM in the theoretical computer science literature.\footnote{In  \citet{mitzenmacher-upfal}, a Monte Carlo algorithm is  a randomized algorithm that may fail or return an incorrect answer but whose time complexity is deterministic and does not depend on the particular sampling. This contrasts with a Las Vegas algorithm which always returns the correct answer but whose time complexity is random. See also \citet{ipsen-etal:11}.}  The Monte-Carlo aspect is easily understood if we take $A$ and $B$ to be $n\times 1 $ vectors. Then $A^T B=\sum_{i=1}^n A^T_{(i)} B_{(i)}=\sum_{i=1}^n f(i)\approx \int_0^n f(x) dx=f(a) n$ where the last step follows from mean-value theorem for $0<a<n$. Approximating $f(a)$ by $\frac{1}{m}\sum_{s=1}^m f(k_s)$ gives  $\frac{n}{m} \sum_{s=1}^m f(k_s)$ as the Monte Carlo estimate of $\int_0^n f(x)dx$. 

 Two properties of $\tilde C$ produced by AMM are noteworthy.    Under independent sampling,   
 \[
 \mathbb E\bigg[ \frac{1}{m} \sum_{s=1}^m \frac{A_{i k_s} B _{k_s j}}{p_{k_s}} \bigg]= 
 \frac{1}{m} \sum_{s=1}^m \sum_{k=1}^n p_k \frac{A_{ik} B_{kj} }{p_k}=[A^T B]_{ij}.
 \]
  Hence regardless of the sampling distribution, $\tilde C$ is unbiased. 
The  variance of $\tilde C$  defined  in terms of the Frobenius norm is
  \begin{eqnarray*}
  \mathbb{E} \Big[ \| \tilde C - C \|_F^2 \Big] 
  &=&\frac{1}{m} \sum_{k=1}^ n \frac{1}{p_k}  \| A^T_{(k)} \| _2^2  \| B_{(k)} \| _2^2 - \frac{1}{m} \| C \|_F^2
  \end{eqnarray*}
  which  depends on the sampling distribution $p$. \citet[Theorem 1]{drineas-kannan-mahoney:06} shows that minimizing $\sum_{k=1}^n \frac{1}{p_k} \|A^T_{(k)}\|_2^2 \|B_{(k)}\|_2^2 $ with respect to $p$ subject to the constraint  $\sum_{k=1}^n p_k=1$ gives\footnote{The first order condition satifies $0=-\frac{1}{p_k^2} \|A_{(k)}^T\|_2^2\|B_{(k)}\|_2^2+\lambda$. Solving for $\sqrt{\lambda}$ and imposing the constraint gives the result stated.  \citet{ipsen-etal:11} derives probabilities that minimize expected variance for given distribution of the matrix elements.} 
\begin{eqnarray*}
\label{eq:optimal-p}
 p_k = \frac{ \|  A_{(k)} \| _2  \| B_{(k)} \| _2}
{\sum_{s=1}^n  \|  A_{(s)} \| _2  \| B_{(s)} \| _2}.
\end{eqnarray*}
This optimal $p$ yields a variance of
\begin{align*}
\mathbb{E} \Big[ \| \tilde C - C \|_F^2 \Big] &\le \frac{1}{m} \Big[ \sum_{k=1}^n   \| A_{(k)} \| _2  \| B_{(k)} \| _2 \Big]^2 \le \frac{1}{m}  \| A \| _F^2 \| B \| _F^2.
\end{align*}
It follows from Markov's inequality that for given error of size $\varepsilon$ and failure probability $\delta>0$,
\[
P\bigg( \| \tilde C-C \| _F^2>\varepsilon^2  \| A \| _F^2  \| B \| _F^2\bigg) < \frac{\mathbb{E} \Big[ \| \tilde C - C \|_F^2 \Big] }{\varepsilon^2 \| A \| _F^2  \| B \| _F^2 }< \frac{1}{m\varepsilon^2},
\]
implying that  to have an approximation error no larger than $\varepsilon$ with probability $1-\delta$,
 the  number of rows used in the approximation must satisfy
$m=\Omega(\frac{1}{\delta\varepsilon^2}).$

The approximate matrix multiplication result $\|  A^T B-\tilde A^T\tilde B \| _F\le \varepsilon  \| A \| _F \| B \| _F
$ is  the building block of  many of the theoretical results to follow. The result also holds under the spectral norm since it is upper bounded by the Frobenius norm. Since $A_{(i)}^T B_{(i)}$ is a rank one matrix, $\|A_{(i)}^TB_{(i)}\|_2=\|A_{(i)}^T \|_2 \|B_{(i)}\|_2$. Many of the results to follow are in  spectral norm because it is simpler to work with a  product of two Euclidean vector norms.  Furthermore, putting $A=B$, we have
 \[ P(\|(\Pi A)^T (\Pi A) -A^T A)\|_2\ge \epsilon \|A\|_2^2)<\delta.\]
 One may think of the goal of AMM as  preserving the second moment properties of $A$. The challenge in practice is to understand the  conditions that validate the approximation. For example, even though uniform sampling is the simplest of sampling schemes,  it cannot be used blindly.   Intuitively,  uniform sampling treats all data points equally, and  when information in the rows are not uniformly dispersed,  the influential rows will likely be omitted.  From the above derivations, we see that
   $\text{var}(\tilde C)= O(\frac{n}{m})$ when $p_k=\frac{1}{n}$,  which   can be prohibitively large.  The   Mincer equation in the Introduction illustrates the pitfall with uniform sampling when $m$ is too small, but that the problem can by and large be alleviated when  $m>2000$. Hence, care must be taken in using the algorithmic sampling schemes.  We will provide some guides below.

\subsection{Subspace Embedding}
To study the properties of the least squares estimates  using sketched data, we first need to  make clear what features of $A$  need to be preserved in $\tilde A$.
 Formally, the requirement is that  $\Pi$   has a `subspace embedding' property. An embedding is a linear transformation of the data that has the Johnson-Lindenstrauss (JL) property, and a subspace embedding is a matrix generalization of an embedding. Hence it is useful to start with the celebrated JL Lemma.


The JL Lemma,   due to \citet{johnson-lindenstrauss},  is usually written for linear maps that reduce the number of columns from an $n\times d$ matrix (i.e., $d$) to $k$.  Given that our interest is ultimately in reducing the number of rows from $n$ to $m$ while keeping $d$ fixed,  we state the JL Lemma as follows:

\begin{lemma}[JL Lemma]
\label{lem:jllemma}
Let $0<\epsilon<1$ and $\{a_1,\ldots,a_d\}$ be a set of $d$ points in $\mathbb R^n$ with $n>d$.  Let $m\ge 8 \log d /\epsilon^2$.  There exists a linear map $\Pi: \mathbb R^n\rightarrow\mathbb R^m$ such that $ \forall a_i,a_j$
\[
 (1-\epsilon)||a_i-a_j||^2_2 \le || \Pi a_i-\Pi a_j||^2_2\le (1+\epsilon) ||a_i-a_j||^2_2  .\]
\end{lemma}

In words, the Lemma states that  every set of $d$ points in Euclidean space of dimension $n$ can be represented by a Euclidean space of dimension $m=\Omega(\log d/\epsilon^2)$   with all pairwise distances preserved up to a $1\pm \epsilon$ factor. Notice that  $m$ is logarithmic in $d$ and does not depend on $n$. A sketch of the proof is given in the Appendix.


The JL Lemma establishes that $d$ vectors in $\mathbb R^n$ can be embedded into $m=\Omega(\log d/\epsilon^2)$ dimensions. But there are situations when we need to preserve the information in the $d$ columns jointly. This leads to the notion of  `subspace embedding' which requires that the norm of vectors in the column space of $A$ be approximately preserved by $\Pi$ with high probability.

 \begin{definition}[Subspace-Embedding] Let $A$ be an $n\times d$ matrix. An  $L_2$ subspace embedding for the column space of $A$ is an $m(\epsilon,\delta,d)\times n$ matrix $\Pi$ such that $\forall x \in\mathbb R^d$,
\begin{equation}
\label{eq:subspace}
 (1-\epsilon)  \| Ax \| _2^2 \le  \| \Pi Ax \| _2^2 \le (1+\epsilon)  \| Ax \| _2^2.
 \end{equation}
\end{definition}
\medskip
    
  Subspace embedding is an important  concept and it is useful to understand it  from different perspectives. Given that $ \| Ax \| _2^2=x^T A^T Ax$,  preserving the column space of $A$ means preserving the information in $A^T A$. The result can analogously be written as
 \[ \| \Pi Ax \| _2^2 \in\bigg[  (1- \epsilon) \| Ax \| _2^2, (1+\epsilon) \|Ax\|_2^2\bigg].\]
 Since $Ax=U\Sigma V^T x=Uz$  where $z=\Sigma V^T x\in\mathbb R^d$ and $U$ is orthonormal, a change of basis gives:
\begin{eqnarray*}
  \| \Pi U z \| _2^2 &\in& \bigg[(1-\epsilon)  \| Uz \| _2^2, (1+\epsilon)  \| Uz \| _2^2 \bigg]\\
 &=& \bigg[(1-\epsilon)  \| z \| _2^2, (1+\epsilon) \| z \| _2^2\bigg]\\
&\Leftrightarrow&   \|  (\Pi U)^T (\Pi U)-U^T U  \| _2\le \epsilon\\
&\Leftrightarrow& z^T \bigg( (\Pi U)^T (\Pi U)-I_d\bigg ) z\le  \epsilon.
 \end{eqnarray*}
The following Lemma defines  subspace embedding in terms of  singular value  distortions.
\begin{lemma}
\label{lem:subspaceembedding}
Let $U\in \mathbb R^{n\times d}$ be a unitary matrix  and
$\Pi$ be a subspace embedding for the column space of $A$.  Let $\sigma_k$ is the $k$-th singular value of $A$. Then (\ref{eq:subspace}) is equivalent to
\[ \sigma_k^2(\Pi U) \in [1-\epsilon,1+\epsilon]\quad \forall k \in [1,d].\]

\end{lemma}
 To understand Lemma \ref{lem:subspaceembedding},  consider the  Rayleigh quotient form of $\Pi U$:\footnote{For a Hermitian matrix $M$, the Rayleigh quotient is $\frac {c^T M c}{c^T c}$ for a nonzero vector $c$. By Rayleigh-Ritz Theorem,
$ \min(\sigma (M))  \le \frac {c^T M c}{c^T c} \le \max(\sigma (M)) $ with equalities when $c$ is the eigenvector corresponding to  the smallest and largest eigenvalues of $M$, respectively. See, e.g. \citet[Section 8.2]{hogben-07}.}
\[
\omega_k((\Pi U)^T (\Pi U))= \frac{v_k^T (\Pi U)^T (\Pi U) v_k}{v_k^T v_k}
\]
for some vector $v_k\ne 0$. As   $\omega_k(A^TA)=\sigma_k^2(A)$,
\begin{eqnarray*}
\omega_k((\Pi U)^T(\Pi U))&=& \frac{v_k^T v_k -v_k^T \bigg(I_d-(\Pi U)^T (\Pi U)\bigg) v_k}{v_k^T v_k}\\
&=& 1-\omega_k\bigg(I_d- (\Pi U)^T (\Pi U)\bigg).
\end{eqnarray*}
This implies that $|1-\sigma_k^2(\Pi U)|=|\omega_k(I_d- (\Pi U)^T (\Pi U))|=\sigma_k(I_d-(\Pi U)^T (\Pi U))$.  It follows that
 \begin{eqnarray*}
 |1-\sigma_k^2 (\Pi U)|&=& \bigg|\sigma_k\bigg(U^T U-(\Pi U)^T (\Pi U)\bigg)\bigg|\\
 &\le & \sigma_{max}(U^TU-(\Pi U)^T(\Pi^TU))\\
 &=&  \| U^TU-(\Pi U)^T (\Pi U) \| _2
 \le \epsilon\\
\Leftrightarrow \sigma_k^2(\Pi U)&\in &[1-\epsilon, 1+\epsilon] \quad \forall k\in[1,d].
 \end{eqnarray*}
 Hence the condition $\|(\Pi U)^T \Pi U-I_d\|_2\le \epsilon$ is equivalent to $\Pi$ generating small singular value distortions.
\citet{nelson-nguyen:13} relates  this condition to similar results in random matrix theory.\footnote{ Consider a $T\times N$  matrix of random variables with mean zero and unit variance with $c=\lim_{N,T\rightarrow\infty} \frac{N}{T}$.  In random matrix theory, the largest and smallest eigenvalues of the sample covariance matrix have been shown to  converge to $(1+\sqrt{c})^2, (1-\sqrt{c})^2$, respectively.  See, e.g., \citet{ybk:88} and \citet{bai-yin:93}. }

 But where to find these embedding matrices? We can look for  data dependent or  data oblivious ones. We say that  $\Pi$ is a  data oblivious embedding if it can be designed without knowledge of the input matrix.  The idea of oblivious subspace-embedding first appeared in \citet{sarlos-06} in which it is suggested that  $\Pi$ can be drawn from a distribution with the  JL properties.

 \begin{definition}
\label{lem:lemma0}

A random matrix $\Pi\in \mathbb R^{m\times n}$ drawn from a distribution $F$ forms a JL transform with parameters $\epsilon,\delta,d$  if there exists a function $f$ such that for any $0\le \epsilon,\delta\le 1$ and $m=\Omega(\log (\frac{d}{\epsilon^2} f(\delta)))$, $(1-\epsilon)\|x\|_2^2 \le \|\Pi x\|_2^2 \le (1+\epsilon)\|x\|_2^2$  holds with probability at least $1-\delta$ for all $d$-vector $x\subset \mathbb R^n$.
\end{definition}
A JL transform is often written  JLT($\epsilon,\delta,d)$ for short.
Embedding matrices $\Pi$ that are  JLT  guarantee good approximation to matrix products in terms of Frobenius norm.  This means that for such $\Pi$s,  an  $m$ can be chosen such  that for conformable matrices $A,B$ having $n$ rows:
\begin{equation}
\label{eq:fronorm}
P\bigg( \|(\Pi A)^T (\Pi B)-A^T B\|_F \le \epsilon \|A\|_F \|B\|_F\bigg)\ge 1-\delta.
\end{equation}
 The Frobenius norm bound has many uses. If $A=B$,  then $\|\Pi A\|_F^2=(1\pm \epsilon) \|A\|_F^2$ with high probability. The result also holds in the spectral norm, \citet[Corollary 11]{sarlos-06}.


\section{Random Sampling, Random Projections, and the Countsketch}
There are two classes of $\Pi$s with the JL property:  random sampling  which reduces the row dimension by randomly picking rows of $A$, and  random projections  which form  linear combinations from the rows of $A$. A scheme known as  a countsketch that is not a JL transform can also  achieve subspace embedding efficiently. We will use a pen and pencil example with  $m=3$ and $n=9$ to provide a better understanding of the three types of $\Pi$s.  In this example, $A$ has 9 rows given by $A_1,\ldots, A_9$.


\subsection{Random Sampling (RS)} 
 Let $D$ be a  diagonal {\em rescaling} matrix with $\frac{1}{\sqrt{m p_i}}$ in the $i$-th diagonal and $p_i$ is the probability that row $i$ is chosen. Under random sampling,
\begin{eqnarray*}
\Pi &=& D  S,
\end{eqnarray*}
where  $S_{jk}=1$  if row $k$ is selected in the $j$-th draw and zero otherwise so that
  the $j$-th row of the  selection matrix $S$   is the $j$th-row of an $n$ dimensional indentity matrix.  Examples of sampling schemes are:
\begin{itemize}
    \item[RS1.] Uniform sampling without replacement: $\Pi\in \mathbb R^{ m\times n}$, $D\in\mathbb R^{m\times m}$, $p_i=\frac{1}{n}$ for all $i$.
 Each row is sampled at most once.

\item[RS2.] Uniform sampling with replacement: $\Pi\in \mathbb R^{ m\times n}$, $D\in\mathbb R^{m\times m}$, $p_i=\frac{1}{n}$ for all $i$. Each row can be sampled more than once.

\item[RS3.] Bernoulli sampling uses an $n\times n$ matrix $\Pi=DS$, where $D=\sqrt{\frac{n}{m}} I_n$,
$S$ is initialized to be $0_{n\times n}$ and the $j$-th diagonal entry is updated by
\[ S_{jj}=\begin{cases} 1 & \quad \text{with probability }\frac{m}{n}\\
 0 & \text{with probability } 1-\frac{m}{n}
 \end{cases}
 \]
Each row is sampled at most once,  and $m$ is the expected number of sampled rows. 
\item[RS4.] Leverage score sampling:  the sampling probabilities are taken from  {\em importance sampling distribution}  
\begin{equation}
\label{eq:leverageprob}
 p_i=\frac{\ell_i}{\sum_{i=1}^n \ell_i}=\frac{\ell_i}{d},
 \end{equation}
\end{itemize}
where for  $A$  with \textsc{svd}$(A)=UDV^T$, 
\[ \ell _i= \| U_{(i)} \| ^2_2= \|  e_i^T U \| _2^2,\]
is the leverage score for row $i$, $\sum_i \ell_i= \| U \| _F^2=d$, and $e_i$ is a standard basis vector.   

Notably, the rows of the sketch produced by random sampling are the rows of the original matrix $A$. For example,
If rows 9,5,1 are randomly chosen by uniform sampling,   RS1 would give
\begin{eqnarray*}
\tilde A=D\begin{pmatrix}
0 & 0 &0 & 0 & 0 & 0 &  0 & 0  &1 \\
0 & 0 & 0 & 0 & 1 & 0 & 0 & 0  &0\\
1 & 0 & 0 & 0 & 0 & 0 & 0 & 0 &0 \end{pmatrix}
A= \frac{\sqrt{9}}{\sqrt{3}}
\begin{pmatrix}  A_{9}\\
 A_5 \\  A_{1}\end{pmatrix}.
\end{eqnarray*}

\citet[Section 3.3]{ipsen-wentworth:14} shows that sampling schemes RS1-RS3 are similar in terms of the condition number and rank deficiency in the matrices that are being subsampled.
 Unlike these three sampling schemes, leverage score sampling  is not  data oblivious and warrants further explanation.  
 
 As noted above, uniform sampling may not be efficient.  More precisely,  uniform sampling does not work well when the data have high coherence, where coherence  refers to the  maximum of the row leverage scores $\ell _i$ defined above. Early work suggests to use sampling weights that depend on the  Euclidean norm, $ p_i=\frac{\|A_i\|_2^2}{\|A\|_F^2}.$ See, e.g., \citet{drineas-kannan-mahoney:06} and \citet{drineas-mahoney:05}. Subsequent work finds that a better approach is to sample according to  the leverage scores which  measure the correlation between the left singular vectors of $A$ with the standard basis, and thus indicates whether or not information is spread out. The idea of leverage-sampling,   first used  in \citet{jolliffe-72}, is to sampling a row more frequently if it  has more information.\footnote{There are other variations of leverage score sampling. \citet{mcwilliams-buhmann:14} considers subsampling in linear regression models when the observations of the covariances may be corrupted by an additive noise. The  influence of observation $i$  is defined by
$ d_i = \frac{e_i^2 \ell _i}{(1-\ell_i)^2},$ where $e_i$ is the OLS residual and $\ell_i$ is the leverage score.
Unlike leverage scores, $d_i$  takes into account the relation between the predictor variables and the $y$.}   Of course, $\ell_i$ is  simply the $i$-th diagonal element of the hat matrix $A (A^TA)^{-1}A^T$,  known  to contain information about influential observations. In practice, computation of the leverage scores  requires an eigen decomposition which is itself expensive.  \citet{drineas-etal:12} and \citet{cohen-15} suggest fast approximation of  leverage scores.

\subsection{Random Projections (RP)}   
Some examples of random projections are:

\begin{itemize}

\item[RP1.]  Gaussian:  $\Pi\in \mathbb R^{m\times n}$ where 
$ \Pi_{ij} =\frac{1}{\sqrt{m}}N(0,1).$
\item[RP2.]  Rademacher  random variables: with entries of $\{+1,-1\}$, \citet{sarlos-06}, \citet{achlioptas}.

\item[RP3.] Randomized Orthogonal Systems:
 $\Pi = \sqrt{\frac{n}{m} } P H D$
  where $D$ is an $n\times n$ is  diagonal Rademacher matrix with entries of   $\pm 1$, $P$ is a sparse matrix, and $H$ is an orthonormal matrix.

\item[RP4.] Sparse Random Projections (SRP)
\[\Pi = D S\]
where  $D\in\mathbb R^{m\times m}$ is a diagonal matrix of $\sqrt{\frac{s}{m}}$ and $S\in \mathbb R^{m\times n}$
\[   S_{ij}=\begin{cases}
-1 \quad & \text{with probability } \frac{1}{2s}\\
0 \quad & \text{with probability } 1-\frac{1}{s}\\
1 \quad & \text{with probability } \frac{1}{2s}
\end{cases}
\]
\end{itemize}

RP1 and RP2 form sub-Gaussian random projections:\footnote{A mean-zero vector $s\in \mathbb R^n$ is sub-Gaussian if for any $u\in\mathbb R^n$ and for all $\epsilon>0$, 
$P\{ |u^T s| \ge \epsilon \|u\|_2 \} \le 2 e^{-\epsilon^2/K^2}$ for some absolute constant $K > 0$.} The rows of the sketch produced by random projections are linear combinations of the rows of the original matrix. For example, RP4 with $s=3$ could give
\begin{eqnarray*}
\tilde A =
D
\begin{pmatrix*}[r]
0 & 0 & 1 & 1 & 0 & -1 & 0 & 0 & 0 \\
1 & 0 & -1 & 0 & -1 & 0 & 0 & 0 & 1 \\
0 & 1 & 0 & 0 & 0 & 0 & 1 &0 & -1 & \\
\end{pmatrix*}A =\frac{\sqrt{3}}{\sqrt{9}}\begin{pmatrix} A_3+A_4-A_6\\ A_1-A_3-A_5 + A_9\\
A_2 +A_7-A_{9}\end{pmatrix}
\end{eqnarray*}

Early work on random projections such as   \citet{dasgupta-kumar-sarlos} uses $\Pi$s that are dense, an example being  RP1. Subsequent work favors  sparser $\Pi$s, an example being RP4. \citet{achlioptas} initially considers $s=3$.
 \citet{li-hastie-church} suggests to  increase $s$ to $\sqrt{n}$.
 Given that uniform sampling is algorithmically inefficient when information is concentrated, the idea of randomized orthogonal systems  is to first randomize the data by the matrix $H$ to destroy uniformity, so that sampling in a data oblivious manner using $P$ and rescaling by $D$ remains appropriate.   The randomization step is sometimes referred to as `preconditioning'.  Common choices of  $H$ are the Hadamard matrix as in the SRHT of \citet{ailon-chazelle-09}\footnote{The Hadamard matrix is defined recursively by
  $ H_n=\begin{pmatrix} H_{n/2} & H_{n/2} \\ H_{n/2} & -H_{n/2} \end{pmatrix}, \quad H_2=\begin{pmatrix} 1 & 1 \\ 1 & -1\end{pmatrix}
  $. A constraint is  that $n$ must be in powers of two.} and the discrete Fourier transform as in FJLT of \citet{liberty:08}.

\subsection{Countsketch}
While sparse $\Pi$s reduce computation cost, \citet[Theorem 2.3]{kane-nelson:14} shows that each column of $\Pi$ must have  $\Theta(d/\epsilon)$ non-zero entries to create an $L_2$ subspace embedding. This would seem to suggest that $\Pi$ cannot be too sparse. However, \citet{clarkson-woodruff:13} argues that if the non-zero entries of $\Pi$ are carefully chosen, $\Pi$ need not be a JLT and a very sparse subspace embedding is actually possible.  Their  insight is that $\Pi$ need not preserve the norms of an arbitrary subset of vectors in $\mathbb R^n$, but only those that sit in the $d$-dimensional subspace of $\mathbb R^n$.
The sparse embedding matrix considered in \citet{clarkson-woodruff:13} is the countsketch.\footnote{The  definition is taken from \citet{dahiya-konomis-woodruff}.  Given input $j$, a count-sketch matrix can also be characterized by a hash function $h(j)$  such that $\forall j,j', j\ne j'\rightarrow h(j)\ne h(j')$. Then
$ \Pi_{h(j),j}= \pm 1$  with equal probability  $1/2$.}

A countsketch of sketching dimension $m$ is a random linear map $\Pi=PD:\mathbb R^n\rightarrow \mathbb R^m$ where $D$ is an $n\times n$ random diagonal matrix with  entries chosen independently to be $+1$ or $-1$ with equal probability. Furthermore,    $P\in\{0,1\}$ is an $m\times n $ 
binary matrix such that $P_{h(i),i}=1$  and $P_{j,i}=0$ for all $j \neq h(i)$, and  $h:[n]\rightarrow [m]$ is a random map such that for each $i\in[n]$, $h(i)=m'$ for $m'\in[m]$ with probability $\frac{1}{m}$. As an example, a countsketch  might be 
\[
\tilde A=
\begin{pmatrix*}[r]
0 & 0 & 1 & 0 & 1 &-1 & 0& 0 & 1\\
-1 &0 &   0 &-1 &   0 & 0 &   0&  -1  &  0\\
0 &-1 &   0 & 0 &   0 & 0 &   1 & 0  &  0
\end{pmatrix*}A =\begin{pmatrix} A_3+A_5-A_6+A_9\\ -A_1-A_4 - A_8\\
-A_2 +A_7\end{pmatrix}.
\]
Like  random projections, the rows of a countsketch are also a linear combinations of the rows of $A$.

Though the countsketch is not a JLT,   \citet{nelson-nguyen:13b} and \citet{meng-mahoney:13} show that  the following spectral norm bound holds for the countsketch with appropriate choice of $m$:
 \begin{equation}
\label{eq:jlmoment}
 P\bigg( \|(\Pi U)^T (\Pi U) -I_d\|_2 > 3\epsilon  \bigg) \le
 \delta 
\end{equation} which implies  that  countsketch   provides a $1+\varepsilon$ subspace embedding for the column space of $A$ in spite of not being a JLT, see \citet[Theorem 2.6]{woodruff-book}.

The main appeal of the countsketch is  that  the run time needed to compute $\Pi A$  can be reduced to $O(\textsc{nnz}(A))$, where \textsc{nnz}(A) denotes the number of non-zero entries of $A$. The efficiency gain  is due to extreme sparsity of a countsketch $\Pi$ which  only  has  one non-zero element per column.  Still, the $\Pi$ matrix can be costly to store when $n$ is large.
Fortunately, it is  possible to compute the sketch without constructing $\Pi$.

The streaming version of the  countsketch  is a  variant of the frequent-items algorithm where we recall that having to compute summaries such as the most frequent item in  the data that stream by was instrumental to the development of sketching algorithms. The streaming algorithm proceeds by initializing $\tilde A$ to an $m\times n$ matrix of zeros. Each row $A_{(i)}$ of $A$ is updated as
\[ \tilde A_{h(i)} =\tilde A_{h(i)}+g(i) A_{(i)}\]
where $h(i)$  sampled uniformly at random from $[1,2,\ldots m]$  and $g_i$  sampled from $\{+1,-1\}$ are independent.
 Computation can be done one row at a time.\footnote{See \citet{liberty-woodruff:16}. Similar schemes have been proposed in \citet{charika:02, cormode-muthukrishnan}.} The Appendix  provides the streaming implementation of the example above.

\subsection{Properties of the $\Pi$s}

To assess the actual performance of the different $\Pi$s, we conduct  a small Monte Carlo experiment with 1000 replications.  For each replication $b$, we simulate an $n\times d$ matrix $A$ and construct the seven JL embeddings considered above.  For each embedding,  we count the number of times that   $\||\Pi (a_i-a_j)||_2^2$ is within  $(1\pm \epsilon)$ of $||a_i-a_j||^2_2$ for all $d(d+1)/2$ pairs of distinct $(i,j)$. The success rate for the replication is the total count divided by  $d((d+1)/2$. We also  record $||\frac{\sigma(\Pi A)}{\sigma(A)}-1||_2 $ where $\sigma(\Pi A)$ is a vector of $d$  singular values of $\Pi A$.   According to theory, the pairwise distortion of the vectors should be small if $m\ge C \log d/\epsilon^2$.  We set $(n,d)=(20,000)$ and $\epsilon=0.1$. Four values of $C=\{1,2,3,4,5,6,8,16\}$ are considered. We draw $A$ from the (i) normal distribution, and (ii)  the exponential distribution.  In \textsc{matlab}, these are generated as  \textsc{X=randn(n,d)} and \textsc{X=exprnd(d,[n d])}. Results for the Pearson distribution using  \textsc{X=pearsrnd(0,1,1,5,n,d)} are similar and not reported.

Table \ref{tbl:table1} reports the results averaged over 1000 simulations. With probability around 0.975, the pairwise distance between columns with 1000 rows is close to the  pairwise distance between columns with 20000 rows. The average singular value distortion also levels off with about 1000 rows of data.  Hence, information in $n$ rows can be well summarized by a smaller matrix with $m$ rows.  However, note that more rows are generally needed for uniform sampling, while fewer rows are needed for leverage score sampling,  to have the same error as the remaining methods. The takeaway from Table \ref{tbl:table1} is that the performance of the different $\Pi$s are quite similar, making computation cost and analytical tractability two important factor in deciding which ones to use.

Choosing a $\Pi$ is akin to choosing a kernel function in nonparametric regression and many will work well, but there are analytical differences.  Any $\Pi^T \Pi $ can be written as $I_n+ R_{11} + R_{12}$ where $R_{11}$ is a generic diagonal and $R_{12}$ is a generic $n\times n$ matrix with zeros in each diagonal entry. Two features will be particularly useful.
\begin{subequations}
\begin{align}
 \Pi^T \Pi &= I_n+R_{11} \quad \label{prop1}\\
 \Pi \Pi^T &= \frac{n}{m} I_m \label{prop2}
\end{align}
\end{subequations}
 Property (\ref{prop1}) imposes that  $\Pi^T \Pi$ is a diagonal matrix, allowing $R_{11}\ne 0$ but restricting $R_{12}=0$. Each $\Pi\Pi^T$ can also be written as $\Pi\Pi^T=\frac{n}{m} I_m+ R_{21}+ R_{22}$ where $R_{21}$ is a generic diagonal and $R_{22}$ is a generic $m\times m$ matrix with zeros in each diagonal entry. Property (\ref{prop2}) requires that $\Pi\Pi^T$ is proportional to an identity matrix, and hence that $R_{21}=R_{22}=0_{m\times m}$.

For the $\Pi$s  previously considered, we summarize their properties as follows:
\begin{center}
\begin{tabular}{l|llll} \hline
 & (\ref{prop1}) & (\ref{prop2}) &  \\ \hline
RS1 (Uniform,w/o)& yes & yes&  \\
RS2 (Uniform,w)& yes & no & \\
RS3 (Bernoulli)& yes &no &  \\
RS4 (Leverage) & yes & no \\
RP1 (Gaussian)& no & no &  \\
RP2 (Rademacher)& no & no &  \\
RP3 (SRHT) & yes & yes &  \\
RP4 (SRP) & no &no & \\
CS (Countsketch) & no & no & \\
\hline
\end{tabular}
\end{center}
Property (\ref{prop1}) holds for all three random sampling methods but  of all the random projection methods considered, the property only holds for SRHT. This is because SRHT effectively performs uniform sampling of the randomized data.  For property (\ref{prop2}),  it is easy to see that  $R_{21}=0$ and $R_{22}=0_{m\times m}$ when uniform
 sampling is done without replacement  and $\Pi\Pi^T=\frac{n}{m}I_m$.   By implication, the condition   also holds for SRHT  if  sampling is done without replacement since
  $H$ and $D$  are orthonormal matrices.  But uniform sampling is computationally much cheaper than SRHT and has the distinct advantage over the SRHT that the rows of the sketch are those of the original matrix and hence interpretable. For this reason, we will subsequently focus on uniform sampling and use its special structure to  obtain  precise statistical results. Though neither condition holds for the  countsketch, the computation advantage due to its extreme sparse structure makes it worthy of further investigation.

\section{Algorithmic Results for the Linear Regression Model}

 The linear regression model with $K$ regressors is $y=X\beta+e$.  The least squares estimator minimizes 
 $\|y-X b\|_2$ with respect to $b$ and is defined by
\[ \hat\beta  =(X^T X)^{-1} X^T y=V \Sigma^{-1} U^T y.\]
 We are familiar with the statistical properties of $\hat\beta$  under  assumptions about $e$ and $X$. But even without specifying a probabilistic structure, $\|y-Xb||_2$ with $n>K$ is an over-determined system of equations and can be solved by algebraically. The \textsc{svd} solution gives $\beta^*=X^- y$ where  $\textsc{svd}(X)=U\Sigma V^T$,  the  pseudoinverse is $X^-=V\Sigma^{-1} U^T$. The 'Choleski' solution  starts with the normal equations  $X^TX \beta=X^Ty$ and factorizes $X^T X$. The algebraic properties of these solutions  are well studied in the matrix computations literature when all data are used.

 Given an embedding matrix $\Pi$ and sketched data $(\Pi y, \Pi X)$,   minimizing
$\| \Pi (y- Xb)\|_2^2$ with respect to $b$ gives the sketched estimator  
\[ \tilde\beta=\bigg((\Pi X)^T \Pi X\bigg)^{-1} (\Pi X)^T \Pi y.\]
Let  $\widehat{\textsc{ssr}}=\|y-X\hat\beta\|^2_2$ be the full sample sum of squared residuals. For  an embedding matrix $\Pi\in \mathbb R^{m\times n}$, let   $\widetilde{\textsc{ssr}}=\|\tilde y-\tilde X\tilde\beta\|_2^2$  be the sum of squared residuals from using the sketched data. Assume that the following two conditions hold with  probability $1-\delta$ for $0<\epsilon<1$:
\begin{subequations}
\begin{eqnarray}
 |1- \sigma^2_k(\Pi U)| &\le& \frac{1}{\sqrt{2}} \quad \forall k=1,\ldots, K; \label{eq:condition1}\\
\|  (\Pi U)^T \Pi (y-X\hat\beta)\|_2^2 &\le& \epsilon \; \widehat{\textsc{ssr}}^2/2. \label{eq:condition2}
\end{eqnarray}
\end{subequations}
Condition (\ref{eq:condition1}) is  is equivalent to $\|(\Pi U)^T (\Pi U)-U^T U\|_2\le \frac{1}{\sqrt{2}}$ as discussed above. Since $\sigma_i(U)=1$ for all $k\in[1,K]$, the condition requires the smallest singular value, $\sigma_K(\Pi U)$, to be positive so that $\Pi X$ has the same rank as $X$. A property of the least squares estimator is for the least squares residuals to be orthogonal to $X$, ie.  $U^T (y-X\hat\beta)=0$. Condition (\ref{eq:condition2}) requires near orthogonality  when both quantities are multiplied by $\Pi$. The  two algorithmic features of sketched least squares estimation  are summarized below.

  \begin{lemma}
\label{lem:sarlos} Let the sketched data be $(\Pi y,\Pi X)=(\tilde y,\tilde X)$ where $\Pi\in\mathbb R^{m\times n}$ is a subspace embedding matrix.  Let $\sigma_{\min}(X)$ be the smallest singular value of $X$. Suppose that conditions (\ref{eq:condition1})
 and (\ref{eq:condition2}) hold. Then with probability at least $(1-\delta)$ and for suitable choice of $m$,
  \begin{itemize}
 \item[(i).] 
$\widetilde{\textsc{ssr}} \le(1+\epsilon)  \widehat{\textsc{ssr}}$;
    \item[(ii).]   $\|\tilde\beta-\hat\beta\|_2\le \epsilon \cdot \widehat{\textsc{ssr}} /\sigma_{\min}$.
\end{itemize}
\end{lemma}
\citet{sarlos-06} provides the proof for random projections, while 
\citet{dmm:06} analyzes the case of leverage score sampling. The desired $m$ depends on the result and the sampling scheme.

 Part (i) is based on  subspace embedding arguments. By optimality of  $\tilde\beta$  and  JL Lemma,
 \begin{eqnarray*}
\widetilde{\textsc{ssr}}&=& \|\Pi (y-X\tilde\beta)|_2\\
&\le& \| \Pi (y-X\hat\beta)\|_2 \quad \text{by optimality of }\tilde\beta\\
&\le&  (1+\epsilon) \| y-X\hat\beta\|_2\quad \text{by subspace embedding}\\ 
&=&(1+\epsilon) \widehat{\textsc{ssr}}.
\end{eqnarray*}
 Part (ii) shows that the sketching error  is data dependent. Consider a  reparameterization of $X\hat\beta =U\Sigma V^T \hat\beta=U\hat \theta$ and $X\tilde\beta=U\Sigma V^T \tilde\beta=U\tilde\theta$.  As shown in the Appendix, 
$ \|\tilde\theta-\hat\theta\|_2\le \sqrt{\epsilon}\; \widehat{\textsc{ssr}}.$ Taking norms on both sides of $X(\tilde\beta-\hat\beta)=U(\tilde\theta-\hat\theta)$ and since $U$ is orthonormal,
    \begin{eqnarray*}
       \|\tilde\beta-\hat\beta\|_2 
       &\le & \frac{\|U(\tilde\theta-\hat\theta)\|_2}{\sigma_{\min}}.
    \end{eqnarray*}    
 Notably, difference between  $\hat\beta$ and  $\tilde\beta$ depends on  the  minimum singular value of $X$. Recall that for consistent estimation, we also require that  the minimum eigenvalue to diverge.

 The non-asymptotic  worse case error bounds  in Lemma \ref{lem:sarlos} are  valid for any subspace embedding matrix $\Pi$, though more precise statements  are available for certain  $\Pi$s. For leverage score sampling, see \citet{dmm:06}, for uniform sampling and  SRHT, see  \citet{dmms}; and  for the countsketch,  \citet[Theorem 2.16]{woodruff-book},
\citet[Theorem 1]{meng-mahoney:13}, \citet{nelson-nguyen:13}. These  algorithmic results  are derived without reference to the probabilistic structure of the data. Hence the results do not convey information such as bias and  sampling uncertainty. An interesting  question is  whether optimality from an algorithmic perspective implies optimality from a statistical perspective. Using Taylor series expansion,  \citet{ma-mahoney-yu} shows that leverage-based sampling does not dominate uniform sampling in terms of bias and variance, while \citet{raskutti-mahoney} finds that prediction efficiency requires $m$ to be quite large. \citet{pilanci-wainwright:15} shows that the solutions from sketched least squares regressions have larger variance than the oracle solution that uses the full sample. \citet{pilanci-wainwright:15} provides a result that relates $m$ to the rank of the matrix. \citet{wang-gittens-mahoney:18} studies four sketching methods in the context of ridge regressions that nests least squares as a special case. It is reported that sketching schemes with near optimal algorithmic properties may have   features that not statistically optimal. \citet{chi-ipsen:18} decomposes the variance of $\tilde\beta$ into a model induced component and an algorithm induced component. 

 \section{Statistical Properties of $\tilde\beta$}

  We  consider the  linear regression model with $K$ regressors:
\[ y=X^T\beta+e, \quad\quad e_i \sim (0,\Omega_e)\]
where $y$ is  the dependent variable,
$X$ is the $n \times K$ matrix of regressors,  $\beta$ is the $K \times 1$ vector of regression coefficients whose true value is $\beta_0$.   It should be noted that  $K$ is the number of predictors which  is generally larger than   $d$, which is the number of covariates available  since the predictors  may include transformation of the $d$ covariates.  In the Mincer example, we have data  for \textsc{edu}, \textsc{exp} collected into $A$ with $d=2$ columns. 
From these two covariates,  $K=4$ regressors are constructed for  regression (\ref{eq:mincer1}), while  $K=14$ regressors are  constructed for regression (\ref{eq:mincer2}).

The full sample estimator using  data $(y,X)$  is  $\hat\beta= (X^T X)^{-1} X^T y$. For a given $\Pi$,  the estimator using sketched data  $(\tilde y,\tilde X)=(\Pi y, \Pi X)$ is
\[\tilde \beta = (\tilde X^T \tilde X)^{-1} \tilde X^T \tilde y.\]

\paragraph{Assumption OLS:}
\begin{itemize}
\item[(i)] the regressors  $X$ are  non-random, has \textsc{svd}  $X=U\Sigma V^T$, and  $X^T X$ is non-singular;
\item[(ii)] $\mathbb{E} [e_i] = 0$ and $\mathbb E[ee^T]=\Omega_e$ is a diagonal positive definite matrix.  
\end{itemize}

\paragraph{Assumption PI:}
\begin{itemize}
\item[(i)] $\Pi$ is independent of $e$;
\item[(ii)] for given singular value distortion parameter $\varepsilon_\sigma\in (0,1)$, there exists failure parameter $\delta_\sigma \in (0,1)$ such that
$
P \Big(|1-\sigma^2_k(\Pi U)|\le \varepsilon_\sigma
 \;
  \text{for all $k=1,\ldots K$}
\Big)
\ge   1-\delta_\sigma.
$
\item[(iii)]  $\Pi^T \Pi$ is an $n \times n$ diagonal matrix and  $\Pi\Pi^T =\frac{n}{m} I_m$.

\end{itemize}
Assumption OLS is standard in  regression analyses. The errors are allowed to be possibly heteroskedastic but not cross-correlated. Under Assumption OLS, $\hat \beta$ is  unbiased, i.e. $\mathbb E[\hat\beta]=\beta_0$ with a sandwich variance
\begin{eqnarray*}
\mathbb V(\hat\beta)=  (X^T X)^{-1}(X^T\Omega_e X) (X^T X)^{-1}.
\end{eqnarray*}
Assumption PI.(i) is needed for $\tilde\beta$ to be unbiased. Assumption PI.(ii) restricts attention to $\Pi$ matrices that have subspace embedding property.  As previously noted, the condition is equivalent to  $ \|  I_d-(\Pi U)^T (\Pi U) \| _2\le \varepsilon_\sigma$ holding with probability $1-\delta_\sigma$.     We use PI2 to refer Assumptions PI (i) and (ii) holding jointly. Results under PI2 are not specific to any $\Pi$.  

Assumption  PI.(iii)  simplifies the expression for $\mathbb V(\tilde\beta|\Pi)$, 
the variance of $\tilde\beta$ conditional on $\Pi$, and we will use   PI3 to denote  Assumptions PI (i)-(iii) holding jointly. PI3  effectively narrows the analysis to  uniform sampling and SRHT without replacement.  We will focus on   uniform sampling without replacement for a number of reasons.  It is simple to implement, and unlike the SRHT, the rows have meaningful interpretation.   In a regression context, uniform sampling has an added advantage  that there is no need to reconstruct $(\tilde y,\tilde X)$  each time we  add or drop a variable in the $X$ matrix. In contrast,  most other $\Pi$s   require  $(\tilde y,\tilde X)$ to be reconstructed. This can be cumbersome when variable selection is part of the empirical exercise. Uniform sampling without replacement is an exception since  the columns are unaffected once the rows are randomly chosen. 

For regressions, we need to know not only the error in approximating  $X^TX$, but also the error in approximating $(X^T X)^{-1}$. This is made precise in the next Lemma.


\begin{lemma}
\label{lem:lemma-variance}
Suppose that  PI2 is satisfied. For given non-random matrix $X\in\mathbb R^{n\times K}$ of full rank with $\textsc{svd}(X)=U\Sigma V^T$, consider  any non-zero $K\times 1$ vector $c$. It holds 
with probability at least $1-\delta_\sigma$
that
\[\bigg|\frac{ c^T [( X^T  X)^{-1} - (\tilde X^T\tilde X)^{-1}] c}{c^T (X^T X)^{-1} c} \bigg| \le \frac{\varepsilon_\sigma}{1-\varepsilon_\sigma}.\]
\end{lemma}
The Lemma follows from the fact that
\begin{eqnarray*}
 (X^T X)^{-1}&=&V\Sigma^{-2}V^T\equiv PP^T\\
  \bigg((\Pi X)^T (\Pi X)\bigg)^{-1}&=&P QP^T
  \end{eqnarray*}
  where  $P= V\Sigma^{-1}$ and   $Q^{-1}=(\Pi U)^T (\Pi U)$.
 By the property of Rayleigh quotient, the smallest eigenvalue of $(U^T \Pi^T \Pi U)$ is bounded below by $(1-\varepsilon_\sigma)$. Hence
\begin{eqnarray*}
\left| \frac{c^T( PQP^T - PP^T ) c }{c^T PP^T c}   \right|
&=& \left| \frac{c^T P(I_d - Q^{-1}) Q P^T c}{c^T P P^T c}  \right|
\leq   \| Q \| _2  \| I_d - Q^{-1} \| _2
\leq \frac{\varepsilon_\sigma}{(1-\varepsilon_\sigma)}.
\end{eqnarray*}
The  approximation error  $(X^TX)^{-1}$ is thus larger than that for $X^TX$, which equals $\varepsilon_\sigma$.

Under Assumptions OLS and PI3, $\tilde\beta$  is unbiased and has  sandwich  variance\begin{eqnarray*}
\mathbb V(\tilde\beta|\Pi)
&=&\frac{n}{m} (\tilde X^T\tilde X)^{-1}(\tilde X^T\Omega_e \tilde X)^{-1} (\tilde X^T\tilde X)^{-1}
\end{eqnarray*}
since $\Pi\Pi^T=\frac{n}{m}I_m$.
The variance of $\tilde\beta$  is inflated over that of $\hat\beta$  through the sketching error on the `bread'
  $(\tilde X^T \tilde X)^{-1}$, as well as  on the `meat'  because $X^T\Omega_e X$ is now approximated by $\tilde X^T \Pi\Omega_e\Pi^T \tilde X$.  
  If $e\sim (0,\sigma^2_e I_n)$ is  homoskedastic,   then $\tilde\beta$ has
variance
\begin{eqnarray}
\mathbb V(\tilde \beta|\Pi)
&=&\sigma_e^2\frac{n}{m} (\tilde X^T\tilde X)^{-1}. \label{eq:varbetatilde} 
\end{eqnarray}
Though $\hat\beta$ is the best linear unbiased  estimator under homoskedasticity,  $\tilde\beta$  may not be best in the class of linear estimators using sketched data.

\subsection{Efficiency of $\tilde\beta$  Under Uniform Sampling }

  Suppose we are interested in predicting $y$ at some $x^0$. According to the model, $\mathbb E[y|x=x^0]=\beta^T x^0$. Feasible predictions are obtained upon replacing $\beta$ with $\hat\beta$ and $\tilde\beta$.  Since both estimators are unbiased, their respective variance is also the mean-squared prediction error.

\begin{theorem}
\label{thm:prediction} Suppose that $e_i\sim (0,\sigma^2_e)$ and Assumptions OLS and PI3  hold. Let $\textsc{mse}(x_0^T\hat\beta)$ and $\textsc{mse}(x_0^T\tilde\beta | \Pi)$ be  the mean-squared prediction error  of $y$ at $x_0$ using $\hat\beta$ and $\tilde\beta$ conditional on $\Pi$, respectively.  Then with probability at least $1-\delta_\sigma$, it holds that
$$ \frac{\textsc{mse}(x_0^T\tilde\beta | \Pi) }{\textsc{mse}(x_0^T\hat\beta)} 
\le  \underbrace{\frac{n}{m}}_{\text{sample size}}\underbrace{\bigg(\frac{1}{1-\varepsilon_\sigma}\bigg)}_{\text{sketching error}}
.
$$

\end{theorem}
 The prediction error  has  has two components:
a sample size effect given by $\frac{n}{m}> 1$, and a sketching effect given by $\frac{1}{1-\varepsilon_\sigma}>1$. The result arises because under homoskedasticity,  
 \begin{align*}
 x_0^T  (\tilde X^T \tilde X)^{-1} x_0 - x_0^T (X^T  X)^{-1}x_0
=   \frac{n}{m} x_0^T \left[   \big( X^T \Pi^T \Pi X \big)^{-1} - (X^T X)^{-1}\right] x_0
+  \frac{n-m}{m} x_0^T   (X^T X)^{-1} x_0.
\end{align*}
It follows that
 \begin{align*}
\bigg|\frac{x_0^T \mathbb V(\tilde\beta|\Pi)x_0 - x_0^T \mathbb V(\hat\beta)x_0}{
x_0^T \mathbb V(\hat\beta)x_0} \bigg|
&=  \bigg| \frac{n}{m} \frac{x_0^T [   \big( ( \Pi X)^T  \Pi X \big)^{-1} - (X^T X)^{-1}] x_0 }{x_0^T (X^T X)^{-1}x_0}
+  \frac{n-m}{m}\bigg|\\
&\le  \frac{n}{m}\frac{\varepsilon_\sigma}{1-\varepsilon_\sigma} + \frac{n-m}{m}
.
 \end{align*}

We will  subsequently be interested in the effect of sketching for testing linear restrictions as given by a $K\times 1 $ vector $c$. The estimated linear combination $c^T \tilde\beta$ has variance   $\mathbb V(c^T\tilde\beta | \Pi )=c^T \mathbb V(\tilde\beta |\Pi) c$. When $c$ is a vector of zeros except in the $k$-th entry, $\text{var}(c^T\tilde\beta | \Pi)$ is the variance of $\tilde\beta_k$. When $c$ is a vector of ones, $\text{var}(c^T\tilde\beta |\Pi)$ is the variance of the sum of estimates. A straightforward generalization of Theorem 1 leads to  the following.

 \begin{corollary}
 \label{thm:thm1}
Let $c$ be a known $K\times 1$   vector. Under the Assumptions of Theorem 1, it holds  with probability $1-\delta_\sigma$ that
 \[
\frac{c^T\mathbb V(\tilde\beta|\Pi) c }{c^T\mathbb V(\hat\beta)c} 
\le \frac{n}{m} \bigg(\frac{1}{1-\varepsilon_\sigma}\bigg).
\]
\end{corollary}

 The relative  error is thus primarily determined by the relative sample size.  As $m$ is expected to be much smaller than $n$, the efficiency loss is undeniable.

A lower bound in estimation error can be obtained for embedding matrices  $\Phi\in \mathbb R^{m\times n}$  satisfying $\|\Phi U\|_2^2 \leq 1 + \varepsilon_\sigma$. For a sketch size of $m$ rows,   define a class of OLS estimators as follows:
\begin{align*}\label{def-B-set}
\begin{split}
\mathcal{B}(m, n, \varepsilon_\sigma)
:= \left\{ \breve{\beta} :=  ((X \Phi)^T \Phi X)^{-1}  (X \Phi)^T \Phi y \;
 \right\}.
\end{split}
\end{align*}
For such $\breve\beta$, let ${\mathbb V}(\breve{\beta}| \Phi)$ denote its mse for  given $\Phi$. Assuming that $e_i\sim (0,\sigma^2_e)$,
\begin{align*}
\frac{c^T {\mathbb V}(\breve{\beta}| \Phi) c}{c^T {\mathbb V}(\hat\beta) c}
&= \frac{m^{-1} \sigma^2_e c^T ((\Phi X)^ T \Phi X)^{-1} c}{n^{-1} \sigma^2_e c^T (X^T  X)^{-1} c}
= \frac{n}{m}  \frac{ c^T ( (\Phi X)^T \Phi X)^{-1} c}{c^T (X^T  X)^{-1} c}\\
&= \frac{n}{m} \frac{c^T (V \Sigma U^T \Phi^T \Phi U \Sigma V^T)^{-1} c}{c^T (V \Sigma^2   V^T)^{-1} c}
= \frac{n}{m} \frac{c^T V \Sigma^{-1} (U^T \Phi^T \Phi U)^{-1} \Sigma^{-1} V^T c}{c^T V \Sigma^{-2}   V^T c}\\
&\geq \frac{n}{m}  \sigma_{\min}[ (U^T \Phi^T \Phi U)^{-1} ].
\end{align*}
But  by the definition of spectral norm, $ \| \Phi U \| _2^2=\sigma_{\max}^2(\Phi U)$ for any $\Phi$. Thus the subspace embedding condition $ \|  \Phi U \| _2^2\le 1+\varepsilon_\sigma$ implies $\sigma_{\max}^2(\Phi U) =\sigma_{\max}( (\Phi U)^T \Phi U) \le 1+\varepsilon_\sigma$, and hence
\[ \sigma_{\min} \bigg(( U^T \Phi ^T \Phi U)^{-1}\bigg) \ge \frac{1}{1+\varepsilon_\sigma}.\]
This leads to the following  lower bound for $\breve \beta$:
\begin{align*}
\frac{c^T {\mathbb V}(\breve{\beta}| \Phi) c}{c^T {\mathbb V}(\hat\beta) c}
&\geq
\frac{n}{m} \bigg(\frac{1}{1+\varepsilon_\sigma}\bigg).
\end{align*}
Combining the upper and lower bounds leads to the following:
\begin{theorem}\label{thm-lower-bound} Under OLS and PI3,
   the estimator $\tilde\beta$ with $e_i\sim (0,\sigma^2_e)$ has  mean-squared error relative to the full sample estimator $\hat\beta$ bounded by   
\[\frac{n}{m} \bigg(\frac{1}{1+\varepsilon_\sigma}\bigg)\le  \frac{c^T\mathbb V(\tilde\beta|\Pi) c }{c^T\mathbb V(\hat\beta)c}  \le \frac{n}{m}\frac{1}{1 - \varepsilon_\sigma}.\]
\end{theorem}

These are the upper and lower bounds  for uniform sampling  when implemented by sampling without replacement. 

It is also of interest to know how heteroskedasticity affects the sketching error. Let $\Omega_{e,ii}$ denote the $i$th diagonal element of $\Omega_e$. Under OLS and PI3,  it holds with probability at least  $1-\delta_\sigma$ that
\begin{equation*}
\label{eq:relativeV-general-NR}
\frac{\textsc{mse}(x_0^T\tilde\beta | \Pi) }{\textsc{mse}(x_0^T\hat\beta)}
\le    \bigg(
\frac{\max_i \; \Omega_{e,ii}}{\min_i \; \Omega_{e,ii}}\bigg)
\bigg(\frac{n}{m}\bigg)
\frac{(1 + \varepsilon_\sigma)}{(1-\varepsilon_\sigma)^2}.
\end{equation*}
Hence  heteroskedasticity independently interacts with the structure of  $\Pi$ to inflate  the mean-squared prediction error. The  upper and lower bound for $\mathbb V(c^T \tilde\beta|\Phi)$ are larger than under homoskedasticity by a magnitude that depends on the extent of dispersion in $\Omega_{e,ii}$. A formal result is given in the appendix.

  \subsection{Efficiency of $\tilde \beta$ under Countsketch}

  Condition PI.(iii) puts restrictions on  $\Pi^T \Pi$ and holds for  uniform sampling. But the condition does not hold for the countsketch. In its place, we assume the following to obtain a different embedding result for the countsketch:
\paragraph{Assumption CS:} For given $\varepsilon_\Pi > 0$ and for all $U\in\mathbb{R}^{n\times K}$ satisfying $U^TU=I_K$,
 there exists an $n \times n$ matrix $A (\Omega_e, m, n)$, which may depend on $(\Omega_e, m, n)$,
and
 a constant $\delta_\Pi \in (0,1)$ such that
$$
P \Big( \|  U^T \Pi^T \Pi\Omega_e \Pi^T \Pi U -  U^T A (\Omega_e, m, n) U \| _2 \le \frac{n}{m} \varepsilon_\Pi \Big) \ge 1-\delta_\Pi.
$$
The conditions for Assumption CS are verified in Appendix B. The Assumption is enough to provide a subspace embedding result for  $\Pi^T \Pi \Omega_e \Pi^T \Pi$ because as shown in the Appendix, the following holds for Countsketch, 
\begin{align}\label{pi4-all-ineq-new}
\begin{split}
&
\norm{ U^T \Pi^T \Pi \Omega_{e} \Pi^T \Pi U - \frac{n}{m} U^T \Omega_{e} U }_2 \\
&\leq  \norm{ U^T \Pi^T \Pi \Omega_{e} \Pi^T \Pi U - U^T A (\Omega_{e}, m,n) U }_2
+  \norm{  U^T A (\Omega_{e}, m,n) U -  \frac{n}{m}  U^T  \Omega_{e}    U }_2 \\
&\leq \frac{n}{m} \left[\varepsilon_\Pi   +  \norm{   \frac{m}{n} A(\Omega_{e}, m, n)   -    \Omega_{e}   }_2 \right]
\end{split}
\end{align}
where
\[ A(\Omega_e,m,n)=\Omega_e+\frac{1}{m} \bigg(\text{tr}(\Omega_e) I_n-\Omega_e\bigg).\]
Hence under OLS, PI2, and CS  it holds with probability at least  $1-\delta_\Pi-\delta_\sigma$ that
\begin{equation*}
\label{eq:relativeV-general}
\frac{\textsc{mse}(x_0^T\tilde\beta | \Pi) }{\textsc{mse}(x_0^T\hat\beta)}
\le    \bigg(
\frac{\max_i \; \Omega_{e,ii}}{\min_i \; \Omega_{e,ii}}\bigg)
\bigg(\frac{n}{m}\bigg)
\bigg(\frac{1}{1-\varepsilon_\sigma}\bigg)
\frac{\left[1 + \varepsilon_\Pi   +  \|   \frac{m}{n} A (\Omega_e, m, n)  -    \Omega_e   \|_2 \right]}{(1-\varepsilon_\sigma)}.
\end{equation*}
The  prediction error of the countsketch  depends on  the quantity $A(\Omega_e,m,n)$. 
But if $\Omega_e = \sigma_e^2 I_n$,
$\norm{ \frac{m}{n} A (\Omega_e, m,n) -  \Omega_e }_2
= \sigma_e^2 ( \frac{m-1}{n} ),
$ which will be negligible if $m/n = o(1)$. Hence to a first approximation,  Theorem 1 also holds  under  the countsketch.  This result is of interest  since the countsketch is computationally inexpensive.

\subsection{Hypothesis Testing}
The statistical implications of sketching in a regression setting  have primarily focused on properties of the point estimates. The implications  for  inference are largely unknown.   We  analyze the problem from view point of hypothesis testing.

Consider the goal of  testing $q$ linear restrictions formulated as $H_0: R\beta=r$ where $R$ is a $q\times K$ matrix of restrictions with no unknowns.
In this subsection, we further assume that 
 $e_i\sim N(0,\sigma_e^2)$.  Under normality, the $F$ test is exact and has the property that at the true value of $\beta=\beta^0$,
\begin{eqnarray*}
F_n &=&R(\hat \beta - \beta_0) ^T
\bigg(\hat{\mathbb V}(R\hat\beta)\bigg)^{-1} R(\hat\beta-\beta_0)\sim \mathbb F_{q,n-d}.
\end{eqnarray*}
Under the null hypothesis that $\beta_0$ is the true value of $\beta$, $F_n$ has a Fisher distribution with $q$ and $n-d$ degrees of freedom. The power of a test given a data of size $n$ against a fixed alternative $\beta_1\ne \beta_0$ depends on $\mathbb V(\hat\beta)$ only through non-centrality parameter $\phi_n$,\footnote{The definition of non-centrality is not universal, sometimes the factor of two is omitted. See, for example, \citet{cramer-joe:87} and \citet{rudd-book}.}  defined in \citet{wallace:72} as
\[\phi_n=\frac{(R\beta_0-r)^T \mathbb V(R\hat\beta)^{-1} (R\beta-r)}{2}.
\]
 The non-centrality parameter is increasing in $|R\beta_0-r|$ and the sample size through the variance, but decreasing in $\sigma_e^2$. In the case of one restriction  $(q=1)$,
\[ \phi_n=\frac{(R\beta_0-r)^2}{\mathbb V(R\hat\beta)}> \frac{(R\beta_0-r)^2}{\mathbb  V(R\tilde\beta | \Pi )}= \phi_m. \]
This leads to  the relative non-centrality
\[ \frac{\phi_n}{\phi_m}=\frac{\mathbb V(R\tilde\beta  | \Pi )}{\mathbb V(R\hat\beta)}
\le \frac{n}{m}\frac{1}{(1-\varepsilon_\phi)}.\]
which also has a sample size effect and an effect due to sketching error. The effective size of the subsample from the viewpoint of power can be thought of as $m(1-\varepsilon_\phi)$.

A loss in power is to be expected when $\tilde\beta$ is used. But by how much? Insights can be performed from some back of the envelope calculations. Recall that if $U$ and $V$ are independent $\chi^2$ variables with $\nu_1$ and $\nu_2$ degrees of freedom, $V$ is central and $U$ has non-centrality parameter $\phi$,
\[
 \mathbb E[F]=\mathbb  E \bigg[\frac{ (U/\nu_1)}{ (V/\nu_2)}\bigg] =\frac{\nu_2(\nu_1+\phi_n)}{\nu_1(\nu_2-2)}.\]
In the full sample case, $\nu_1=q$ and $\nu_2=n-d$ and hence
\begin{eqnarray*}
 \mathbb E[F_n] &\approx& \frac{(n-d)(q+\phi_n)}{q(n-d-2)}.
 \end{eqnarray*}
For the subsampled estimator,  $\nu_1=q$ and $\nu_2=m-d$, giving
\begin{eqnarray*}
 \mathbb E[ F_m  | \Pi ] &\approx& \frac{(m-d)(q+\phi_m)}{q(m-d-2)}.
 \end{eqnarray*}
  While $q$ and $\phi_m$ affect absolute power, the relative power of testing a hypothesis against a fixed alternative is mainly driven by the relative sample size, $\frac{m}{n}$. However, the power loss from using $\tilde\beta$ to test hypothesis can be made  negligible because in a big data setting,    we have the luxury of allowing $m$ to be as large as we wish, irrespective of $q$. We will return to the choice of $m$.



\section{Econometrically Motivated Refinements}
As more and more data are being collected, sketching continues to be an active area of research. For any sketching scheme, a solution of higher accuracy can be obtained by iteration. The idea is to approximate  the deviation from an initial estimate  $\Delta=\hat \beta-\tilde\beta^{(1)}$  by solving, for example, 
$\hat\Delta^{(1)}= \argmin_\Delta  \| y-(X(\tilde\beta^{(1)}+\Delta) \|_2^2$
and update $\tilde\beta^{(2)}=\tilde\beta^{(1)}+\hat\Delta^{(1)}$. 
  \citet{pilanci-wainwright:16} starts with the observation that since the least squares objective function is $\|y-X\beta\|_2^2=\|y\|_2^2+\|X\beta\|_2^2-2y^T X\beta$, it is possible  to sketch the quadratic term $\|X\beta\|_2^2$ but not the linear term $y^T X\beta$. The result is  a Hessian sketch of $\beta$, defined as $((\Pi X)^T (\Pi X))^{-1} X^T y$.    \citet{wlmks:17} suggests that this can be seen as a type of Newton updating with the true Hessian replaced by the sketched Hessian, and the iterative Hessian sketch is also a form of iterative random projection.

  In the rest of this section, we consider statistically motivated ways to improve upon $\tilde\beta$. Subsection \ref{sec:avg} considers pooling estimates from multiple sketches. Subsection \ref{sec:choice-m} suggests an $m$ that is motivated by hypothesis testing.

\subsection{Combining Sketches}\label{sec:avg}
The main result of the previous section is that the least squares estimates using sketched data has two errors, one due to a smaller sample size, and one due to sketching. The efficiency loss is hardly surprising, and  there are different ways to improve upon it. \citet{dlfu:13} proposes  a two-stage algorithm that uses $m$ rows of $(y,X)$ to obtain an initial estimate of $\Sigma_{XX}$ and $\Sigma_{Xy}$. In the second stage, the remaining rows are used to estimate the bias of the first stage estimator. The final estimate is a weighted average of the two estimates. An error bound of $O(\frac{\sqrt{K}}{\sqrt{n}})$ is obtained. This bound is independent of the amount of subsampling provided $m> O(\sqrt{K/n})$. \citet{cvsk:16} suggests to choose sample indices from  an importance sampling distribution that is proportional to a sampling score computed from the data. They show that the optimal $p_i$ depends on whether minimizing mean-squared error of $\tilde\beta$ or of $X\tilde\beta$ is the goal, though $\mathbb E[e_i^2]$ plays a role in both objectives.

 The sample size effect is to be expected, and is the cost we pay for not being able to use the full data. But if it is computationally simple to create a sample of size $m$, the possibility arises that we can better exploit information in the data without hitting the computation bottleneck by generating many subsamples and subsequently pool estimates constructed from the subsamples.  \citet{breiman:99} explored an idea  known as {\em pasting bites} that,  when applied to regressions, would repeatedly form  training samples of size $m$ by random sampling from the original data,  then make prediction by fitting the model to the training data. The final prediction is the average of the predictions.  Similar ideas are considered in \citet{chbk:04} and \citet{csh:07}.
Also related  is  distributed computing which takes advantage of  many nodes in the computing cluster. Typically, each machine only sees a subsample of the full data and the parameter of interest is updated.
\citet{heinze-mcwillians-neinshausen} studies a situation when the data are distrubuted across workers according to features of $X$ rather than the sample and show that their \textsc{dual loco} algorithm has bounded approximation error that depends only weakly on the number of workers.

Consider  $\tilde\beta_1,\ldots,\tilde \beta_J$  computed from $J$ subsamples each of size $K$.  As mentioned above, uniform sampling with too few rows is potentially vulnerable to omitting influential observations. Computing multiple sketches also provides the user with an opportunity to check the rank across sketches.  Let  $\tilde t_j=\frac{\tilde\beta_j-\beta_0}{se(\tilde\beta_j)}$ be the $t$ test from sketch $j$.  For a given $m$, we consider sampling without replacement and $J$ is then at most $n/m$. Define the average quantities
\begin{align*}
\bar{\beta} &= \frac{1}{J} \sum_{j=1}^J \tilde{\beta}_j,
\quad\quad &
\text{se}(\bar{\beta}) &=  \sqrt{  \frac{1}{J(J-1)} \sum_{j=1}^J \left[ \text{se} ( \tilde{\beta}_j ) \right]^2 }, \\
\bar{t}_2 &= \frac{1}{J} \sum_{j=1}^J \tilde{t}_j, 
\quad\quad &
\text{se}(\bar{t}_2 ) &= \sqrt{\frac{1}{J-1} \sum_{j=1}^J \left( \tilde{t}_j - \bar{t}_2  \right)^2 }.
\end{align*}
Strictly speaking, pooling requires that  subsamples are non-overlapping and observations are independent across different subsamples. Assuming independence across $j$,
the pooled estimator $\bar{\beta}$ has
$\text{var} ( \bar{\beta} ) = \frac{1}{J^2} \sum_{j=1}^J \text{var} ( \tilde{\beta}_j ).$ Thus,
$
\text{se} ( \bar{\beta} ) \approx \sqrt{ \frac{1}{J^2} \sum_{j=1}^J \left[ \text{se} ( \tilde{\beta}_j ) \right]^2 }\ge \frac{1}{\sqrt{J}} \left[ \frac{1}{J} \sum_{j=1}^J  \text{se} ( \tilde{\beta}_j ) \right]
$
because of Jansen's inequality.
Our estimator for the standard error $\bar{\beta}$ uses
 $J-1$  in the denominator to allow for a correction when  $J$ is relatively small. 
 
 Consider two pooled $t$ statistics:
\begin{subequations}
  \begin{eqnarray}
\label{avg-t1:ttest}
  \bar T_1&=& \frac{\bar\beta-\beta_0}{se(\bar\beta)}\\
  \label{avg-t2:ttest}
  \bar T_2&=& \sqrt{J} \frac{\bar t_2}{se(\bar t_2)}.
\end{eqnarray}
\end{subequations}
Critical values from the standard normal distribution can be used for $\bar T_1$. For example,  for the 5\%-level test, we reject $H_0$ if $|\bar T_1| > 1.96$. For $\bar T_2$,  we recommend using critical values from the $t$ distribution with $J-1$ degrees of freedom.
For example,  for the 5\%-level test, we reject $H_0$ if $|\bar{T}_2| > 2.776$ for $J=5$.

\paragraph{Assumption PI-Avg:}
\begin{itemize}
\item[(i)] $(\Pi_1,\ldots,\Pi_J)$ is independent of $e$;

\item[(ii)]  for all $j, k$ such that $j \neq k$, 
$ \Pi_j  \Pi_j^T = \frac{n}{m} I_m $  and  $ \Pi_j  \Pi_k^T =  O_m$ where
  $O_m$ is an $m \times m$ matrix of zeros.
\item[(iii)] for given singular value distortion parameter $\varepsilon_\sigma\in (0,1)$, there exists failure parameter $\delta_\sigma \in (0,1)$ such that 
$
P \Big(|1-\sigma^2_k(\Pi_j U)|\le \varepsilon_\sigma
\text{ for all $ k=1,\ldots K$  and for all   $j=1,\ldots,J$}
\Big)
\ge   1-\delta_\sigma.
$
\end{itemize}
\medskip
Condition (ii) is crucial; it  is satisfied, for example, if $\Pi_1, \ldots, \Pi_J$ are non-overlapping subsamples and each of them is sampled uniformly  without replacement. 
Assumptions OLS and PI-Avg (i) and (ii) along with the homoskedastic error assumption  ensure that 
\begin{align*}
c^T {\mathbb V}(\bar{\beta}|\Pi_1,\ldots,\Pi_J) c =  \sigma_e^2 \frac{n}{mJ^2} \sum_{j=1}^J
c^T \big( X^T \Pi_j^T \Pi_j X \big)^{-1}  c.
\end{align*}
Condition (iii) is equivalent to the statement that   
$
1-\varepsilon_\sigma \leq \sigma_k (U^T \Pi_j^T \Pi_j U)  \leq 1+\varepsilon_\sigma \;\; \forall \; k=1,\ldots,K,  \;\; \forall \; j=1,\ldots,J.
$

\begin{theorem}\label{thm1-avg}
    Consider $J$ independent sketches  obtained by uniform sampling. Suppose that $e_i\sim (0,\sigma^2_e)$, Assumptions OLS and PI-Avg hold.  Then with probability at leat $1-\delta_\sigma$,  the mean squared error of $c^T \bar{\beta}$ conditional on $(\Pi_1,\ldots,\Pi_J)$ satisifies
\begin{align*}
\frac{c^T {\mathbb V}(\breve\beta | \Pi_1,\ldots,\Pi_J) c}{c^T {\mathbb V}(\hat\beta) c}
&\leq
\frac{n}{mJ}  \frac{1}{(1-\varepsilon_\sigma)}.
\end{align*}

\end{theorem}
The significance of the Theorem is that by choice of $m$ and $J$, the pooled estimator can be almost as efficient as the full sample estimator. If we set $J=1$, the theorem reduces to Theorem~\ref{thm:thm1}. 

 We use a small Monte Carlo experiment to assess the effectiveness of combining statistics computed from different sketches. The data are generated as $y=X\beta+e$ where $e$ is normally distributed, and $X$ is drawn from a non-normal distribution. In \textsc{matlab}, we have \textsc{ X=pearsrnd(0,1,1,5,n,K)}. With $n=1e6$, we consider different values of $m$ and $J$. Most of our results above were derived for uniform sampling, so it is also of interest to evaluate the properties of the $\tilde\beta$ using data sketched by $\Pi$s that  do not satisfy PI.(iii). Four sketching schemes are considered: uniform sampling without replacement labeled as \textsc{rs1}, \textsc{srht},  the countsketch labeled \textsc{cs},  and leverage score sampling, labeled \textsc{lev}. It should be mentioned that results for \textsc{shrt} and \textsc{lev} took significantly longer time to compute than \textsc{rs1} and \textsc{cs}.

The top panel of Table \ref{tbl:table3} reports results for $\hat\beta_3$ when $K=3$. All sampling schemes precisely estimate $\beta_3$ whose true value is one. The standard error is larger the smaller is $m$, which is the sample size effect.  But averaging $\tilde\beta_j$ over $j$ reduces variability. The \textsc{lev} is slightly more efficient. The size of the $t$ test for $\beta_3=1.0$ is accurate, and the power of the test against  $\beta_3<1$ when $\beta_3=0.98$ is increasing in the amount of total information used. Combining $J$ sketches of size $m$ generally gives a more powerful test than a test based on a sketch size of $mJ$. The bottom panel of Table \ref{tbl:table3} reports for $K=9$, focusing on uniform sampling without replacement and the countsketch. The results  are similar to those for $K=3$. The main point to highlight is that  while there is a sample size effect from sketching, it can be alleviated by pooling across sketches.

\subsection{The Choice of $m$}\label{sec:choice-m}

The JL Lemma shows that  $m=O(\log d\epsilon^{-2})$ rows are needed for $d$ vectors from $\mathbb R^n$ to be embedded into an $m$ dimensional subspace. A rough and ready guide for embedding a $d$ dimensional subspace is   $m=\Omega(d\log d\epsilon^{-2})$. This is indeed the generic condition given in, for example, \citet{sarlos-06}, though more can be said for certain $\Pi$s.\footnote{The result for  SRHT is proved in    Lemma 4.1 of \citet{boutidis-gittens:13}. The result for count sketch is from Theorem 2 of \citet{nelson-nguyen:13}.}  Notably, these desired $m$ for random projections  depend only on $d$  but not on $n$.

As shown in \citet[Lemma 4.3]{boutidis-gittens:13},  subspace embedding by  uniform sampling without replacement  requires that
\begin{align}\label{details-uniform}
 m \geq 6 \varepsilon_\sigma^{-2} n \ell_{\max}   \log(2J\cdot K/\delta_\sigma).
\end{align}
where  $\ell_{\max}=\max_i \ell_i$ is the maximum leverage score, also known as coherence. When  coherence is  large, the information in the data is  not well spread out, and  more rows are required for uniform sampling to provide subspace embedding. Hence unlike random projections, the desired $m$ for uniform sampling is not data oblivious.

But while this choice of $m$ is algorithmically desirable, statistical analysis often cares about the variability of the estimates in repeated sampling, and  a larger $m$ is always desirable for $\mathbb V(\tilde\beta)$. The question arises as to whether $m$ can be designed to take both  algorithmic and statistical considerations into account. We suggest two ways to fine-tune the algorithmic  condition.  Now
\begin{align*}
\ell_i=\| U_{(i)} \|^2 
&= {X}_{(i)}^T (X^T X)^{-1} {X}_{(i)} 
= \frac{1}{n}{X}_{(i)}^T S_X^{-1} {X}_{(i)} \\
&\leq  \sigma_1 (S_X^{-1})   \frac{1}{n}  
 \| {X}_{(i)} \|_2^2 = \sigma_K^{-1} (S_X) \frac{1}{n}\|{X}_{(i)}\|_2^2.
\end{align*}
where $\sigma_K(S_X)$ is the minimum eigenvalue of $S_X=n^{-1} X^T X$,  $ {X}_{(i)}^T$ denotes the $i$-th row of $X$, and $X_{(i,j)}$ the $(i,j)$ element of $X$. But
\begin{align*}
\norm{ {X}_{(i)} }_2^2 = \sum_{j=1}^K \left[ X_{(i,j)} \right]^2 \leq K \cdot \max_{j=1,\ldots,K} \left[ X_{(i,j)} \right]^2= K \cdot X_{\max}^2,
\end{align*}
where $X_{\max}= \max_{i=1,\ldots,n} \max_{j=1,\ldots,d}   \left| X_{(i,j)} \right|$. This 
implies $n \cdot \ell_{\max}
\leq  \sigma_K^{-1}(S_X) \cdot K \cdot
X_{\max} ^2$. Recalling that $p_i=\frac{\ell_i}{K}$ defines the importance sampling distribution, we can now restate the algorithmic condition for $m$
when $J=1$  as 
\begin{align}
  \label{eq:maxmaxU}
 m=\Omega\bigg( n K \log (K)\cdot  p_{\max}\bigg) 
 \quad \text{where} \quad
   p_{\max} 
\leq 
\frac{ \sigma^{-1}_K( S_X)   X_{\max}^2}{n}.
\end{align}
It remains to  relate $X_{\max}$   with  quantities of statistical interest.

\paragraph{Assumption M:} 
\begin{itemize}
  \item[a.]   $\sigma_K(S_X)$ is  bounded below by $c_X$ with probability approaching one as $n \rightarrow \infty$; 
 
  \item[b.]    $\mathbb E[ | X_{(i,j)} |^r ]\le C_X$ for some $C_X$ and some $r\ge 2$. 
\end{itemize}
Condition (a)  is a standard identification condition for   $S_X$  to be positive definite so that it will converge in probability  to $\mathbb{E} \left[ {X}_{(i)} {X}_{(i)}^T \right]$. Condition (b)  requires the existence of $r$ moments so as to bound extreme values.\footnote{Similar conditions are used to obtain results for the hat matrix. See, for example,   Section 6.23 of  Hansen (2019)'s online textbook.} If the condition holds,
\[ X_{\max} = o_p ( (nK)^{1/r}).\]

\begin{proposition}\label{prop:prop1}
 Suppose that Assumption M holds. A deterministic rule for  sketched linear regressions by uniform sampling is  
 \[
 \begin{cases}
     m_1&=\Omega\bigg(  (nK)^{1+2/r} \log K/n\bigg)  \quad \text{if } \quad  r<\infty\\
     m_1&=\Omega(K\log(nK)) \quad\quad\quad \quad \quad \text{ if } \quad \mathbb{E} [  \exp( t X_{(i,j)} ) ]  \leq C_X \quad \text{  holds additionally}.
\end{cases}
\]
\end{proposition}

 Proposition \ref{prop:prop1} can be understood as follows.  Suppose that $r=6$  moments are known to exist.  Proposition \ref{prop:prop1} suggests a sketch of $m_1=\Omega (\log K(nK)^{4/3}/n )$ rows which  will generally be larger than $\Omega(K\log K)$, which is the sketch size suggested for data with thin tails. For such data,  the moment generating function is uniformly bounded and
  $\mathbb{E} [  \exp( t X_{(i,j)} ) ]  \leq C_X$ for some constant $C_X$ and some $t > 0$ so that  $X_{\max} = o_p (\log (nK))$.  In both cases, the desired $m$
 increases with the row sample size $n$, the number of regressors $K$, as well as  $n\cdot K$, which is the number of data points in the regressor matrix $X$. This contrasts with the algorithmic condition for $m$ which does not depend on $n$.
 
   To use Proposition \ref{prop:prop1}, we can either (i) fix $r$ to determine $m_1$ or (ii) target  `observations-per-regressor ratio'.  As an example, suppose $n=1e7$ and $K=10$. If  $r = 6$, Proposition 1 suggests to sample $m_1= 10,687$ rows, implying  $\frac{m_1}{K}\approx 1000$. If instead we fix  $\frac{m}{K}$ at  100 and uniformly sample $m_1=1000$ rows, we must be ready to defend  the existence of 
$r= \frac{2\log (nK)}{\log (\frac{m}{K}) -\log(\log K)}\approx 10$ moments.  There is a clear trade-off between $m_1$  and $r$.

Though  $m_1$   depends on $n$, it is still a deterministic rule. To obtain a rule that is data dependent,  consider again  $c^T \tilde \beta$, where $c$ is  a $K\times 1$ vector, and assume that $e_i\sim N(0,\sigma^2)$ so that 
$ \text{var}(\tilde\beta)= \frac{n}{m}\sigma^2_e(\tilde X^T \tilde X)^{-1}$
where $\beta_0$ is  the  true (unknown) value of $\beta$. Define 
\[
  \tau_0(m)=\frac{c^T (\tilde{\beta} - \beta_0)}{
\textsc{se}(c^T \tilde\beta)}=
\frac{c^T (\tilde{\beta} - \beta^0) + c^T (\beta^0 - \beta_0)}{
\textsc{se}(c^T \tilde\beta)}.
\]
It holds that  
$
P_{\beta_0} 
( \tau_0
> z
| \Pi, X_1,\ldots,X_n
)
= \Phi (-z)
$ for some $z$, where $\Phi(\cdot)$ is the cdf of the standard normal distribution. Now consider a one-sided test $\tau_1$ based on $\tilde\beta$ against an alternative, say, $\beta^0$.  The test $\tau_1$ is related to $\tau_0$ by
\[
\tau_1(m) =  \frac{c^T (\tilde\beta-\beta_0)}{\textsc{se}(c^T \tilde\beta)} - \frac{c^T(\beta^0-\beta_0)}{\textsc{se}(c^T \tilde\beta)}=\tau_0(m)+\tau_2(m). 
\]
The power of $\tau_1$ at nominal size $\alpha$ is  then
\begin{align*}
P_{\beta_0} 
( 
\tau_0+\tau_2
> \Phi^{-1}_{(1-\alpha)}
| \Pi, X_1,\ldots,X_n
)
&= \Phi 
\left(-\Phi^{-1}_{(1-\alpha) }
+ \tau_2
 \right)\equiv
 \gamma.
\end{align*}
 Define \[S(\alpha,\gamma)=\Phi^{-1}_{ \gamma}+\Phi^{-1} _{1-\alpha}.\]
Common values of $\alpha$ and $\gamma$ give the following:

\begin{center}
Selected values of $S(\alpha,\gamma)$

  \begin{tabular}{l|lllll} 
    & \multicolumn{5}{c}{$\gamma$} \\ \hline
$\alpha$    & 0.500& 0.600& 0.700& 0.800& 0.900\\ \hline
0.010& 2.326& 2.580& 2.851& 3.168& 3.608\\
0.050& 1.645& 1.898& 2.169& 2.486& 2.926\\
0.100& 1.282& 1.535& 1.806& 2.123& 2.563\\
\hline
  \end{tabular}
\end{center}

\begin{proposition}
  \label{prop:prop2}
Suppose that $e_i\sim N(0,\sigma^2_e)$ and the Assumptions of Theorem 1 hold.  Let $\bar\gamma$ be the  target power of a one-sided test $\tau_1$ and $\bar\alpha$ be the nominal size of the test.  
\begin{itemize}
  \item Let  $\tilde\beta$ be obtained from a sketch of size $m_1$.  For a given  effect size of $\beta^0-\beta_0$,  a data dependent `inference conscious' sketch size is 
\begin{equation}
  \label{m2rule1} m_2(m_1)=S^2(\bar\alpha,\bar\gamma) \frac{m_1 \text{var}(c^T \tilde\beta)}{[c^T (\beta^0-\beta_0)]^2}=m_1 \frac{S^2(\bar\alpha,\bar\gamma)}{\tau^2_2(m_1)}.
\end{equation}
\item For a pre- specified $\tau_2(\infty)$, a data oblivious `inference conscious' sketch size is 
\begin{equation}
  \label{m2rule2}
  m_3=n\frac{S^2(\bar\alpha,\bar\gamma)}{\tau_2^2(\infty)}. 
\end{equation}
\end{itemize}
\end{proposition}
 Inference considerations suggests to adjust $m_1$ by a factor that depends on $S(\bar\alpha,\bar\gamma)$ and $\tau_2$. 
 For instance, when $\bar\gamma \geq 0.8$ and/or $\bar\alpha \leq 0.01$,
   $m_1$ will be adjusted upwards when the $\tau_2$ is less than two. The precise adjustment depends on the choice of $\tau_2$.

The proposed $m_2$ in Part (i) requires an estimate of $\textsc{var}(c^T\tilde\beta)$ from a preliminary sketch. Table \ref{tbl:table4} provides an illustration for one draw of simulated data with
$n=1e7$ and $K=10$. We consider three values of $\sigma_e, \bar\gamma$, as well as different effect size $\beta_1^0-\beta_{10}$.
Assuming $r=10$ and $\sigma_e=0.5$,  an effect size of 0.025 gives an $m_1$ of roughly 1000. Using a sketch of this size to obtain an estimate of $\tau_2$  will almost hit the target power of 0.5. However, a target power of 0.9 would require $m_2=3,759$, almost four times as many rows as a target power of 0.5.
 The larger is $\sigma_e$, the less precise is $\tilde\beta$ for a given $m$, and  more rows are needed. It is then up to the user how  to trade of computation cost and power of the test.

The proposed $m_3$ in Part (ii)  is motivatead by the fact that setting $m_1$ to $n$  gives $m_2(n)=n\frac{S^2(\bar\alpha,\bar\gamma)}{\tau_2^2(n)}$. Though computation of $\tau_2(n)$ is infeasible,  $\tau_2(n)$ is asymptotically normal as $n\rightarrow\infty$. Now
if the full sample $t$-statistic  cannot reject the null hypothesis, a  test based on sketched data will unlikely reject the null hypothesis.    But when full sample $t$ statistic  is expected to be relatively large (say, 5), the result can be used in conjunction with $S(\bar\alpha,\bar \gamma)$ to give $m_3$.  
  Say if  $S(\bar\alpha,\bar\gamma)$ is 2,  $m_3=(2/5)^2 n$. This  allows us to gauge the sample size effect since  $\frac{n}{m}= \frac{\tau_2^2(\infty)}{S^2(\bar\alpha,\bar\gamma)}.$ Note that $m_3$  only requires the choice of $\bar \alpha, \bar\gamma,$ and $\tau_2(\infty)$ which, unlike $m_2$, can be computed without a preliminary sketch.

Though Propositions \ref{prop:prop1} and \ref{prop:prop2} were derived for uniform sampling, they can still be used for other choice of $\Pi$. The one exception in which some caution is warranted is  the countsketch.  The rule given for the countsketch in \citet[Theorem 5]{nelson-nguyen:13} of
$m \geq  \varepsilon^{-2} K(K+1)\delta^{-1}$.
 Though such an $m$ is data oblivious, it is generally larger than the rule given by \citet{boutidis-gittens:13}  for uniform sampling. The larger $m$ required for countsketch  can be seen as a cost of specifying a sparse $\Pi$. Thus,  one might want to first  use a small $r$  to obtain a conservative $m_1$ for the countsketch. One can then  use  Proposition \ref{prop:prop2} to obtain  an `inference conscious'  guide.

 To illustrate how to use $m_1,m_2$ and $m_3$, we consider  \citet{belenzon-chatterji-daley:17} which studies  firms' performance  from naming the company after its owners, a phenomenon known as   eponymy.  
 The parameter of interest is $\alpha_1$  in a `return on assets' regression
 \[ \text{roa}_{it} = \alpha_0 + \alpha_1 \text{eponymous}_{it}+Z_{it}^T \beta + \eta_i + \tau_t + c_i + \epsilon_{it}.
 \]
  The  coefficient gives the effect of the  eponymous dummy   after controlling for time varying firm specific variables $Z_{it}$, SIC dummies $\eta_i$, country dummies $c_i$, and year dummies $\tau_t$. The panel of data includes 1.8 million companies from 2002-2012, but we only use data for one year. An interesting aspect of this regression is that even in the full sample with $n=562160$,  some dummies are sparse while others are collinear, giving an effective number of $K=423$ regressors. We will focus on the four covariates: the indicator variable for being eponymous, the log of assets, the log number of shareholders, and equity dispersion. 

Given  the values of $(n,K)$ for this data, any assumed value of $r$ less than 8 would give an $m_1$ larger than $n$ which is not sensible.\footnote{{This is based on 
$m_1 = (nK)^{1+2/r} \log K/n$. A smaller $r$ is admissible if $m_1 = c_1 (nK)^{1+2/r} \log K/n$ for some constant $0 < c_1 < 1$. We limit our attention to $c_1 = 1$.}}
This immediately restricts us to $r\ge 6$. As point of reference, $(r,m_1)=(8,317657)$ and $(r,m_1)=(15,33476)$. The smallest $m_1$ is obtained by assuming that the data have thin tails, resulting in $m_1=K\log (nK)=8158$.  

Table \ref{tbl:table_EPO} presents the estimation results for several values of $m$.   The top panel presents results for uniform sampling. Note that more than 50 covariates for uniform sampling are omitted  due to colinearity, even with a relatively large sketch size. The first column shows the full sample estimates  for comparison. Column (2) shows that the point estimates given by the smallest sketch with $m_1= 8158$  are not too different from those in column (1), but the precision estimates are much worse. To solve this problem, we compute $m_2 (m_1)$ by plugging in the $t$ statistic for  equity dispersion (ie. $\hat \tau_0= 0.67$) as $\tau_2$. This  gives an $m_2$ of 112358.  A similar sketch size can be obtained  by assuming $r = 10$ for $m_1$, or by plugging in  $\tau_2(\infty) = 5$ for $m_3$. As seen from  Table \ref{tbl:table_EPO}, the point estimates of all sketches are similar, but the  inference conscious sketches are larger in size and give larger test statistics. 

The bottom panel of Table \ref{tbl:table_EPO} presents results for the countsketch.  Compared to uniform sampling in the top panel, only one or two covariates are now dropped. Though the estimate of $\alpha_1$  is almost almost identical to the one for the full sample and for uniform sampling, the estimated coefficient for  equity dispersion is somewhat different.   This might be due to the fact that uniform sampling drops much more covariates than countsketch.

\section{Concluding Remarks}
This paper provides an gentle introduction to sketching and studies its implications for prediction and inference using a linear model. Sample codes for constructing the sketches are avaialble in \textsc{matlab}, \textsc{R}, and \textsc{stata}. Our main findings are as follows:
\begin{enumerate}[itemsep=0mm]
\item  Sketches incur an approximation error that is small relative to the sample size effect. 
\item For speed and parallelization, it is best to use  countsketch.
\item For simple implementation, it is best to use uniform sampling.
\item For improved estimates, it is best to  average over multiple sketches.

\item Statistical analysis may require larger sketch size than what is algorithmically desirable.  We propose two  inference conscious  rules for the sketch size.
\end{enumerate}

Sketching has also drawn attention of  statisticians in recent years.  \citet{ahfock-astle-richardson} provides  an inferential framework  to obtain distributional results for a large class of sketched estimators. \citet{geppert-etal:17} considers random projections in Bayesian regressions and provides sufficient conditions for   a Gaussian likelihood based on sketched data to have an error of  $1+O(\epsilon)$  in terms of $L_2$ Wasserstein distance.  In the design of experiments literature, the goal is to reveal as much information as possible given a fixed budget.\footnote{A criterion that uses the trace norm for ordering matrices is $A$-optimality. A criterion that uses the determinant to order matrices is a $D$-optimal design.} Since sketching is about forming random samples, it is natural to incorporate the principles in design of experiments. \citet{wang-zhu-ma:18}  considers the design of subsamples for logistic regressions. $A$-optimality  and practical considerations suggest to use
$p_i=\frac{|\hat e_i| \| x_i\|}{\sum_i|\hat e_i| \| x_i\|},$
which may be understood  as score based sampling.
\citet{wang-yang-stufken:19} considers the homoskedastic normal linear regression model. The principle of  $D$-optimality suggests to  recursively selecting data according to extreme values of covariances. The algorithm is suited for distributed storage and parallel computing.

Though analysis of the linear regression is an absolute first step in understanding the use and implications of sketching, the real benefits are expected to be in estimation problems that are complex. For example,  \citet{deaton-ng} uses `binning methods' and uniform sampling  to speed up estimation of non-parametric average derivatives.  \citet{portnoy-koenker:97} uses a fast interior point method for quantile regression by preprocessing a random subsample of data to reduce the effective sample size. More generally, if the full-sample Hessian matrix is difficult to estimate, one can consider an approximation of it by subsampling.  This paper provides a new perspective to evaluating the effectiveness of matrix approximations and aims to better understand their implications for inference.

 While using sketches to overcome the computation burden  is a step forward, sometimes we need more than a basic sketch.   We have been silent about how to deal with data that are  dependent  over time or across space, such as due to network effects. We may want our sketch to preserve,  say, the size distribution of firms in the original data. The sampling algorithms considered in this review must then satisfy additional conditions. When the data have a probabilistic structure, having  more data is not always desirable, \citet{boivin-ng-joe}. While discipline-specific problems require discipline-specific  input,  there is also a lot to learn from what has already been done in other literatures. Cross-disciplinary work is a promising path towards  efficient  handling of large volumes of data.


\newpage
\appendix

\section*{Appendices}

\section{Proof of Lemma \ref{lem:sarlos}}


\paragraph{Proof of Lemma \ref{lem:sarlos}}
By an orthogonal decomposition of the least squares residuals,
 \begin{eqnarray}
  \|y-X\tilde\beta\|_2^2&=&\| y-X\hat \beta\|_2^2+ \| X\tilde\beta-X\hat\beta\|_2^2 \nonumber\\
&=&  \| y-U\hat \theta\|_2^2+\| U(\tilde \theta-\hat\theta)\|_2^2 \nonumber \\
&=& \widehat{\textsc{ssr}}^2+ \|U(\tilde\theta-\hat\theta)\|\nonumber \\
&=& \widehat{\textsc{ssr}}^2+ \|\tilde\theta-\hat\theta\|_2, \label{eq:result2}
\end{eqnarray}
where
  \begin{eqnarray*}
 \| \tilde\theta-\hat\theta\|_2&=& \| (\Pi U)^T \Pi U(\tilde\theta-\hat\theta)+ (\tilde\theta-\hat\theta)+ (\Pi U)^T \Pi U (\hat\theta-\tilde\theta)\|_2 \nonumber\\
  &\le & \| (\Pi U)^T \Pi U (\tilde\theta-\hat\theta)\|_2+ \| (\Pi U)^T \Pi U (\hat\theta-\tilde\theta)-(\hat\theta-\tilde\theta)\|_2\nonumber\\
 &\le & \| (\Pi U)^T \Pi U (\tilde\theta-\hat\theta)\|_2+ \| (\Pi U)^T \Pi U -I_d\|_2 \|(\tilde\theta-\hat\theta)\|_2\\
 &\le&  \| (\Pi U)^T (\Pi U) (\tilde\theta-\hat\theta)\|_2+\frac{1}{\sqrt{2}} \|\tilde\theta-\hat\theta\|_2\\
  &\le &\sqrt{2}\|(\Pi U)^T \Pi U (\tilde\theta-\hat\theta)\|_2
 \end{eqnarray*}
 by triangle inequality, Cauchy-Schwarz inequality,
 condition (\ref{eq:condition1}), and rearranging terms.
 Now  the normal equations implies
$ (\Pi U)^T (\Pi U)\tilde\theta= (\Pi U)^T \Pi (y-X\tilde\beta)$.
Hence
 \begin{eqnarray*}
 \|\tilde\theta-\hat\theta\|_2 &\le &\sqrt{2}\|  (\Pi U) ^T  \Pi  (y-U\hat\theta)\|_2
 \\ &\le &\sqrt{\epsilon_0} \; \widehat{\textsc{ssr}}
\end{eqnarray*}
by condition (\ref{eq:condition2}) and for some failure probability $\delta_0$. It follows from (\ref{eq:result2}) that $\widetilde{\textsc{ssr}}^2  \le (1+\epsilon_2)  \widehat{\textsc{ssr}}^2$
holds with probability $1-\delta_2$ where $\epsilon_2=\epsilon_0^2$ and $\delta_2<2\delta_0$. This probability can be made higher with suitable choice of $\epsilon_0$ and $m$, which can be controlled by the researcher.

\paragraph{Proof of Theorem \ref{thm1-avg}}
Note that
\begin{eqnarray*}
c^T (\bar{\beta} - \beta) &=&  \frac{1}{J} \sum_{j=1}^J
c^T \big( X^T \Pi_j^T \Pi_j X \big)^{-1} \big(  X^T \Pi_j^T \Pi_j e \big).
\end{eqnarray*}
Thus,
\begin{align*}
E[ c^T (\bar{\beta} - \beta) |\Pi_1,\ldots,\Pi_J]
&= \frac{1}{J} \sum_{j=1}^J c^T \big( X^T \Pi^T \Pi X \big)^{-1} \big(  X^T \Pi^T \Pi E[e|\Pi_1,\ldots,\Pi_J] \big) =0.
\end{align*}
Write
\begin{align*}
&E[   \{ c^T (\bar{\beta} - \beta) \}^2 |\Pi_1,\ldots,\Pi_J ] \\
&= \frac{1}{J^2} \sum_{j=1}^J \sum_{k=1}^J
c^T \big( X^T \Pi_j^T \Pi_j X \big)^{-1} \big(  X^T \Pi_j^T \Pi_j E[ e e^T |\Pi_1,\ldots,\Pi_J ] \Pi_k^T \Pi_k X \big) \big( X^T \Pi_k^T \Pi_k X \big)^{-1} c \\
&= \sigma_e^2 \frac{1}{J^2} \sum_{j=1}^J
c^T \big( X^T \Pi_j^T \Pi_j X \big)^{-1} \big(  X^T \Pi_j^T \Pi_j  \Pi_j^T \Pi_j X \big) \big( X^T \Pi_j^T \Pi_j X \big)^{-1} c \\
&+ \sigma_e^2 \frac{1}{J^2} \sum_{j=1}^J \sum_{k=1, k \neq J}^J
c^T \big( X^T \Pi_j^T \Pi_j X \big)^{-1} \big(  X^T \Pi_j^T \Pi_j  \Pi_k^T \Pi_k X \big) \big( X^T \Pi_k^T \Pi_k X \big)^{-1} c \\
&= \sigma_e^2 \frac{n}{mJ^2} \sum_{j=1}^J
c^T \big( X^T \Pi_j^T \Pi_j X \big)^{-1}  c.
\end{align*}
Then, 
the desired result is obtained by arguments identical to those used in proving Theorem~\ref{thm:prediction}.
In particular,   we can show that
\begin{align}\label{important-step-0-best-more-avg}
\frac{n}{m} \frac{ c^T \big( X^T \Pi_j^T \Pi_j X \big)^{-1}  c}{c^T   (X^T X)^{-1}  c}
&\leq    \frac{n}{m} \frac{1}{(1-\varepsilon_\sigma)}
\end{align}
 jointly for all $j=1,\ldots,J$ with probability at least $1-\delta_\sigma$. 
 \;$Q.E.D.$

\section{Verification of Assumption CS for Countsketch}

In this part of the appendix, we verify Assumption CS for the countsketch.
Here, we use $d$ to denote the column dimension of $\Pi$.

\begin{lemma}\label{thm-sub-emb-count-more}
Let $\Pi \in \mathbb{R}^{n \times d}$ be a random matrix such that
(i) the $(i,j)$ element $\Pi_{ij}$ of $\Pi$ is $\Pi_{ij} = \delta_{ij} \sigma_{ij}$, where
$\sigma_{ij}$'s are i.i.d. $\pm 1$ random variables and $\delta_{ij}$ is an indicator random variable for the event $\Pi_{ij} \neq 0$;
(ii)  $\sum_{i=1}^{m} \delta_{ij} = 1$ for each $j = 1,\ldots,n$;
(iii) for any $S \subset [n]$, $\mathbb{E} \left( \Pi_{j \in S} \delta_{ij} \right) = m^{-|S|}$;
(iv) the columns of $\Pi$ are i.i.d.
Furthermore, there is a universal constant $C_e$
such that
$\max_{i=1,\ldots,n} \Omega_{e,ii} \leq C_e$.
Suppose that
\begin{align}\label{sub-emb-count-use-growth-rate}
  \frac{ (d^2+1)m}{n} + \frac{(d^2+d)}{m}
\leq \frac{\delta_\Pi \varepsilon_{\Pi}^{2}}{ 8 C_e^2}.
\end{align}
Let
\begin{align}\label{A-choice-thm}
A (\Omega_{e}, m, n) =  \Omega_{e} +  \frac{1}{m} \left\{ \text{tr}(\Omega_{e}) I_n -  \Omega_{e}
\right\}.
\end{align}
Then, we have that
\begin{align*}
\mathbb{P} \left( \left\| U^T \Pi^T \Pi \Omega_{e} \Pi^T \Pi U -  U^T A (\Omega_{e}, m,n) U \right\|_2 > \frac{n}{m} \varepsilon_{\Pi} \right)
&\leq \delta_\Pi.
\end{align*}
\end{lemma}

This lemma states that  Condition CS is satisfied for countsketch, provided that
all the diagonal elements of $\Omega_{e}$ are bounded by a universal constant,
$d^2 m/n = o(\delta_\Pi \varepsilon_{\Pi}^{2})$ and $d^2 /m = o(\delta_\Pi \varepsilon_{\Pi}^{2})$.
The rate conditions in \eqref{sub-emb-count-use-growth-rate} are not stringent.
 When $n$ is very large, $d$ is of a moderate size
and $\delta_\Pi$ and $\varepsilon_{\Pi}$ are given,  there is a range of $m$ that satisfies \eqref{sub-emb-count-use-growth-rate}.

\noindent
\paragraph{Proof of Lemma \ref{thm-sub-emb-count-more}}
Since we assume that each diagonal element of $\Omega_{e}$ is  bounded by a universal constant $C_e$,
\begin{align*}
\text{tr}(\Omega_{e}^2)  \leq C_e^2 n, \;
\text{tr}(\Omega_{e}) = C_e n, \; \text{ and } \;
\| \Omega_{e} \|_2 = C_e.
\end{align*}
Then Lemma F.1, which is given in Appendix F, implies that
\begin{align*}
&\mathbb{P} \left( \left\| U^T \Pi^T \Pi \Omega_{e} \Pi^T \Pi U -  U^T A (\Omega_{e}, m,n) U \right\|_2 > \epsilon \right)  \\
&\leq
  2 \epsilon^{-2}  \bigg\{ \frac{2 d^2 (m-1)}{m^2} \text{tr}(\Omega_{e}^2)
+ \frac{2 d^2}{m^2} \| \Omega_{e} \|_2  \text{tr}(\Omega_{e})
+  \frac{d^2}{m^3}    \left\{ \left[ \text{tr}(\Omega_{e})  \right]^2 + 2 \| \Omega_{e} \|_2^2  \right\} \\
&\;\;\;\;\;\;\;\;\;\;\;\; + \frac{2}{m}  \text{tr}(\Omega_{e}^2)
+ \frac{2}{m^2} \text{tr}(\Omega_{e}) +  \frac{1}{m^3}    \left\{ d \left[  \text{tr}(\Omega_{e})  \right]^2  + 2 \text{tr}(\Omega_{e}^2)  \right\} \bigg\} \\
&\leq 2 \epsilon^{-2}  \bigg\{ \frac{2 d^2 (m-1)}{m^2} C_e^2 n
+ \frac{2 d^2}{m^2} C_e^2 n
+  \frac{d^2}{m^3}    \left\{ C_e^2 n^2 + 2 C_e^2  \right\} \\
&\;\;\;\;\;\;\;\;\;\;\;\; + \frac{2}{m}  C_e^2 n
+ \frac{2}{m^2} C_e n +  \frac{1}{m^3}    \left\{ d C_e^2 n^2  + 2 C_e^2 n  \right\} \bigg\} \\
&\leq 8 C_e^2 \epsilon^{-2}  \bigg(   \frac{ (d^2+1)n}{m} + \frac{(d^2+d)n^2}{m^3}     \bigg).
\end{align*}
If we take $\epsilon = \frac{n}{m} \varepsilon_{\Pi}$, then
\begin{align*}
\mathbb{P} \left( \left\| U^T \Pi^T \Pi \Omega_{e} \Pi^T \Pi U -  U^T A (\Omega_{e}, m,n) U \right\|_2 > \frac{n}{m} \varepsilon_{\Pi} \right)
&\leq 8 C_e^2 \varepsilon_{\Pi}^{-2}
\bigg(   \frac{ (d^2+1)m}{n} + \frac{(d^2+d)}{m}     \bigg).
\end{align*}
To satisfy the probability above is bounded by $\delta_\Pi$, we need to assume that
\begin{align*}
8 C_e^2 \varepsilon_{\Pi}^{-2}
\bigg(   \frac{ (d^2+1)m}{n} + \frac{(d^2+d)}{m}     \bigg) \leq \delta_\Pi,
\end{align*}
which is imposed by \eqref{sub-emb-count-use-growth-rate}.
\emph{Q.E.D.}
\medskip

\section{Proof of JL Lemma}
The original proof assumes that $\Pi$ is Gaussian.\footnote{Subsequent proofs by \citet{dasgupta-gupta:03}, \citet{indyk-motwani}, \citet{matousek:08} use different proof  techniques  to obtain tighter bounds. \citet{fedoruk-etal:18} summaries the evolution of the Lemma. Others consider non-Gaussian  $\Pi$ matrices that are less costly to store.}  We highlight  arguments  for this  case  using properties of sub-exponential random variables. For arbitrary $(i,j)$, define  $u=(a_i-a_j)\in\mathbb R^n$ and $v=\Pi a_i-\Pi a_j=\frac{1}{\sqrt{m}}\Pi u\in\mathbb R^m$. Note that $v_i=\frac{1}{\sqrt{m}} \sum_j \Pi_{ij} u_j\sim N(0,\frac{||u||_2^2}{m})$. The Lemma  then says
$\bigg| \frac{||v||^2_2}{||u||^2_2}-1\bigg| \le \epsilon.$
The proof of this statement proceeds in three steps.

\begin{itemize}
\item[(i)] show that the expected value of the Euclidean distance of the random projection is equal to the Euclidean distance of the original subspace:
\begin{eqnarray*}
\mathbb E[||v||_2^2]&=& \mathbb E[\sum_{i=1}^m v_i^2]=\sum_{i=1}^m E[v_i^2] \\
&=& \sum_{i=1}^m \frac{1}{m} \mathbb E\bigg[ (\sum_j \Pi_{ij}u_i)^2\bigg]
=\sum_{i=1}^m \frac{1}{m}\sum_{1\le j,k\le n} u_ju_k \Pi_{ij}\Pi_{ik}\\
&=& \sum_{i=1}^m \frac{1}{m} \sum_{1\le j,k\le n} u_j u_k \delta_{jk}=\sum_{i=1}^m\frac{1}{m}\sum_{1\le j\le n}u_j^2\\
&=& \sum_{j\le n} u_j^2=||u||_2^2.
\end{eqnarray*}

 \item[(ii)] show that (i) holds with high probability. For this, let  $Z_i=\frac{\Pi_i u}{||u||} $. Then $\sum_{i=1}^m Z_i^2=\frac{m || v||_2^2}{||u||_2^2}$  is $\chi^2_m$, hence sub-exponential.\footnote{A random variable $X$ with mean $\mu$ is sub-exponential with parameters $(\nu,b)$ if $\mathbb E[e^{\lambda(X-\mu)}]\le e^{\nu^2\lambda^2/2}$. If $Z_k\sim N(0,1)$, $ Z_k^2\sim \chi^2_m$ is sub-exponential with $(\nu,b)=(2,4)$ and has two-sided tail bound
$ P\bigg( \bigg| \frac{1}{\sqrt{m}}\sum_{k=1}^m Z_k^2-1\bigg| \ge \epsilon\bigg)\le 2e^{-m \epsilon^2/8}, \quad \epsilon\in(0,1).$ See, for example,
  \citet[Chapter 2]{wainwright-book} or \citet[lecture 1]{vershynin:17}.}:
A two-sided tail bound yields
  \[P\bigg(\frac{1}{m}\bigg|\sum_{k=1}^m Z_k^2-1\bigg| \ge \epsilon\bigg)=P\bigg(\bigg| \frac{||v||_2^2}{||u||_2} -1\bigg|  > \epsilon \bigg) \le  2e^{-m\epsilon^2 /8}.
  \]

\item[(iii)] apply union bound: step (ii) holds for all $d^2$ pairs of $(i,j)$, so
 the overall failure probability is at most  $2\begin{pmatrix} d \\ 2\end{pmatrix} e^{-m\epsilon^2 /8}$. Given a error rate $\epsilon$, this failure probability can be driven below $\delta$ by making   $m\ge \frac{C}{\epsilon^2}
 (\log d/\delta)$ for large enough $C$.

\end{itemize}

\newpage

\section{Streaming Implementation of Countsketch}

{\bf CountSketch, Streaming:}

 \begin{algorithm}[H]
\DontPrintSemicolon

\KwInput{Given $A\in\mathbb R^{n\times d}$ and $m<n$}
\KwOutput{$\tilde A\in\mathbb R^{m\times d}$, initialized to zero}
\For{$s=1:n$}
{
    {sampled $h$ (from $[1,\ldots m]$)}

    { sample $g$ from $[+1, -1]$}

    {$\tilde A_{(h(s))}:=\tilde A_{(h(s))}+ g\times A_{(h(s))}$}

}
\end{algorithm}

\noindent The example considered in the main text is
 the same as $C=\Pi A$, where $n=9$,
\begin{eqnarray*}
\Pi&=& \begin{pmatrix}
0 & 0 & 1 & 0 & 1 &-1 & 0& 0 & 1\\
-1 &0 &   0 &-1 &   0 & 0 &   0&  -1  &  0\\
0 &-1 &   0 & 0 &   0 & 0 &   1 & 0  &  0 \end{pmatrix}
\; \text{ and } \;
A
=\begin{pmatrix} A_3+A_5-A_6+A_9\\ A_1-A_4 - A_8\\
A_2 +A_7\end{pmatrix}.
\end{eqnarray*}
Hence if $d=4$,
\begin{eqnarray*}
\tilde A=
\begin{pmatrix}
    A_{31}+A_{51}-A_{61}+A_{91} & A_{32}+A_{52}-A_{62}+A_{92}& A_{33}+A_{53}-A_{63} +A_{93}& A_{34}+A_{54}-A_{64}+A_{94}\\
   -A_{11}-A_{41}-A_{81}   &-A_{12}-A_{42}-A_{82}   &-A_{13}-A_{43}-A_{83}   &-A_{14}-A_{44}-A_{84}\\
   -A_{21}+A_{71}     &-A_{22}+A_{72} & -A_{23}+A_{73} & -A_{24}+A_{74}
\end{pmatrix}.
\end{eqnarray*}

\noindent Consider now the streaming approach when the random draws of $h$ and $g$ are
\begin{eqnarray*}
\begin{pmatrix} h \\ g\end{pmatrix} &=& \begin{pmatrix}
2 &  3 &  1 &  2 &  1 &  1 &  3 &  2 &  1\\
-1 & -1 & +1 & -1 & +1 & -1 & +1 & -1 & 1
\end{pmatrix}
\end{eqnarray*}
\begin{enumerate}
\item $s=1$: $A_{(1)} = \begin{pmatrix}   A_{11} & A_{12} & A_{13} & A_{14 }\end{pmatrix}$, $h=2,g=-1$. Updating $A_{h(1)}=A_2$ gives
\begin{eqnarray*}
\tilde A &=& \begin{pmatrix}
            0     &    0  & 0     &  0     \\
A_{11}\times -1 &   A_{12}\times -1 & A_{13}\times -1  & A_{14}-1\\
            0       &     0  &              0  &   0 \end{pmatrix}.
\end{eqnarray*}
\item $s=2$: $A_{(2)} = \begin{pmatrix}
 A_{21}&  A_{22} & A_{23} & A_{24}\end{pmatrix}$, $h=3,g=-1$. Updating $A_{h(2)}=A_3$ gives
\begin{eqnarray*}
\tilde A&=&   \begin{pmatrix}
           0 &        0  &       0 &      0\\
    -A_{11} &  -A_{12} & -A_{13} & -A_{14}\\
    A_{21}\times -1  &  A_{22}\times -1 & A_{23}\times -1 &  A_{24}\times -1\
    \end{pmatrix}
\end{eqnarray*}
\item $s=3$:  $A_{(3)} = \begin{pmatrix}
 A_{31}&  A_{32} & A_{33} & A_{34}\end{pmatrix}$, $h=1,g=1$. Updating $A_{h(3)}=A_1$ gives
\begin{eqnarray*}
\tilde A&=&      \begin{pmatrix}
A_{31} &   A_{32}  &     A_{33}  &  A_{34}\\
    -A_{11} &  -A_{12} &  -A_{13} & -A_{14}\\
    -A_{21} &  -A_{22}& -A_{23} &  -A_{24}
\end{pmatrix}
\end{eqnarray*}
\item $s=4$:  $A_{(4)} = \begin{pmatrix}
A_{41}& A_{42}& A_{43} & A_{44} \end{pmatrix}$, $h=2,g=-1$. Updating $A_{h(4)}=A_2$ gives
\begin{eqnarray*}
\tilde A&=&\begin{pmatrix}
A_{31}\times 1    &        A_{32}\times 1    &       A_{33}\times 1&           A_{34}\times 1 \\
 -A_{11}+(A_{41}\times -1) &  -A_{12}+(A_{42}\times -1) &  -A_{13}+(A_{43}\times -1) &  -A_{14}+(A_{44}\times -1) \\
            -A_{21}   &      -A_{22}     &   -A_{23}    & -A_{24}
            \end{pmatrix}
\end{eqnarray*}
\item $s=5$: $A_{(5)}=\begin{pmatrix}
A_{51}& A_{52}& A_{53} & A_{54} \end{pmatrix}$, $h=1,g=1$. Updating $A_{h(5)}=A_1$ gives
\begin{eqnarray*}
\tilde A&=&\begin{pmatrix}
A_{31}+A_{51}    &        A_{32}+A_{52}    &       A_{33}+A_{53}&           A_{34}+A_{54} \\
 -A_{11}-A_{41} &  -A_{12}-A_{42} &  -A_{13}-A_{43} &  -A_{14}-A_{44}
  \\
            A_{21}\times -1   &           A_{22}\times -1     &     A_{23}\times -1    &      A_{24}\times -1
            \end{pmatrix}
\end{eqnarray*}
\item $s=6$: $A_{(6)}=\begin{pmatrix}
A_{61}& A_{62}& A_{63} & A_{64} \end{pmatrix}$, $h=1,g=-1$. Updating $A_{h(6)}=A_1$ gives
\begin{eqnarray*}
\tilde A&=&\begin{pmatrix}
A_{31}+A_{51}-A_{61}    &   A_{32}+A_{52}-A_{62}    &   A_{33}+A_{53}-A_{63}&           A_{34}+A_{54}-A_{64} \\
-A_{11}-A_{41} &  -A_{12}-A_{42} &  -A_{13}-A_{43} &  -A_{14}-A_{44}\\
            A_{21}\times -1   &           A_{22}\times -1     &     A_{23}\times -1    &      A_{24}\times -1
            \end{pmatrix}
\end{eqnarray*}
\item $s=7$: $A_{(7)}=\begin{pmatrix}
A_{71}& A_{72}& A_{73} & A_{74} \end{pmatrix}$, $h=3,g=1$. Updating $A_{h(7)}=A_3$ gives
\begin{eqnarray*}
\tilde A&=&\begin{pmatrix}
A_{31}+A_{51}-A_{61}    &   A_{32}+A_{52}-A_{62}    &   A_{33}+A_{53}-A_{63}&           A_{34}+A_{54}-A_{64} \\
-A_{11}-A_{41} &  -A_{12}-A_{42} &  -A_{13}-A_{43} &  -A_{14}-A_{44}\\
    -A_{21}+A_{71}   &   -A_{22}+A_{72}     &   -A_{23}+A_{73}    & -A_{24}+A_{74}
    \end{pmatrix}
\end{eqnarray*}
\item $s=8$: $A_{(8)}=\begin{pmatrix}
A_{81}& A_{82}& A_{83} & A_{84} \end{pmatrix}$, $h=2,g=-1$. Updating $A_{h(8)}=A_2$ gives
\begin{eqnarray*}
\tilde A&=&\begin{pmatrix}
A_{31}+A_{51}-A_{61}    &   A_{32}+A_{52}-A_{62}    &   A_{33}+A_{53}-A_{63}&           A_{34}+A_{54}-A_{64} \\
 -A_{11}-A_{41}-A_{81} &  -A_{12}-A_{42}-A_{82} &  -A_{13}-A_{43}-A_{83} &  -A_{14}-A_{44}-A_{84} \\
    -A_{21}+A_{71}   &   -A_{22}+A_{72}     &   -A_{23}+A_{73}    & -A_{24}+A_{74}
    \end{pmatrix}
    \end{eqnarray*}
\item $s=9$: $A_{(9)}=\begin{pmatrix}
A_{91}& A_{92}& A_{93} & A_{94} \end{pmatrix}$, $h=1,g=1$. Updating $A_{h(9)}=A_1$ gives
{\footnotesize
\begin{eqnarray*}
\tilde A&=&\begin{pmatrix}
A_{31}+A_{51}-A_{61}+A_{91}    &   A_{32}+A_{52}-A_{62} +A_{92}   &   A_{33}+A_{53}-A_{63}+A_{93}&           A_{34}+A_{54}-A_{64} +A_{94}\\
 -A_{11}-A_{41}-A_{81} &  -A_{12}-A_{42}-A_{82} &  -A_{13}-A_{43}-A_{83} &  -A_{14}-A_{44}-A_{84} \\
    -A_{21}+A_{71}  &   -A_{22}+A_{72}    &   -A_{23}+A_{73}& -A_{24}+A_{74}\\
    \end{pmatrix}.
\end{eqnarray*}
}
This is precisely $\Pi A$.
\end{enumerate}

\newpage

\section{General Version of Theorem 1} 

In this section, 
we give a general version of Theorem 1 under heteroskedasticity. 

 \begin{theorem}
 \label{thm:thm1-general}
Let $c$ be a known $K\times 1$   vector. 
Assume that $e_i\sim (0,\Omega_{e,ii})$,
where  $\Omega_{e,ii}$ denote the $i$th diagonal element of $\Omega_{e}$. Under OLS and PI3,  it holds with probability at least  $1-\delta_\sigma$ that
\begin{equation*}
\label{eq:relativeV-general-NR-general}
\frac{\textsc{mse}(x_0^T\tilde\beta | \Pi) }{\textsc{mse}(x_0^T\hat\beta)}
\le    \bigg(
\frac{\max_i \; \Omega_{e,ii}}{\min_i \; \Omega_{e,ii}}\bigg)
\bigg(\frac{n}{m}\bigg)
\frac{(1 + \varepsilon_\sigma)}{(1-\varepsilon_\sigma)^2}.
\end{equation*}
\end{theorem}

\paragraph{Proof of Theorem \ref{thm:thm1-general}}

We have $ \mathbb V(\tilde\beta|\Pi) = (\tilde X^T \tilde X)^{-1} \tilde X^T \Pi \Omega_{e} \Pi^T \tilde X (\tilde X^T \tilde X)^{-1}$.
The quantity of interest is
\begin{align*}
& c^T \mathbb V(\tilde\beta|\Pi) c- \frac{n}{m} c^T \mathbb V(\hat\beta) c \\
 &= c^T A_1 B_1 A_1 c - c^T A_2 B_2 A_2 c \\
&=
c^T (A_1 - A_2) B_1 (A_1 - A_2) c + 2 c^T A_2 B_1 (A_1 - A_2) c + c^T A_2 (B_1 - B_2) A_2 c,
 \end{align*}
where $A_1 = \big( X^T  \Pi^T  \Pi X \big)^{-1}$, $B_1 = X^T  \Pi^T  \Pi  \Omega_{e} \Pi^T  \Pi X$, $A_2 = \big( X^T   X \big)^{-1}$,
and $B_2 = \frac{n}{m}   \big( X^T \Omega_{e}  X \big)$.
Write
\begin{align*}
& \left| c^T \mathbb V(\tilde\beta|\Pi) c- \frac{n}{m} c^T \mathbb V(\hat\beta) c \right| \\
&\leq \left| c^T (A_1 - A_2) B_2 (A_1 - A_2) c \right|
+ \left|c^T (A_1 - A_2) (B_1 - B_2) (A_1 - A_2) c \right| \\
&+ 2 \left| c^T A_2 B_2 (A_1 - A_2) c \right|
+ 2 \left| c^T A_2 (B_1 - B_2) (A_1 - A_2) c \right|
+ \left| c^T A_2 (B_1 - B_2) A_2 c \right|.
\end{align*}
Let 
$$
C(\Omega_{e})
=
\frac{\max_i \; \Omega_{e,ii}}{\min_i \; \Omega_{e,ii}}.
$$
We will bound each of the terms above by establishing the following lemma.

\begin{lemma}\label{claim-00-general}
The following statements hold
with probability at least $1-\delta_\sigma:$
\begin{subequations}
\begin{align}
 \left| \frac{ c^T A_2 B_2 (A_1 - A_2)  c}{c^T \mathbb V(\hat\beta) c} \right|
&\leq \frac{n}{m} C(\Omega_{e})^{1/2} \frac{\varepsilon_\sigma}{(1-\varepsilon_\sigma)},
 \label{eq:sub-1} \\
 \left| \frac{ c^T (A_1 - A_2) B_2 (A_1 - A_2) c}{c^T \mathbb V(\hat\beta) c} \right|
&\leq \frac{n}{m} C(\Omega_{e}) \frac{\varepsilon_\sigma^2}{(1-\varepsilon_\sigma)^2}, \label{eq:sub-2}
 \\
 \left| \frac{ c^T (A_1 - A_2) (B_1 - B_2)  (A_1 - A_2) c}{c^T \mathbb V(\hat\beta) c} \right|
&\leq \frac{n}{m} C(\Omega_{e}) \frac{\varepsilon_\sigma^2}{(1-\varepsilon_\sigma)^2}   \varepsilon_\sigma,  \label{eq:sub-3}\\
 \left| \frac{ c^T A_2 (B_1 - B_2)  (A_1 - A_2) c}{c^T \mathbb V(\hat\beta) c} \right|
&\leq \frac{n}{m} C(\Omega_{e})^{1/2} \frac{\varepsilon_\sigma}{(1-\varepsilon_\sigma)}   \varepsilon_\sigma, \label{eq:sub-4} \\
 \left| \frac{ c^T A_2 (B_1 - B_2)  A_2 c}{c^T \mathbb V(\hat\beta) c} \right|
&\leq    \frac{n}{m}  \varepsilon_\sigma \label{eq:sub-5}.
\end{align}
\end{subequations}
\end{lemma}

\paragraph{Proof of Lemma \ref{claim-00-general}}
To show part \eqref{eq:sub-1},
define $W = (U^T \Omega_{e}  U)^{1/2}  \Sigma^{-1} V^T$.
Note  that
\begin{align*}
\mathbb V(\hat\beta)
&=  (X^T X)^{-1}(X^T\Omega_{e} X) (X^T X)^{-1} \\
&=  V\Sigma^{-1} U^T \Omega_{e} U  \Sigma^{-1}V^T \\
& = W^T W.
\end{align*}
Write
\begin{align*}
A_2 B_2 (A_1 - A_2)
&=
\frac{n}{m} \big( X^T   X \big)^{-1}   \big( X^T \Omega_{e}  X \big)
\left[ \big( X^T  \Pi^T  \Pi X \big)^{-1} - \big( X^T   X \big)^{-1} \right] \\
&=  \frac{n}{m} V \Sigma^{-1}  U^T \Omega_{e} U
\left[(U^T \Pi^T \Pi U)^{-1} - I_n \right] \Sigma^{-1}  V^T \\
&= \frac{n}{m} V \Sigma^{-1}  U^T \Omega_{e} U
\left[(U^T \Pi^T \Pi U)^{-1} - I_n \right] (U^T \Omega_{e}  U)^{-1/2} (U^T \Omega_{e}  U)^{1/2} \Sigma^{-1}  V^T \\
&= \frac{n}{m} W^T (U^T \Omega_{e}  U)^{1/2}
\left[(U^T \Pi^T \Pi U)^{-1} - I_n \right] (U^T \Omega_{e}  U)^{-1/2} W.
\end{align*}
Then
\begin{align*}
\left| \frac{ c^T A_2 B_2 (A_1 - A_2)  c}{c^T \mathbb V(\hat\beta) c} \right|
&=
\frac{n}{m} \left| \frac{c^T W^T (U^T \Omega_{e}  U)^{1/2}
\left[(U^T \Pi^T \Pi U)^{-1} - I_n \right] (U^T \Omega_{e}  U)^{-1/2} W c}{c^T W^T W c} \right| \\
&\leq \frac{n}{m} \norm{(U^T \Omega_{e}  U)^{1/2}}_2 \norm{(U^T \Pi^T \Pi U)^{-1} - I_n}_2 \norm{(U^T \Omega_{e}  U)^{-1/2}}_2 \\
&\leq \frac{n}{m} C(\Omega_{e})^{1/2} \frac{\varepsilon_\sigma}{(1-\varepsilon_\sigma)},
\end{align*}
which proves part \eqref{eq:sub-1}.

For the second part, write
\begin{align*}
&(A_1 - A_2) B_2 (A_1 - A_2) \\
&= \frac{n}{m} \left[ \big( X^T  \Pi^T  \Pi X \big)^{-1} - \big( X^T   X \big)^{-1} \right]
\big( X^T \Omega_{e}  X \big)
\left[ \big( X^T  \Pi^T  \Pi X \big)^{-1} - \big( X^T   X \big)^{-1} \right] \\
&= \frac{n}{m} V \Sigma^{-1} \left[(U^T \Pi^T \Pi U)^{-1} - I_n \right]
U^T \Omega_{e} U \left[(U^T \Pi^T \Pi U)^{-1} - I_n \right] \Sigma^{-1}  V^T \\
&=  \frac{n}{m} W^T (U^T \Omega_{e}  U)^{-1/2} \left[(U^T \Pi^T \Pi U)^{-1} - I_n \right]
U^T \Omega_{e} U \left[(U^T \Pi^T \Pi U)^{-1} - I_n \right] (U^T \Omega_{e}  U)^{-1/2} W.
\end{align*}
Thus,
\begin{align*}
\left| \frac{ c^T (A_1 - A_2) B_2 (A_1 - A_2)  c}{c^T \mathbb V(\hat\beta) c} \right|
&\leq \frac{n}{m} C(\Omega_{e}) \frac{\varepsilon_\sigma^2}{(1-\varepsilon_\sigma)^2},
\end{align*}
which proves part \eqref{eq:sub-2}.

For the third part,  write
\begin{align*}
& (A_1 - A_2) (B_1 - B_2)  (A_1 - A_2) \\
&= V \Sigma^{-1} \left[(U^T \Pi^T \Pi U)^{-1} - I_n \right]
\left[ U^T \Omega_{e}^{1/2}  \Pi^T \Pi  \Pi^T \Pi \Omega_{e}^{1/2} U - \frac{n}{m} U^T \Omega_{e}^{1/2} \Omega_{e}^{1/2} U\right]
\left[(U^T \Pi^T \Pi U)^{-1} - I_n \right] \Sigma^{-1}  V^T \\
&= W^T   (U^T \Omega_{e}  U)^{-1/2} \left[(U^T \Pi^T \Pi U)^{-1} - I_n \right] (U^T \Omega_{e}  U)^{1/2} \\
&\times (U^T \Omega_{e}  U)^{-1/2}
\left[ U^T \Omega_{e}^{1/2}  \Pi^T \Pi  \Pi^T \Pi \Omega_{e}^{1/2} U - \frac{n}{m} U^T \Omega_{e}^{1/2} \Omega_{e}^{1/2} U\right] (U^T \Omega_{e}  U)^{-1/2} \\
&\times (U^T \Omega_{e}  U)^{1/2}
\left[(U^T \Pi^T \Pi U)^{-1} - I_n \right] (U^T \Omega_{e}  U)^{-1/2} W.
\end{align*}
Define $U_{e} = \Omega_{e}^{1/2} U ( U^T \Omega_{e} U)^{-1/2}$, so that
\begin{align*}
& (U^T \Omega_{e}  U)^{-1/2}
\left[ U^T \Omega_{e}^{1/2}  \Pi^T \Pi  \Pi^T \Pi \Omega_{e}^{1/2} U - \frac{n}{m} U^T \Omega_{e}^{1/2} \Omega_{e}^{1/2} U\right] (U^T \Omega_{e}  U)^{-1/2} \\
&= U_{e}^T  \Pi^T \Pi  \Pi^T \Pi U_{e}
- \frac{n}{m} U_{e}^T U_{e}.
\end{align*}
Now observe that
\begin{align}\label{pi4-all-ineq}
\begin{split}
&\norm{  U_{e}^T  \Pi^T \Pi  \Pi^T \Pi U_{e}
- \frac{n}{m} U_{e}^T U_{e} }_2 \\
&\leq \norm{  U_{e}^T  \Pi^T \Pi  \Pi^T \Pi U_{e} - \frac{n}{m}  U_{e}^T \Pi^T  \Pi  U_{e} } _2
+ \frac{n}{m}  \norm{  U_{e}^T \Pi^T  \Pi  U_{e} - U_{e}^T  U_{e} }_2  \\
&\leq  \frac{n}{m} \varepsilon_\sigma.
\end{split}
\end{align}
Thus,
\begin{align*}
\left| \frac{ c^T (A_1 - A_2) (B_1 - B_2)  (A_1 - A_2) c}{c^T \mathbb V(\hat\beta) c} \right|
\leq  C(\Omega_{e}) \frac{\varepsilon_\sigma^2}{(1-\varepsilon_\sigma)^2} \frac{n}{m} \varepsilon_\sigma,
\end{align*}
which proves part \eqref{eq:sub-3}.

Similarly,
\begin{align*}
&A_2 (B_1 - B_2)  (A_1 - A_2)  \\
&= V \Sigma^{-1}
\left[ U^T \Omega_{e}^{1/2}  \Pi^T \Pi  \Pi^T \Pi \Omega_{e}^{1/2} U - \frac{n}{m} U^T \Omega_{e}^{1/2} \Omega_{e}^{1/2} U\right]
\left[(U^T \Pi^T \Pi U)^{-1} - I_n \right] \Sigma^{-1}  V^T.
\end{align*}
Then
\begin{align*}
\left| \frac{ c^T A_2 (B_1 - B_2)  (A_1 - A_2) c}{c^T \mathbb V(\hat\beta) c} \right|
\leq C(\Omega_{e})^{1/2} \frac{\varepsilon_\sigma}{(1-\varepsilon_\sigma)} \frac{n}{m}  \varepsilon_\sigma,
\end{align*}
which proves part \eqref{eq:sub-4}.

Finally,
\begin{align*}
A_2 (B_1 - B_2) A_2
&= V \Sigma^{-1}
\left[ U^T \Omega_{e}^{1/2}  \Pi^T \Pi  \Pi^T \Pi \Omega_{e}^{1/2} U - \frac{n}{m} U^T \Omega_{e}^{1/2} \Omega_{e}^{1/2} U\right] \Sigma^{-1}  V^T.
\end{align*}
Therefore,
\begin{align*}
\left| \frac{ c^T A_2 (B_1 - B_2) A_2 c}{c^T \mathbb V(\hat\beta) c} \right|
\leq  \frac{n}{m}  \varepsilon_\sigma,
\end{align*}
which proves part \eqref{eq:sub-5}.
\; \emph{Q.E.D.}

We now return to the proof of the theorem.
Applying (\ref{eq:sub-1})-(\ref{eq:sub-5}) and simplifying,
\begin{align*}
 \left| \frac{c^T \mathbb V(\tilde\beta|\Pi) c- \frac{n}{m} c^T \mathbb V(\hat\beta) c}{c^T \mathbb V(\hat\beta) c} \right| &\leq
 C(\Omega_{e}) \frac{n}{m} \left[   \frac{\varepsilon_\sigma(2-\varepsilon_\sigma)}{(1-\varepsilon_\sigma)^2}
 ( 1 +  \varepsilon_\sigma) +  \varepsilon_\sigma
  \right] \\
 \left| \frac{c^T \mathbb V(\tilde\beta|\Pi) c-  c^T \mathbb V(\hat\beta) c}{c^T \mathbb V(\hat\beta) c} \right|
&\leq  C(\Omega_{e}) \frac{n}{m} \left[   \frac{\varepsilon_\sigma(2-\varepsilon_\sigma)}{(1-\varepsilon_\sigma)^2}
 ( 1  + \varepsilon_\sigma) +  \varepsilon_\sigma)
 \right] + \frac{n-m}{m}
\end{align*}
with probability at least $1-\delta_\sigma-\delta_\Pi$. This in turn implies that with the same probability,
\[
\frac{c^T \mathbb V(\tilde\beta|\Pi) c}{c^T \mathbb V(\hat\beta)c}\leq   C(\Omega_{e}) \frac{n}{m}
  \frac{(1 + \varepsilon_\sigma)}{(1-\varepsilon_\sigma)^2}.
\]
\; \emph{Q.E.D.}

\newpage
\clearpage

\section{Subspace embedding for  $\Pi^T \Pi \Psi \Pi^T \Pi$}\label{sec:sub:emb:count}

In this appendix, we consider $\Pi^T \Pi \Psi \Pi^T \Pi$ for a diagonal matrix $\Psi$ in the case of countsketch.

\begin{lemma}\label{sub-emb-count-more}
Let $\Pi \in \mathbb{R}^{n \times d}$ be a random matrix such that
(i) the $(i,j)$ element $\Pi_{ij}$ of $\Pi$ is $\Pi_{ij} = \delta_{ij} \sigma_{ij}$, where
$\sigma_{ij}$'s are i.i.d. $\pm 1$ random variables and $\delta_{ij}$ is an indicator random variable for the event $\Pi_{ij} \neq 0$;
(ii)  $\sum_{i=1}^{m} \delta_{ij} = 1$ for each $j = 1,\ldots,n$;
(iii) for any $S \subset [n]$, $\mathbb{E} \left( \Pi_{j \in S} \delta_{ij} \right) = m^{-|S|}$;
(iv) the columns of $\Pi$ are i.i.d.
Choose
\begin{align}\label{A-choice}
A (\Psi, m, n) =  \Psi +  \frac{1}{m} \left\{ \text{tr}(\Psi) I_n -  \Psi
\right\}.
\end{align}
Then
\begin{align*}
&\mathbb{P} \left( \left\| U^T \Pi^T \Pi \Psi \Pi^T \Pi U -  U^T A (\Psi, m,n) U \right\|_2 > \epsilon \right)  \\
&\leq
  2 \epsilon^{-2}  \bigg\{ \frac{2 d^2 (m-1)}{m^2} \text{tr}(\Psi^2)
+ \frac{2 d^2}{m^2} \| \Psi \|_2  \text{tr}(\Psi)
+  \frac{d^2}{m^3}    \left\{ \left[ \text{tr}(\Psi)  \right]^2 + 2 \| \Psi \|_2^2  \right\} \\
&\;\;\;\;\;\;\;\;\;\;\;\; + \frac{2}{m}  \text{tr}(\Psi^2)
+ \frac{2}{m^2} \text{tr}(\Psi) +  \frac{1}{m^3}    \left\{ d \left[  \text{tr}(\Psi)  \right]^2  + 2 \text{tr}(\Psi^2)  \right\} \bigg\}.
\end{align*}
\end{lemma}

\paragraph{Proof of Lemma \ref{sub-emb-count-more}}
To show the desired result, we follow Nelson and Nguyen (2013). That is,
we start with the following moment inequality:
\begin{align*}
&\mathbb{P} \left( \left\| U^T \Pi^T \Pi \Psi \Pi^T \Pi U -  U^T A (\Psi, m,n) U \right\|_2 > \epsilon \right)  \\
&\leq
\epsilon^{-2} \mathbb{E} \left[ \left\| U^T \Pi^T \Pi \Psi \Pi^T \Pi U -  U^T A (\Psi, m,n) U \right\|_F^2  \right].
\end{align*}

For countsketch, we have one non-zero entry for each column. This implies that
$\Pi \Psi \Pi^T$ is a diagonal matrix.
Hence,
\begin{align*}
[\Pi \Psi \Pi^T]_{ii} = \sum_{k=1}^{n} \Psi_{kk} \Pi_{ik}^2  = \sum_{k=1}^{n} \Psi_{kk} \delta_{ik}.
\end{align*}
Now
\begin{align*}
[\Pi^T \Pi  \Psi \Pi^T \Pi]_{\ell \ell'}
&=   \sum_{i=1}^{m}  \sum_{k=1}^{n}     \Pi_{i \ell} \Psi_{kk} \delta_{ik}    \Pi_{i \ell'}
=  \sum_{i=1}^{m} \sum_{k=1}^{n}    \Psi_{kk}  \delta_{ik}  \delta_{i \ell}    \delta_{i \ell'} \sigma_{i \ell} \sigma_{i \ell'},
\end{align*}
where the last equality comes from the fact that
$\Pi_{ij} = \delta_{ij} \sigma_{ij}$.
Then
\begin{align*}
[U^T \Pi^T \Pi \Psi \Pi^T \Pi U]_{uv}
&=   \sum_{\ell=1}^n  \sum_{\ell'=1}^n  [\Pi^T \Pi \Psi \Pi^T \Pi]_{\ell \ell'} U_{\ell u} U_{\ell' v} \\
&=    \sum_{\ell=1}^n  \sum_{\ell'=1}^n
\sum_{i=1}^{m} \sum_{k=1}^{n}   \Psi_{kk}   \delta_{ik}  \delta_{i \ell}    \delta_{i \ell'} \sigma_{i \ell} \sigma_{i \ell'}
 U_{\ell u} U_{\ell' v}.
\end{align*}
Write
\begin{align*}
[U^T \Pi^T \Pi \Psi \Pi^T \Pi U]_{uv}
&=   \sum_{\ell=1}^n
\sum_{i=1}^{m} \sum_{k=1}^{n}   \Psi_{kk}   \delta_{ik}  \delta_{i \ell}
 U_{\ell u} U_{\ell v} \\
&+    \sum_{\ell=1}^n  \sum_{\ell'=1, \ell' \neq \ell}^n
\sum_{i=1}^{m} \sum_{k=1}^{n}  \Psi_{kk}    \delta_{ik}  \delta_{i \ell}    \delta_{i \ell'} \sigma_{i \ell} \sigma_{i \ell'}
 U_{\ell u} U_{\ell' v} \\
 &=
\sum_{\ell=1}^n  \sum_{i=1}^{m}    \Psi_{\ell \ell}     \delta_{i \ell}
 U_{\ell u} U_{\ell v} \\
&+
\sum_{\ell=1}^n
\sum_{i=1}^{m} \sum_{k=1, k \neq \ell}^{n}   \Psi_{kk}   \delta_{ik}  \delta_{i \ell}
 U_{\ell u} U_{\ell v} \\
&+    \sum_{\ell=1}^n  \sum_{\ell'=1, \ell' \neq \ell}^n
\sum_{i=1}^{m} \sum_{k=1}^{n}  \Psi_{kk}    \delta_{ik}  \delta_{i \ell}    \delta_{i \ell'} \sigma_{i \ell} \sigma_{i \ell'}
 U_{\ell u} U_{\ell' v} \\
 &=
 \sum_{\ell=1}^n      \Psi_{\ell \ell}       U_{\ell u} U_{\ell v} \\
&+
\sum_{\ell=1}^n
\sum_{i=1}^{m} \sum_{k=1, k \neq \ell}^{n}   \Psi_{kk}   \delta_{ik}  \delta_{i \ell}
 U_{\ell u} U_{\ell v} \\
&+    \sum_{\ell=1}^n  \sum_{\ell'=1, \ell' \neq \ell}^n
\sum_{i=1}^{m} \sum_{k=1}^{n}  \Psi_{kk}    \delta_{ik}  \delta_{i \ell}    \delta_{i \ell'} \sigma_{i \ell} \sigma_{i \ell'}
 U_{\ell u} U_{\ell' v}
\end{align*}
since $\sum_{i=1}^{m} \delta_{i\ell} = 1$ for each $\ell$.
Recall that we have assumed that for any $S \subset [n]$, $\mathbb{E} \left( \Pi_{j \in S} \delta_{ij} \right) = m^{-|S|}$.
Then, note that
\begin{align*}
& \mathbb{E} \left(
\sum_{\ell=1}^n  \sum_{i=1}^{m} \sum_{k=1, k \neq \ell}^{n}   \Psi_{kk}   \delta_{ik}  \delta_{i \ell}
 U_{\ell u} U_{\ell v} \right) \\
& =  \sum_{\ell=1}^n   \sum_{i=1}^{m} \sum_{k=1, k \neq \ell}^{n}   \Psi_{kk}  \mathbb{E} \left( \delta_{ik}  \delta_{i \ell}   \right)
 U_{\ell u} U_{\ell v}  \\
&= \frac{1}{m} \sum_{\ell=1}^n    \sum_{k=1, k \neq \ell}^{n}   \Psi_{kk}   U_{\ell u} U_{\ell v}  \\
  &= \frac{1}{m} \sum_{\ell=1}^n \left[ \sum_{k=1}^{n}   \Psi_{kk}   - \Psi_{\ell \ell} \right]
   U_{\ell u} U_{\ell v} \\
 &= \frac{\text{tr}(\Psi)}{m} 1( u = v) - m^{-1} \sum_{\ell=1}^{n}   \Psi_{\ell \ell}    U_{\ell u} U_{\ell v}.
\end{align*}
Since
\begin{align*}
\sum_{k=1}^{n}   \Psi_{kk}    U_{k u} U_{k v}
= [U^T \Psi U]_{uv},
\end{align*}
$U^T \Pi^T \Pi \Psi \Pi^T \Pi U$
should center on
\begin{align*}
\frac{\text{tr}(\Psi)}{m} U^T  U + \left( 1-\frac{1}{m} \right) U^T \Psi U
&= U^T \left[ \Psi +  \frac{1}{m} \left\{ \text{tr}(\Psi) I_n -  \Psi
\right\} \right] U.
\end{align*}
If $\Psi = \sigma_e^2 I_n$, then its form is
\begin{align*}
\sigma_e^2 \frac{n}{m} U^T  U + \sigma_e^2 \left( 1-\frac{1}{m} \right) U^T  U
&= \sigma_e^2 \left[ \frac{n+m-1}{m} \right] U^T U.
\end{align*}
Now
\begin{align*}
&\left[U^T \Pi^T \Pi \Psi \Pi^T \Pi U - U^T \left( \Psi +  \frac{1}{m} \left\{ \text{tr}(\Psi) I_n -  \Psi
\right\} \right) U \right]_{uv} \\
&=
 \sum_{\ell=1}^n   \sum_{k=1, k \neq \ell}^{n}   \Psi_{kk}   \left[ \sum_{i=1}^{m} \delta_{ik}  \delta_{i \ell}   - \frac{1}{m} \right]
 U_{\ell u} U_{\ell v} \\
&+    \sum_{\ell=1}^n  \sum_{\ell'=1, \ell' \neq \ell}^n
 \sum_{k=1}^{n}  \Psi_{kk}   \left[ \sum_{i=1}^{m}  \delta_{ik}  \delta_{i \ell}    \delta_{i \ell'} \sigma_{i \ell} \sigma_{i \ell'} \right]
 U_{\ell u} U_{\ell' v} \\
 &=: T_{1, uv} + T_{2, uv}.
\end{align*}

First consider $T_{1, uv}$. Define
\[
k_m(k,\ell) := \sum_{i=1}^{m} \delta_{ik}  \delta_{i \ell}   - \frac{1}{m}.
\]
Write
\begin{align*}
&k_m(k,\ell) k_m(k',\ell')  \\
&= \sum_{i=1}^{m}  \delta_{ik}  \delta_{i \ell} \delta_{ik'}  \delta_{i \ell'}
+\sum_{i=1}^{m} \sum_{i'=1, i' \neq i}^m \delta_{ik}  \delta_{i \ell} \delta_{i'k'}  \delta_{i' \ell'}
   - \frac{1}{m} \sum_{i=1}^{m} \delta_{ik}  \delta_{i \ell} - \frac{1}{m} \sum_{i'=1}^{m} \delta_{i'k'}  \delta_{i' \ell'}
   + \frac{1}{m^2}.
\end{align*}
Then for any $\ell \neq k$ and $\ell' \neq k'$,
\begin{align*}
\sum_{i=1}^{m}  \delta_{ik}  \delta_{i \ell} \delta_{ik'}  \delta_{i \ell'} &=
\begin{cases}
\sum_{i=1}^{m}  \delta_{ik}  \delta_{i \ell}  \quad & \text{if $\ell = \ell'$, $k = k'$} \\
\sum_{i=1}^{m}  \delta_{ik}  \delta_{i \ell} \delta_{ik'}   \quad & \text{if $\ell = \ell'$, $k \neq k'$} \\
\sum_{i=1}^{m}  \delta_{ik}  \delta_{i \ell}   \delta_{i \ell'} \quad & \text{if $\ell \neq \ell'$, $k = k'$} \\
\sum_{i=1}^{m}  \delta_{ik}  \delta_{i \ell}  \quad & \text{if $\ell \neq \ell'$, $k \neq k'$, $k = \ell'$, $\ell = k'$} \\
\sum_{i=1}^{m}  \delta_{ik}  \delta_{i \ell} \delta_{ik'}   \quad & \text{if $\ell \neq \ell'$, $k \neq k'$, $k = \ell'$, $\ell \neq k'$} \\
\sum_{i=1}^{m}  \delta_{ik}  \delta_{i \ell}   \delta_{i \ell'} \quad & \text{if $\ell \neq \ell'$, $k \neq k'$, $k \neq \ell'$, $\ell = k'$} \\
\sum_{i=1}^{m}  \delta_{ik}  \delta_{i \ell} \delta_{ik'}  \delta_{i \ell'} \quad & \text{if $\ell \neq \ell'$, $k \neq k'$, $k \neq \ell'$, $\ell \neq k'$}.
\end{cases}
\end{align*}
Then
\begin{align*}
\mathbb{E} \left( \sum_{i=1}^{m}  \delta_{ik}  \delta_{i \ell} \delta_{ik'}  \delta_{i \ell'} \right) &=
\begin{cases}
m^{-1}  \quad & \text{if $\ell = \ell'$, $k = k'$} \\
m^{-2}  \quad & \text{if $\ell = \ell'$, $k \neq k'$} \\
m^{-2}  \quad & \text{if $\ell \neq \ell'$, $k = k'$} \\
m^{-1}  \quad & \text{if $\ell \neq \ell'$, $k \neq k'$, $k = \ell'$, $\ell = k'$} \\
m^{-2}  \quad & \text{if $\ell \neq \ell'$, $k \neq k'$, $k = \ell'$, $\ell \neq k'$} \\
m^{-2} \quad & \text{if $\ell \neq \ell'$, $k \neq k'$, $k \neq \ell'$, $\ell = k'$} \\
m^{-3} \quad & \text{if $\ell \neq \ell'$, $k \neq k'$, $k \neq \ell'$, $\ell \neq k'$}.
\end{cases}
\end{align*}
In addition, since there is one non-zero entry for each column in count sketch,
\begin{align*}
\sum_{i=1}^{m} \sum_{i'=1, i' \neq i}^m \delta_{ik}  \delta_{i \ell} \delta_{i'k'}  \delta_{i' \ell'} &=
\begin{cases}
0 \quad & \text{if $\ell = \ell'$, $k = k'$} \\
0 \quad & \text{if $\ell = \ell'$, $k \neq k'$} \\
0 \quad & \text{if $\ell \neq \ell'$, $k = k'$} \\
0  \quad & \text{if $\ell \neq \ell'$, $k \neq k'$, $k = \ell'$, $\ell = k'$} \\
0  \quad & \text{if $\ell \neq \ell'$, $k \neq k'$, $k = \ell'$, $\ell \neq k'$} \\
0 \quad & \text{if $\ell \neq \ell'$, $k \neq k'$, $k \neq \ell'$, $\ell = k'$} \\
\sum_{i=1}^{m} \sum_{i'=1, i' \neq i}^m \delta_{ik}  \delta_{i \ell} \delta_{i'k'}  \delta_{i' \ell'} \quad & \text{if $\ell \neq \ell'$, $k \neq k'$, $k \neq \ell'$, $\ell \neq k'$}.
\end{cases}
\end{align*}
Hence,
\begin{align*}
\mathbb{E} \left[ \sum_{i=1}^{m} \sum_{i'=1, i' \neq i}^m \delta_{ik}  \delta_{i \ell} \delta_{i'k'}  \delta_{i' \ell'} \right] &=
\begin{cases}
0 \quad & \text{if $\ell = \ell'$, $k = k'$} \\
0 \quad & \text{if $\ell = \ell'$, $k \neq k'$} \\
0 \quad & \text{if $\ell \neq \ell'$, $k = k'$} \\
0  \quad & \text{if $\ell \neq \ell'$, $k \neq k'$, $k = \ell'$, $\ell = k'$} \\
0  \quad & \text{if $\ell \neq \ell'$, $k \neq k'$, $k = \ell'$, $\ell \neq k'$} \\
0 \quad & \text{if $\ell \neq \ell'$, $k \neq k'$, $k \neq \ell'$, $\ell = k'$} \\
\frac{m-1}{m^3} \quad & \text{if $\ell \neq \ell'$, $k \neq k'$, $k \neq \ell'$, $\ell \neq k'$}.
\end{cases}
\end{align*}
Combining the results above yields
\begin{align*}
\mathbb{E} \left[ k_m(k,\ell) k_m(k',\ell') \right] &=
\begin{cases}
\frac{m-1}{m^2} \quad & \text{if $\ell = \ell'$, $k = k'$} \\
0  \quad & \text{if $\ell = \ell'$, $k \neq k'$} \\
0  \quad & \text{if $\ell \neq \ell'$, $k = k'$} \\
\frac{m-1}{m^2}  \quad & \text{if $\ell \neq \ell'$, $k \neq k'$, $k = \ell'$, $\ell = k'$} \\
0  \quad & \text{if $\ell \neq \ell'$, $k \neq k'$, $k = \ell'$, $\ell \neq k'$} \\
0 \quad & \text{if $\ell \neq \ell'$, $k \neq k'$, $k \neq \ell'$, $\ell = k'$} \\
0 \quad & \text{if $\ell \neq \ell'$, $k \neq k'$, $k \neq \ell'$, $\ell \neq k'$}.
\end{cases}
\end{align*}

Note that
\begin{align*}
\mathbb{E} \left[ T_{1, uv}^2 \right] &= \sum_{\ell=1}^n  \sum_{\ell'=1}^n \sum_{k=1, k \neq \ell}^{n} \sum_{k'=1, k' \neq \ell'}^{n}
\Psi_{kk} \Psi_{k'k'} \mathbb{E} \left[ k_m(k,\ell) k_m(k',\ell') \right] U_{\ell u} U_{\ell v} U_{\ell' u} U_{\ell' v} \\
&= \frac{m-1}{m^2} \sum_{\ell=1}^n   \sum_{k=1, k \neq \ell}^{n}
\Psi_{kk}^2   U_{\ell u}^2 U_{\ell v}^2
+ \frac{m-1}{m^2} \sum_{\ell=1}^n   \Psi_{\ell \ell} U_{\ell u} U_{\ell v} \sum_{\ell'=1, \ell' \neq \ell}^n \Psi_{\ell' \ell'} U_{\ell' u} U_{\ell' v} \\
&= \frac{m-1}{m^2} \sum_{\ell=1}^n   \left[ \text{tr}(\Psi^2) -   \Psi_{\ell \ell}^2 \right]  U_{\ell u}^2 U_{\ell v}^2 \\
&+ \frac{m-1}{m^2} \sum_{\ell=1}^n   \Psi_{\ell \ell} U_{\ell u} U_{\ell v}
\left[ \sum_{\ell'=1}^n \Psi_{\ell' \ell'} U_{\ell' u} U_{\ell' v} - \Psi_{\ell \ell} U_{\ell u} U_{\ell v} \right] \\
&\leq \frac{m-1}{m^2} \left\{ \text{tr}(\Psi^2) + [U^T \Psi U]_{uv}^2 \right\}
\end{align*}
Therefore,
\begin{align*}
& \sum_{u=1}^d \sum_{v=1}^d \mathbb{E} \left[ T_{1, uv}^2 \right] \\
&\leq
 \frac{d^2 (m-1)}{m^2} \text{tr}(\Psi^2) + \frac{m-1}{m^2} \sum_{u=1}^d \sum_{v=1}^d [U^T \Psi U]_{uv}^2  \\
&= \frac{d^2 (m-1)}{m^2} \text{tr}(\Psi^2) + \frac{m-1}{m^2}  \left\| U^T \Psi U \right\|_F^2  \\
&=   \frac{2 d^2 (m-1)}{ m^2} \text{tr}(\Psi^2).
\end{align*}

Now consider $T_{2, uv}$. Recall that
\begin{align*}
T_{2, uv}
&=    \sum_{\ell=1}^n  \sum_{\ell'=1, \ell' \neq \ell}^n
 \sum_{k=1}^{n}  \Psi_{kk}   \left[ \sum_{i=1}^{m}  \delta_{ik}  \delta_{i \ell}    \delta_{i \ell'} \sigma_{i \ell} \sigma_{i \ell'} \right]
 U_{\ell u} U_{\ell' v}.
\end{align*}
Define
\[
q_{n,m}(\ell, \ell') :=  \sum_{k=1}^{n}  \Psi_{kk}   \left[ \sum_{i=1}^{m}  \delta_{ik}  \delta_{i \ell}    \delta_{i \ell'} \sigma_{i \ell} \sigma_{i \ell'} \right].
\]
Note that
\begin{align*}
\mathbb{E} \left[ T_{2, uv}^2 \right] &= \sum_{\ell=1}^n  \sum_{\tilde{\ell}=1}^n \sum_{\ell'=1, \ell' \neq \ell}^{n} \sum_{\tilde{\ell}'=1, \tilde{\ell}' \neq \tilde{\ell}}^{n}
 \mathbb{E} \left[ q_{n,m}(\ell, \ell') q_{n,m}(\tilde{\ell}, \tilde{\ell}') \right] U_{\ell u} U_{\ell' v} U_{\tilde{\ell} u} U_{\tilde{\ell}' v}.
\end{align*}
Write
\begin{align*}
&q_{n,m}(\ell, \ell') q_{n,m}(\tilde{\ell}, \tilde{\ell}')  \\
&= \sum_{k=1}^{n} \sum_{k'=1}^{n} \sum_{i=1}^{m} \sum_{i'=1}^{m}
\Psi_{kk}  \Psi_{k'k'}
\delta_{ik}  \delta_{i \ell}    \delta_{i \ell'}
\delta_{i'k'}  \delta_{i' \tilde{\ell}}    \delta_{i' \tilde{\ell}'}
\sigma_{i \ell} \sigma_{i \ell'} \sigma_{i' \tilde{\ell}} \sigma_{i' \tilde{\ell}'} .
\end{align*}
Then for any $\ell \neq \ell'$ and $\tilde{\ell} \neq \tilde{\ell}'$,
\begin{align*}
& \mathbb{E} \left[ q_{n,m}(\ell, \ell') q_{n,m}(\tilde{\ell}, \tilde{\ell}') \right] \\
&=
\begin{cases}
\sum_{k=1}^{n} \sum_{k'=1}^{n} \sum_{i=1}^{m} \Psi_{kk}  \Psi_{k'k'}
\mathbb{E} \left[ \delta_{ik}  \delta_{i \ell}    \delta_{i \ell'} \delta_{ik'}    \right]
\quad & \text{if $\ell = \tilde{\ell}$, $\ell' = \tilde{\ell}'$} \\
\sum_{k=1}^{n} \sum_{k'=1}^{n} \sum_{i=1}^{m} \Psi_{kk}  \Psi_{k'k'}
\mathbb{E} \left[ \delta_{ik}  \delta_{i \ell}    \delta_{i \ell'} \delta_{ik'}   \right]
  \quad & \text{if $\ell = \tilde{\ell}'$, $\ell' = \tilde{\ell}$} \\
0  \quad & \text{otherwise}.
\end{cases}
\end{align*}
Thus,
\begin{align*}
\mathbb{E} \left[ T_{2, uv}^2 \right]
&= \sum_{\ell=1}^n   \sum_{\ell'=1, \ell' \neq \ell}^{n}
 \mathbb{E} \left[ q_{n,m}^2(\ell, \ell')  \right] U_{\ell u}^2 U_{\ell' v}^2 \\
&+ \sum_{\ell=1}^n   \sum_{\ell'=1, \ell' \neq \ell}^{n}
 \mathbb{E} \left[ q_{n,m}(\ell, \ell') q_{n,m}(\ell', \ell) \right] U_{\ell u} U_{\ell' v} U_{\ell' u} U_{\ell v}.
\end{align*}
Further, write
\begin{align*}
&\sum_{k=1}^{n} \sum_{k'=1}^{n} \sum_{i=1}^{m} \Psi_{kk}  \Psi_{k'k'}
\mathbb{E} \left[ \delta_{ik}  \delta_{i \ell}    \delta_{i \ell'} \delta_{ik'}    \right] \\
&= \sum_{k=1}^{n}  \sum_{i=1}^{m} \Psi_{kk}^2
\mathbb{E} \left[ \delta_{ik}  \delta_{i \ell}    \delta_{i \ell'}    \right]
+ \sum_{k=1}^{n} \sum_{k'=1, k' \neq k}^{n} \sum_{i=1}^{m} \Psi_{kk}  \Psi_{k'k'}
\mathbb{E} \left[ \delta_{ik}  \delta_{i \ell}    \delta_{i \ell'} \delta_{ik'}    \right].
\end{align*}
Note that  for any $\ell \neq \ell'$,
\begin{align*}
\mathbb{E} \left[ \delta_{ik}  \delta_{i \ell}    \delta_{i \ell'}    \right]
&=
\begin{cases}
\frac{1}{m^2} \quad  & \text{if $k = \ell$ or $k = \ell'$} \\
\frac{1}{m^3}  \quad & \text{otherwise},
\end{cases}
\end{align*}
so that
\begin{align*}
\sum_{k=1}^{n}  \sum_{i=1}^{m} \Psi_{kk}^2
\mathbb{E} \left[ \delta_{ik}  \delta_{i \ell}    \delta_{i \ell'}    \right]
&= \left[ \frac{2}{m^2} + \frac{m-2}{m^3} \right] \text{tr}(\Psi^2) \\
&= \sum_{k=1}^{n}   \Psi_{kk}^2  \left[ 1(k=\ell \text{ or } k=\ell') \frac{1}{m} + 1(k \neq \ell \text{ and } k \neq \ell') \frac{1}{m^2} \right] \\
&\leq \frac{\text{tr}(\Psi^2)}{m}.
\end{align*}
Similarly,
for any $\ell \neq \ell'$ and $k \neq k'$,
\begin{align*}
\mathbb{E} \left[ \delta_{ik}  \delta_{i \ell}    \delta_{i \ell'} \delta_{ik'}    \right]
&=
\begin{cases}
\frac{1}{m^2} \quad  & \text{if $k = \ell$, $k' = \ell'$} \\
\frac{1}{m^2} \quad  & \text{if $k = \ell'$, $k' = \ell$} \\
\frac{1}{m^3}  \quad & \text{if $k = \ell$, $k' \neq \ell'$} \\
\frac{1}{m^3}  \quad & \text{if $k \neq \ell$, $k' = \ell'$} \\
\frac{1}{m^4}  \quad & \text{if $k \neq \ell$, $k' \neq \ell'$}
\end{cases}
\end{align*}
Thus,
\begin{align*}
&\sum_{k=1}^{n} \sum_{k'=1, k' \neq k}^{n} \sum_{i=1}^{m} \Psi_{kk}  \Psi_{k'k'}
\mathbb{E} \left[ \delta_{ik}  \delta_{i \ell}    \delta_{i \ell'} \delta_{ik'}    \right]  \\
&= \frac{2}{m}   \Psi_{\ell \ell}  \Psi_{\ell' \ell'}
+ \frac{1}{m^2} \Psi_{\ell \ell}   \sum_{k'=1, k' \neq \ell, k' \neq \ell'}^{n}   \Psi_{k'k'}
+  \frac{1}{m^2} \Psi_{\ell' \ell'}   \sum_{k=1, k \neq \ell', k \neq \ell}^{n}   \Psi_{kk}
\\
&+ \frac{1}{m^3} \sum_{k=1, k \neq \ell}^{n} \sum_{k'=1, k' \neq k, k' \neq \ell'}^{n}  \Psi_{kk}  \Psi_{k'k'}
 \\
&=
\frac{2}{m}   \Psi_{\ell \ell}  \Psi_{\ell' \ell'}
+ \frac{1}{m^2} \left( \Psi_{\ell \ell} + \Psi_{\ell' \ell'} \right) \left[ \text{tr}(\Psi) - \Psi_{\ell \ell} - \Psi_{\ell' \ell'} \right] \\
&+  \frac{1}{m^3}    \left\{ \left[ \text{tr}(\Psi) - \Psi_{\ell \ell} \right]  \left[ \text{tr}(\Psi) - \Psi_{\ell' \ell'} \right]  - \left[ \text{tr}(\Psi^2) - \Psi_{\ell \ell}^2 \right] \right\} \\
&=\frac{2}{m}   \Psi_{\ell \ell}  \Psi_{\ell' \ell'}
- \frac{1}{m^2} \left( \Psi_{\ell \ell} + \Psi_{\ell' \ell'} \right)^2  \\
&+  \frac{1}{m^2} \left( \Psi_{\ell \ell} + \Psi_{\ell' \ell'} \right)  \text{tr}(\Psi)
-  \frac{1}{m^3} \left( \Psi_{\ell \ell} + \Psi_{\ell' \ell'} \right)  \text{tr}(\Psi) - \frac{1}{m^3} \text{tr}(\Psi^2)  \\
&+ \frac{1}{m^3}    \left\{ \left[ \text{tr}(\Psi)  \right]^2 + \Psi_{\ell \ell} \Psi_{\ell' \ell'} + \Psi_{\ell \ell}^2 \right\} \\
&\leq \frac{2}{m}   \Psi_{\ell \ell}  \Psi_{\ell' \ell'}
+ \frac{1}{m^2} \left( \Psi_{\ell \ell} + \Psi_{\ell' \ell'} \right)  \text{tr}(\Psi)
+  \frac{1}{m^3}    \left\{ \left[ \text{tr}(\Psi)  \right]^2 + \Psi_{\ell \ell} \Psi_{\ell' \ell'} + \Psi_{\ell \ell}^2 \right\}.
\end{align*}
 Note that
 \begin{align*}
 \sum_{\ell=1}^n   \sum_{\ell'=1, \ell' \neq \ell}^{n}
   \Psi_{\ell \ell}  \Psi_{\ell' \ell'}  U_{\ell u}^2 U_{\ell' v}^2
&\leq   \| \Psi \|_2^2, \\
 \sum_{\ell=1}^n   \sum_{\ell'=1, \ell' \neq \ell}^{n}  \Psi_{\ell \ell} U_{\ell u}^2 U_{\ell' v}^2   &\leq    \| \Psi \|_2, \\
 \sum_{\ell=1}^n   \sum_{\ell'=1, \ell' \neq \ell}^{n}  \Psi_{\ell' \ell'} U_{\ell u}^2 U_{\ell' v}^2   &\leq    \| \Psi \|_2, \\
 \sum_{\ell=1}^n   \sum_{\ell'=1, \ell' \neq \ell}^{n}  \Psi_{\ell \ell}^2 U_{\ell u}^2 U_{\ell' v}^2   &\leq    \| \Psi \|_2^2, \\
\sum_{\ell=1}^n   \sum_{\ell'=1, \ell' \neq \ell}^{n}   U_{\ell u}^2 U_{\ell' v}^2   &\leq    1.
\end{align*}
Using these to obtain
\begin{align*}
&\sum_{\ell=1}^n   \sum_{\ell'=1, \ell' \neq \ell}^{n}
 \mathbb{E} \left[ q_{n,m}^2(\ell, \ell')  \right] U_{\ell u}^2 U_{\ell' v}^2 \\
&\leq
\frac{2}{m}  \| \Psi \|_2^2
+ \frac{2}{m^2} \| \Psi \|_2  \text{tr}(\Psi)
+  \frac{1}{m^3}    \left\{ \left[ \text{tr}(\Psi)  \right]^2 + 2 \| \Psi \|_2^2  \right\}.
\end{align*}
 Now
\begin{align*}
& \sum_{\ell=1}^n   \sum_{\ell'=1, \ell' \neq \ell}^{n}
   \Psi_{\ell \ell}  \Psi_{\ell' \ell'}  U_{\ell u} U_{\ell' v} U_{\ell' u} U_{\ell v} \\
&=   \sum_{\ell=1}^n \Psi_{\ell \ell}  U_{\ell u}  U_{\ell v} \left[ \sum_{\ell'=1, \ell' \neq \ell}^{n} \Psi_{\ell' \ell'} U_{\ell' v} U_{\ell' u} \right] \\
&= \sum_{\ell=1}^n \Psi_{\ell \ell}  U_{\ell u}  U_{\ell v} \left\{  [U^T \Psi U]_{uv}  -  \Psi_{\ell \ell} U_{\ell v} U_{\ell u} \right\} \\
&\leq   [U^T \Psi U]_{uv}^2.
\end{align*}
Similarly,
\begin{align*}
 \sum_{\ell=1}^n   \sum_{\ell'=1, \ell' \neq \ell}^{n}  \Psi_{\ell \ell} U_{\ell u} U_{\ell' v} U_{\ell' u} U_{\ell v}   &\leq   [U^T \Psi U]_{uv}  [U^T U]_{uv}, \\
 \sum_{\ell=1}^n   \sum_{\ell'=1, \ell' \neq \ell}^{n}  \Psi_{\ell' \ell'} U_{\ell u} U_{\ell' v} U_{\ell' u} U_{\ell v}   &\leq    [U^T \Psi U]_{uv}  [U^T U]_{uv}, \\
 \sum_{\ell=1}^n   \sum_{\ell'=1, \ell' \neq \ell}^{n}  \Psi_{\ell \ell}^2 U_{\ell u} U_{\ell' v} U_{\ell' u} U_{\ell v}   &\leq    [U^T \Psi^2 U]_{uv}  [U^T U]_{uv}, \\
\sum_{\ell=1}^n   \sum_{\ell'=1, \ell' \neq \ell}^{n}   U_{\ell u} U_{\ell' v} U_{\ell' u} U_{\ell v}   &\leq    [U^T U]_{uv}.
\end{align*}
Then
\begin{align*}
&\sum_{\ell=1}^n   \sum_{\ell'=1, \ell' \neq \ell}^{n}
 \mathbb{E} \left[ q_{n,m}(\ell, \ell') q_{n,m}(\ell', \ell) \right] U_{\ell u} U_{\ell' v} U_{\ell' u} U_{\ell v} \\
&\leq
\frac{2}{m}  [U^T \Psi U]_{uv}^2
+ \frac{2}{m^2} [U^T \Psi U]_{uv}  [U^T U]_{uv} \\
&+  \frac{1}{m^3}    \left\{ \left[ \text{tr}(\Psi)  \right]^2 [U^T U]_{uv} + [U^T \Psi U]_{uv}^2 +  [U^T \Psi^2 U]_{uv}  [U^T U]_{uv} \right\}.
\end{align*}
Combing the results above together gives
\begin{align*}
&\mathbb{E} \left[ T_{2, uv}^2 \right]  \\
&\leq
\frac{2}{m}  \| \Psi \|_2^2
+ \frac{2}{m^2} \| \Psi \|_2  \text{tr}(\Psi)
+  \frac{1}{m^3}    \left\{ \left[ \text{tr}(\Psi)  \right]^2 + 2 \| \Psi \|_2^2  \right\} \\
&+ \frac{2}{m}  [U^T \Psi U]_{uv}^2
+ \frac{2}{m^2} [U^T \Psi U]_{uv}  [U^T U]_{uv} \\
&+  \frac{1}{m^3}    \left\{ \left[ \text{tr}(\Psi)  \right]^2 [U^T U]_{uv} + [U^T \Psi U]_{uv}^2 +  [U^T \Psi^2 U]_{uv}  [U^T U]_{uv} \right\}.
\end{align*}
Thus,
\begin{align*}
& \sum_{u=1}^d \sum_{v=1}^d \mathbb{E} \left[ T_{2, uv}^2 \right]  \\
&\leq
\frac{2 d^2}{m}  \| \Psi \|_2^2
+ \frac{2 d^2}{m^2} \| \Psi \|_2  \text{tr}(\Psi)
+  \frac{d^2}{m^3}    \left\{ \left[ \text{tr}(\Psi)  \right]^2 + 2 \| \Psi \|_2^2  \right\} \\
&+ \frac{2}{m}  \text{tr}(\Psi^2)
+ \frac{2}{m^2} \text{tr}(\Psi) +  \frac{1}{m^3}    \left\{ d \left[  \text{tr}(\Psi)  \right]^2  + 2 \text{tr}(\Psi^2)  \right\}.
\end{align*}

Choose $A (\Psi, m,n)$ as in \eqref{A-choice}.
Then
\begin{align*}
&\mathbb{P} \left( \left\| U^T \Pi^T \Pi \Psi \Pi^T \Pi U -  U^T A (\Psi, m,n) U \right\|_2 > \epsilon \right)  \\
&\leq
\epsilon^{-2} \mathbb{E} \left[ \left\| U^T \Pi^T \Pi \Psi \Pi^T \Pi U -  U^T A (\Psi, m,n) U \right\|_F^2  \right] \\
&\leq 2 \epsilon^{-2}  \sum_{u=1}^d \sum_{v=1}^d \mathbb{E} \left[ T_{1, uv}^2 + T_{2, uv}^2 \right]   \\
&=  2 \epsilon^{-2}  \bigg\{ \frac{2 d^2 (m-1)}{m^2} \text{tr}(\Psi^2)
+ \frac{2 d^2}{m^2} \| \Psi \|_2  \text{tr}(\Psi)
+  \frac{d^2}{m^3}    \left\{ \left[ \text{tr}(\Psi)  \right]^2 + 2 \| \Psi \|_2^2  \right\} \\
&\;\;\;\;\;\;\;\;\;\;\;\; + \frac{2}{m}  \text{tr}(\Psi^2)
+ \frac{2}{m^2} \text{tr}(\Psi) +  \frac{1}{m^3}    \left\{ d \left[  \text{tr}(\Psi)  \right]^2  + 2 \text{tr}(\Psi^2)  \right\} \bigg\}.
\end{align*}
\emph{Q.E.D.}

\newpage
\begin{table}[ht]
\caption{Assessment of JL Lemma: $n=20,000$, $d=5$.}
\label{tbl:table1}

\begin{center}

\begin{tabular}{l|lll|lllll|l} \hline
$m$ & \multicolumn{3}{c|}{Random Sampling} & \multicolumn{5}{|c}{Random Projections} \\ \hline
  & \textsc{rs1}   & \textsc{rs2} &  \textsc{rs3} & \textsc{rp1}  & \textsc{rp2}  & \textsc{rp3} & \textsc{rp4}   & \textsc{cs} & \textsc{lev}\\ \hline
Normal
   & \multicolumn{6}{c}{Norm approximation} \\ \hline
161 & 0.627 &  0.624 &  0.538 &  0.628 &  0.633 &  0.631 &  0.640 &  0.642 &  0.757 \\
   322 & 0.801 &  0.792 &  0.700 &  0.790 &  0.795 &  0.795 &  0.800 &  0.793 &  0.909 \\
   644 & 0.931 &  0.931 &  0.871 &  0.926 &  0.929 &  0.927 &  0.931 &  0.928 &  0.982 \\
   966 & 0.978 &  0.972 &  0.932 &  0.971 &  0.974 &  0.974 &  0.975 &  0.972 &  0.997 \\
  1288 & 0.990 &  0.987 &  0.973 &  0.990 &  0.991 &  0.989 &  0.990 &  0.991 &  1.000 \\
  2576 & 1.000 &  1.000 &  0.998 &  1.000 &  1.000 &  1.000 &  1.000 &  1.000 &  1.000 \\
      \hline
   & \multicolumn{6}{c}{Eigenvalue distortion} \\ \hline
161 & 0.189 &  0.191 &  0.191 &  0.189 &  0.187 &  0.188 &  0.189 &  0.188 &  0.158 \\
   322 & 0.126 &  0.128 &  0.127 &  0.127 &  0.127 &  0.128 &  0.126 &  0.129 &  0.105 \\
   644 & 0.082 &  0.085 &  0.084 &  0.086 &  0.084 &  0.085 &  0.085 &  0.086 &  0.071 \\
   966 & 0.065 &  0.067 &  0.065 &  0.067 &  0.066 &  0.067 &  0.067 &  0.068 &  0.055 \\
  1288 & 0.055 &  0.056 &  0.055 &  0.056 &  0.056 &  0.056 &  0.055 &  0.055 &  0.045 \\
  2576 & 0.033 &  0.036 &  0.033 &  0.036 &  0.037 &  0.036 &  0.035 &  0.037 &  0.029 \\

  \hline \hline
Exponential   & \multicolumn{6}{c}{Norm approximation} \\ \hline

161 & 0.432 &  0.429 &  0.402 &  0.627 &  0.624 &  0.636 &  0.628 &  0.637 &  0.717 \\
   322 & 0.580 &  0.578 &  0.548 &  0.796 &  0.795 &  0.794 &  0.800 &  0.791 &  0.875 \\
   644 & 0.747 &  0.738 &  0.717 &  0.925 &  0.930 &  0.929 &  0.930 &  0.928 &  0.972 \\
   966 & 0.851 &  0.840 &  0.812 &  0.971 &  0.968 &  0.973 &  0.969 &  0.972 &  0.992 \\
  1288 & 0.899 &  0.894 &  0.866 &  0.990 &  0.988 &  0.989 &  0.991 &  0.989 &  0.998 \\
  2576 & 0.986 &  0.974 &  0.975 &  1.000 &  1.000 &  1.000 &  1.000 &  1.000 &  1.000 \\

   \hline
   & \multicolumn{6}{c}{Eigenvalue distortion} \\ \hline
161 & 0.263 &  0.257 &  0.259 &  0.188 &  0.193 &  0.188 &  0.190 &  0.188 &  0.158 \\
   322 & 0.176 &  0.177 &  0.175 &  0.126 &  0.128 &  0.127 &  0.127 &  0.127 &  0.104 \\
   644 & 0.116 &  0.118 &  0.116 &  0.084 &  0.083 &  0.083 &  0.082 &  0.085 &  0.069 \\
   966 & 0.090 &  0.094 &  0.090 &  0.066 &  0.067 &  0.066 &  0.065 &  0.065 &  0.055 \\
  1288 & 0.076 &  0.079 &  0.075 &  0.055 &  0.055 &  0.055 &  0.054 &  0.055 &  0.045 \\
  2576 & 0.048 &  0.052 &  0.048 &  0.036 &  0.036 &  0.037 &  0.035 &  0.036 &  0.030 \\

  \hline

  \hline

\end{tabular}
\end{center}

\end{table}


   \newpage
\begin{table}[ht]
  \caption{Monte Carlo Experiments: Properties of Combining Sketches, $n=1e6$}
\label{tbl:table3}
\begin{center}
    $K=3$
    
\begin{tabular}{ll|llll||llllll}
$m$ & $J$ &
\textsc{rs1} & \textsc{srht} & \textsc{cs}  & \textsc{lev} &
\textsc{rs1} &    \textsc{srht} & \textsc{cs} & \textsc{lev}\\\hline
&& \multicolumn{4}{c}{$\hat\beta_3$ with $ \beta_3=1.0$}  &
  \multicolumn{4}{c}{se($\hat\beta_3$)} 
 \\  \hline 
 500 &      1 &0.999 &1.002 &0.999 &1.000 &0.046 &0.044 &0.045 &0.039 \\
 500 &      5 &0.999 &1.000 &1.000 &1.001 &0.021 &0.020 &0.020 &0.018 \\
 500 &     10 &1.000 &1.000 &1.000 &1.000 &0.014 &0.015 &0.015 &0.012 \\
 1000 &      1 &1.000 &0.999 &0.998 &1.000 &0.032 &0.032 &0.031 &0.026 \\
 1000 &      5 &1.000 &1.000 &1.000 &1.000 &0.014 &0.015 &0.015 &0.012 \\
 1000 &     10 &1.000 &0.999 &1.000 &1.000 &0.010 &0.010 &0.010 &0.008 \\
 2000 &      1 &1.001 &1.000 &1.001 &1.000 &0.021 &0.022 &0.022 &0.019 \\
 2000 &      5 &1.000 &1.000 &1.000 &1.000 &0.010 &0.010 &0.010 &0.008 \\
 2000 &     10 &1.000 &1.000 &1.000 &1.000 &0.007 &0.007 &0.007 &0.006 \\
 5000 &      1 &1.000 &1.001 &0.999 &1.000 &0.014 &0.014 &0.014 &0.012 \\
 5000 &      5 &1.000 &1.000 &0.999 &1.000 &0.006 &0.006 &0.006 &0.005 \\
 5000 &     10 &1.000 &1.000 &1.000 &1.000 &0.005 &0.005 &0.004 &0.004 \\
\hline
& &\multicolumn{4}{c}{Size} & \multicolumn{4}{c}{Power, $\beta_3=0.98$}\\ \hline
500 &      1 &0.050 &0.040 &0.062 &0.063 &0.081 &0.069 &0.071 &0.104 \\
500 &      5 &0.035 &0.029 &0.021 &0.045 &0.114 &0.115 &0.123 &0.185 \\
500 &     10 &0.039 &0.051 &0.053 &0.037 &0.276 &0.258 &0.265 &0.345 \\
1000 &      1 &0.048 &0.044 &0.050 &0.052 &0.101 &0.101 &0.085 &0.113 \\
1000 &      5 &0.024 &0.046 &0.032 &0.023 &0.221 &0.218 &0.233 &0.320 \\
1000 &     10 &0.041 &0.035 &0.044 &0.042 &0.461 &0.454 &0.452 &0.617 \\
2000 &      1 &0.045 &0.052 &0.058 &0.055 &0.136 &0.142 &0.147 &0.189 \\
2000 &      5 &0.034 &0.022 &0.035 &0.025 &0.436 &0.432 &0.451 &0.545 \\
2000 &     10 &0.040 &0.043 &0.038 &0.053 &0.763 &0.761 &0.767 &0.902 \\
5000 &      1 &0.053 &0.046 &0.040 &0.047 &0.298 &0.322 &0.275 &0.399 \\
5000 &      5 &0.026 &0.018 &0.026 &0.019 &0.835 &0.832 &0.829 &0.930 \\
5000 &     10 &0.045 &0.046 &0.036 &0.054 &0.987 &0.993 &0.989 &0.999 \\
\hline \hline
  \hline
\end{tabular}
\end{center}

\begin{center}
    $ K=9$

\begin{tabular}{ll|llll||llll}
    $m$ & $J$ &
\textsc{rs1} & \textsc{cs}   &
\textsc{rs1} &  \textsc{cs} &
\textsc{rs1} & \textsc{cs}   &
\textsc{rs1} &  \textsc{cs} \\ \hline
& &
\multicolumn{2}{c}{Size} &
\multicolumn{2}{c|}{Power} &
\multicolumn{2}{c}{$\hat\beta_9$}  &
\multicolumn{2}{c}{se($\hat\beta_9$)}
   \\ \hline

500 &      1 &0.049 &0.067 &0.076 &0.079 &0.999 &0.998 &0.046 &0.047 \\
500 &      5 &0.038 &0.029 &0.129 &0.120 &1.000 &1.000 &0.021 &0.020 \\
500 &     10 &0.032 &0.039 &0.241 &0.268 &0.999 &1.000 &0.014 &0.015 \\
1000 &      1 &0.052 &0.041 &0.099 &0.087&1.000 &1.000 &0.032 &0.031 \\
1000 &      5 &0.036 &0.027 &0.219 &0.214&1.000 &1.000 &0.014 &0.014 \\
1000 &     10 &0.033 &0.042 &0.461 &0.484&1.000 &1.000 &0.010 &0.010 \\
2000 &      1 &0.043 &0.050 &0.143 &0.128&1.000 &0.999 &0.022 &0.022 \\
2000 &      5 &0.025 &0.028 &0.411 &0.400&1.000 &1.000 &0.010 &0.010 \\
2000 &     10 &0.041 &0.044 &0.782 &0.773&1.000 &1.000 &0.007 &0.007 \\
5000 &      1 &0.051 &0.057 &0.260 &0.292&0.999 &1.000 &0.014 &0.015 \\
5000 &      5 &0.021 &0.037 &0.839 &0.813&1.000 &1.000 &0.006 &0.007 \\
5000 &     10 &0.033 &0.044 &0.988 &0.990&1.000 &1.000 &0.005 &0.005 \\
\hline
\end{tabular}
\end{center}

\end{table}

\newpage
\begin{table}
  \caption{Inference Conscious Choice of $m$}
  \label{tbl:table4}
\begin{center}
  $n=1e7, r=10, K=10, m_0=1000$, $\bar\alpha=0.05$

  \begin{tabular}{ll|llllll} \hline
&& \multicolumn{5}{c}{$(\beta_1^0-\beta_{10})$} \\     \hline
$\bar\gamma$ & $\sigma_e$  & .005 & .01 & .015 & .02 & .025   \\ \hline
  0.50 &0.50 &   29686 &   7421 &   3298 &   1855 &  1187 \\
  0.80 &0.50 &   67837 &  16959 &   7537 &   4240 &  2713 \\
  0.90 &0.50 &   93965 &  23491 &  10441 &   5873 &  3759 \\
  0.50 &1.00 &   98296 &  24574 &  10922 &   6143 &  3932 \\
  0.80 &1.00 &  224620 &  56155 &  24958 &  14039 &  8985 \\
  0.90 &1.00 &  311136 &  77784 &  34571 &  19446 & 12445 \\
  0.50 &3.00 &  981128 & 245282 & 109014 &  61321 & 39245 \\
  0.80 &3.00 &  2242020 & 560505 & 249113 & 140126 & 89681 \\
  0.90 &3.00 &  3105562 & 776391 & 345062 & 194098 &124222 \\

\hline
  \end{tabular}
\end{center}

\end{table}

\newpage
\begin{table}[ht]
\caption{Example of \citet{belenzon-chatterji-daley:17}}
\label{tbl:table_EPO}
\begin{center}
\begin{tabular}{lcccc} \hline
 & (1) & (2) & (3) & (4) \\
Uniform Sampling & Full Sample & $m_1$ & $m_2(m_1)$ & $m_3$ \\ \hline
& &  $K\log(n K)$ & $(\bar\alpha,\bar\gamma)=(.05,.8)$ & $\tau_2(\infty)=5$ \\ \hline
 &  &  &  \\
Dummy for eponymous & 0.031** & 0.035** & 0.031** & 0.031** \\
 & (0.001) & (0.010) & (0.003) & (0.002) \\
ln(assets) & 0.005** & 0.000 & 0.006** & 0.007** \\
 & (0.001) & (0.004) & (0.001) & (0.001) \\
 ln(no. shareholders) & -0.032** & -0.025** & -0.030** & -0.031** \\
 & (0.001) & (0.007) & (0.002) & (0.002) \\
Equity dispersion & -0.012** & -0.007 & -0.016** & -0.017** \\
 & (0.001) & (0.011) & (0.003) & (0.003) \\
 &  &  &  \\
Omitted Covariates &  & 132 & 58 & 54 \\ 
 &  &  &  \\
Observations & 562,170 & 8,158 & 112,355 & 139,022  
 \\ \hline
\end{tabular}

\[
\]

\begin{tabular}{lcccc} \hline
 & (1) & (2) & (3) & (4) \\
Countsketch & Full Sample & $m_1$ & $m_2$ & $m_3$ \\ \hline
 &  &  &  \\
Dummy for eponymous & 0.031** & 0.035** & 0.030** & 0.032** \\
 & (0.001) & (0.011) & (0.003) & (0.003) \\
ln(assets) & 0.005** & 0.009** & 0.004** & 0.005** \\
 & (0.001) & (0.003) & (0.001) & (0.001) \\
ln(no. shareholders) & -0.032** & -0.028** & -0.032** & -0.032** \\
 & (0.001) & (0.008) & (0.002) & (0.002) \\
Equity dispersion & -0.012** & -0.024 & -0.010* & -0.011** \\
 & (0.001) & (0.014) & (0.004) & (0.004) \\
&  &  &  \\
Omitted Covariates &  & 2 & 1 & 1 \\ 
 &  &  &  \\
Observations & 562,170 & 8,147 & 112,347 & 139,015
 \\ \hline
 \\
\multicolumn{4}{l}{ Robust standard errors in parentheses} \\
\multicolumn{4}{l}{ ** p$<$0.01, * p$<$0.05} \\
\end{tabular}

\end{center}

\end{table}

\clearpage
\newpage
\baselineskip=12.0pt
\bibliography{bigdata,metrics,metrics2,macro,factor}

\end{document}